\documentclass{article}

\usepackage[utf8]{inputenc}
 
\title{Sections and Chapters}
\author{Gubert Farnsworth}
\date{ }
\usepackage{arxiv}
\usepackage{cite}
\usepackage{amsmath,amssymb,amsfonts}
\usepackage{algorithmic}
\usepackage{graphicx}
\usepackage{textcomp}
\usepackage{xcolor}
\usepackage{tabulary}
\usepackage{csquotes}

\newtheorem{remark}{Remark}
\newtheorem{example}{Example}

\newtheorem{lemma}{Lemma}
\usepackage{framed}
\usepackage{nomencl} 
\usepackage{adjustbox}
\usepackage[utf8]{inputenc} % allow utf-8 input
\usepackage[T1]{fontenc}    % use 8-bit T1 fonts
\usepackage{hyperref}       % hyperlinks
\usepackage{url}            % simple URL typesetting
\usepackage{booktabs}       % professional-quality tables
\usepackage{amsfonts}       % blackboard math symbols
\usepackage{nicefrac}       % compact symbols for 1/2, etc.
\usepackage{microtype}      % microtypography
\usepackage{lipsum}
\usepackage{graphicx}
\usepackage{accents}
\newtheorem{theorem}{Theorem}
\newtheorem{prop}{Proposition}
\graphicspath{ {./images/} }
\newcommand*{\dt}[1]{%
   \accentset{\mbox{\large\bfseries .}}{#1}}

\def\BibTeX{{\rm B\kern-.05em{\sc i\kern-.025em b}\kern-.08em
    T\kern-.1667em\lower.7ex\hbox{E}\kern-.125emX}}

\usepackage{xcolor}
\def\BibTeX{{\rm B\kern-.05em{\sc i\kern-.025em b}\kern-.08em
    T\kern-.1667em\lower.7ex\hbox{E}\kern-.125emX}}

\title{
Towards a Constructive Framework for Stabilization and Control of Nonlinear Systems: Passivity and Immersion (P\&I) Approach
}

\author{
 Syed Shadab Nayyer, S. R. Wagh, and N. M. Singh \\
  Control and Decision Research Center (CDRC), EED,\\ VJTI, Mumbai, India 
  \texttt{sisyed\_p19@ee.vjti.ac.in} \\
}

\begin{document}
\maketitle
\begin{abstract}
The varied and complex dynamics of systems encountered in the real world challenge the formulation of a systematic strategy for designing a stabilizing feedback law. Thus far, the control strategies formulated to handle this problem are specific to the systems with a suitable structure rather than a general approach. In view of this, the paper attempts to sketch the outline of a proposed theory which provides a stepping stone to be general enough. Keeping it central to the development of the stabilizing feedback law, it is applicable wherever possible for a general class of systems in standard structured and unstructured forms discussed in the literature. The foundation behind this generalized theory of controller design utilizes the idea of an invariant target manifold giving rise to a non-degenerate two form, through which the definition of certain passive outputs and storage functions leads to a generation arises of control law for stabilizing the system. Because the above ideas connect with the immersion design policy and passivity theory of controller design, the developed methodology is labeled as the “Passivity and Immersion based approach” (P\&I).  The rigorously solved examples demonstrate how the diversified design paradigms can be unified in the proposed  P\&I methodology.  Additionally, the target dynamics are set to achieve the parameter-independent implicit manifolds, passive output, and associated storage function in order to mitigate the impacts of the parametric changes in the system.
\end{abstract}

% keywords can be removed
\tableofcontents
\section{Introduction}
Stabilization problems intend to build a control system that stabilizes the states of a closed-loop system around a specified equilibrium point \cite{khalil2002nonlinear}. 
%Most of the analysis, synthesis, and stabilization problems in nonlinear systems (NLS) are tackled using the theory of Control Lyapunov Functions (CLF) \cite{freemanbook2008robust}.  The feedback control law can be designed for stabilization and control of a broader class of nonlinear systems using identified CLF as stated by \cite{IanManchester} with comparatively straightforward computations.
Being relatively simple and straightforward computations, as stated in \cite{IanManchester, khalil2002nonlinear}, the theory of Control Lyapunov Functions (CLF) has become most preferred for the analysis, synthesis, and stabilization problems in Nonlinear systems (NLS). 
%The Lyapunov functions are substituted with storage functions in systems with Lagrangian or Hamiltonian structures, with passivity being the desired feature \cite{ortega1998passivity}.  The passivity-based control is restricted to systems with a relative degree of one.
%Despite the CLF's widespread acceptance and success, there are a few reasons to consider an alternate framework, especially, two of them are emphasized as, first, lack of universally applicable and computationally efficient methods for determining a CLF and second,  limitations posed because of the CLF structured around a certain equilibrium point.  Considering various performance scenarios, such designed CLF  at a particular equilibrium point may turn out to be inadequate to stabilize the system for a wide range of specified trajectories.
However, despite the widespread acceptance and success of CLF,  it is essential to consider an alternate framework because of: %some of its drawbacks, especially, two of them are emphasized as:
\begin{enumerate}
    \item  the lack of universally applicable and computationally efficient methods for determining a CLF \cite{Sepulchre2011ConstructiveNC, IanManchester}.
    \item the limitations posed by structuring the CLF around a certain equilibrium point.
    \item a simple  change of variables transforms the CLF criteria to a linear matrix inequality (LMI) for Linear Systems (LS); however, when applied to NLS, the set of CLFs is not always convex or even related \cite{IanManchester}.
\end{enumerate}
%The classical Backstepping (BS) methodology \cite{krstic1995nonlinear, JoyKokotovic} for the global stabilization of nonlinear feedback systems enforces the CLF for design of feedback control law. It enables a way of constructing a series of virtual control law at each step.
%To enable a way of constructing a series of virtual control law at each step, The classical Backstepping (BS) methodology \cite{krstic1995nonlinear, JoyKokotovic} enforces the CLF to design the final control law.
As an extended version based on Lyapunov theory, the classical Backstepping  \cite{krstic1995nonlinear, Sepulchre2011ConstructiveNC} is enforced to develop a series of virtual control laws at each step by imposing the identified CLF for the global stabilization of nonlinear feedback systems.
Instead of the global asymptotic stability, the concept of incremental input-to-state stability and contraction analysis is adopted in \cite{zamaniCDC, ZamaniTAC} for the development of the BS methodology. The basic cornerstone of Contraction analysis is the variational dynamics along solutions of an NLS.  The existence of a control contraction metric (CCM) for an NLS is sufficient for universal exponential stabilizability.  
%According to contraction theory, if a system has the Incremental Lyapunov Function (ILF) and Contraction Metrics (CM) of the differential state, all NLS trajectories converge incrementally and exponentially to one single trajectory, irrespective of the initial conditions \cite{tsukamoto2021contraction}. 
However, deriving evaluation criteria and a systematic form of CCM for general NLS is complicated and challenging. The Forwarding approach (i.e., the procedure starts from the input and proceeds forward) \cite{Sepulchre2011ConstructiveNC, Mazenc}  to stabilize the systems in upper-triangular form and is complementary to the BS procedure.  Moreover, computing derivatives of virtual inputs and its successive calculations in BS lead to an explosion of complexity \cite{PANcomposite}, making computer involvement inevitable \cite{Issuesbackstepping}.

The above-mentioned control design approaches 
%(like Feedback linearization, Backstepping, Forwarding, Contraction theory, and Sum-of-Squares Optimization (SOS)) 
are best suited to the well-known class of systems which are available in strict-feedback and strict-feedforward structures  \cite{Sepulchre2011ConstructiveNC}. Several alternative approaches to the above methods based on Lyapunov constructions for cascade systems have been developed by treating a cascade system as a collection of subsystems. The issue, in general, is that while each subsystem may be independently stable, their interconnectedness might cause the entire cascade to become unstable.
%However, due to the complexity of the mathematical calculations required in the computation of the resultant nonlinear control law for stabilization, the involvement of a computer to execute the control signal calculation is inevitable \cite{Issuesbackstepping}. Moreover, the virtual derivatives computation and successive derivations of virtual control inputs in BS, on the other hand, cause the explosion of complexity \cite{PANcomposite}.
%The stabilization of a class of systems in feedforward form is addressed using the classical Forwarding approach, as explained in \cite{Sepulchre2011ConstructiveNC, Mazenc}.
%The Forwarding approach (i.e., the procedure starts from the input and proceeds forward)  to stabilize the systems in upper-triangular form is complementary to the BS procedure. The author in \cite{teel2} proposes a Forwarding algorithm based on the nested saturation for a class of feedforward systems, and it involves a hierarchy of time scales and invariant manifolds. Finding an explicit solution to a partial differential equation (PDE), which may be difficult or even impossible to solve, is a crucial stage in the Forwarding approach \cite{Praly}. 

The I\&I methodology proposed in \cite{Astolfi} is a relatively recent strategy to develop the controllers for nonlinear systems without the requirement of a CLF in the control law design phase. However, the last crucial stage of manifold attractivity in the classical I\&I method lacks a straightforward and systematic procedure, which hampers the implementation of the classical I\&I in practical applications. In I\&I HCP \cite{WangTAC}, the variational dynamics are employed to get the prolonged closed-loop system by using an infinitesimal displacement $\delta \mathrm{x}$ at a fixed time. The contraction theory is applied by substituting the Finsler Lyapunov Function (FLF) for the ILF and a positive definite matrix $\Xi(\mathrm{x})$ for the CM. However, the selection of $\Xi(\mathrm{x})=\Theta ^{\mathrm{T}}(\mathrm{x})\Theta (\mathrm{x})$ (refer Ex. 3 in \cite{WangTAC}) for the FLF is randomly chosen and not defined properly. Moreover, solving a matrix inequality requires certain lengthy and tedious calculations.

According to contraction theory, if a system has the Incremental Lyapunov Function (ILF) and Contraction Metrics (CM) of the differential state, all NLS trajectories converge incrementally and exponentially to one single trajectory, irrespective of the initial conditions \cite{tsukamoto2021contraction, nayyer2022passivitylcss, nayyer2022passivity}. The CM and associated Lyapunov function for systems with suitable structures, such as Lagrangian systems and feedback linearizable systems, can be determined analytically. There are several strategies for identifying CM for general NLS that exploit the LMI aspect. The authors in \cite{IanManchester}  proposed an approach for finding the CM using SOS programming via the solver Mosek and the parser YALMIP. The optimization-based approach is proposed in \cite{TsukamotoJacobi}. It is formulated that the Hamilton-Jacobi inequality for the finite-gain $\pounds_2$ stability condition can be expressed as an LMI when contraction theory is equipped with the extended linearity of the  State-Dependent Coefficient (SDC) formulation. The State Dependent Riccati Equation (SDRE) approach proposed in \cite{SDC} employs the SDC parameterization of NLS for feedback control. A convex optimization-based Steady-state Tracking Error Minimization (CV-STEM) is proposed in \cite{TsukamotoJacobi, Tsukamotocvtem} to identify a CM that minimizes an upper bound of the steady-state distance between unperturbed and perturbed system trajectories. However, deriving evaluation criteria and a systematic form of CM for general NLS is an arduous, tedious, and long-drawn-out process.

\textit{Statement of contributions:} To overcome the above-mentioned issues and challenges, a constructive and systematic strategy with more apparent degrees of freedom to achieve the stabilization and control of the NLS is proposed. The major contribution of the paper is the development of the P\&I approach for accommodating  systems in structured and unstructured forms (systems that are not in a suitable structure). To make this happen \textbf{the design starts with blending the classical I\&I with the concept of the generation of a suitable passive output and corresponding storage function. Moreover, the involvement of  a particular structure of tangent bundle associated with a suitable pseudo-Riemannian structure imposed on the control system makes the P\&I more methodical and systematic}.  By choosing the implicit manifold suitably, the P\&I controller performance is rendered independent of the system parameter. The suitable off-the-manifold  and associated pseudo-Riemannian metric are free from the matched and extended matched parametric uncertainties. As a result, the performance of the closed-loop system with the proposed P\&I controller remains unaffected even after setting all the parameter values in the controller to zero. 

\textbf{As the proposed P\&I approach is an extended and enhanced version of \cite{mehra2017control}, this technique can be easily applied to electro-mechanical under-actuated systems
matching conditions". In the above-mentioned paper \cite{mehra2017control}, the Riemannian metric is obtained from the concept of kinetic energy.  The ideas of kinetic and potential shaping are formulated in terms of the passive outputs and the desired controller is obtained using standard passivity-based techniques. In this manuscript, a concept of the pseudo-Riemannian metric is utilized instead of the Riemannian metric. A pseudo-Riemannian manifold  generalization of a Riemannian manifold in which the requirement of positive definiteness is relaxed. The obtained pseudo-Riemannian metric helps in getting the passive output and corresponding storage function.} Furthermore,  the idea of the proposed theory is extended to show that the proposed P\&I approach can also be applied to solve a new problem,
that is, the generation of attractive periodic solutions. The only modification required is in the definition of the target
dynamics that, instead of having an asymptotically stable equilibrium, should be chosen with attractive periodic
orbits.

\textit{Organization of paper:} Section \ref{appen2} proceeds with the recent stabilization and control methodologies (I\&I approach and CCM) and their issues. Based on their issues, the control objective and problem formulation are stated in Section \ref{sec2}. Section \ref{sec3} explores the detailed procedure of the proposed P\&I approach for the stabilization and control of NLS. The contribution provides a more systematic P\&I control scheme with more apparent degrees of freedom to achieve the stabilization and control objective. The idea of the proposed P\&I approach is applied to the system in strict feedback form and the results are shown in Section \ref{sec4}. As an alternative approach to the methods based on the generation of CCM, the applicability of the proposed P\&I approach to systems without any particular structure is demonstrated in Section \ref{sec5}. Section \ref{sec6} shows the applicability and limitation of existing control approach as well the proposed P\&I approach for the systems in strict feedforward form and Interlaced systems.
Section \ref{sec7} extends the idea of applying the P\&I approach for orbital stabilization followed by the Conclusion in Section \ref{conclusion}. 
 
\textbf{Notations:} If $\gamma (\mathrm{x}):\mathbb{R}^n\rightarrow \mathbb{R}$  is continuously differentiable in $\mathrm{x}\in $, then $\triangledown_\mathrm{x}\gamma:\mathbb{R}^n\rightarrow \mathbb{R}^n$ is the gradient of $\gamma$ w.r.t $\mathrm{x}$.
$\mathrm{I^{k \times k}}$ denotes a $\mathrm{{k \times k}}$ identity matrix. The transpose of a matrix $\mathrm{A}$ is defined as $\mathrm{A}^T$.
\section{Problem formulation}\label{sec2}
The main objective is to design control law systematically to \textit{stabilize and control} the physical systems. Consider an NLS
\begin{equation}\label{shshshasdasf}
     \dt{\mathrm{x}}=\mathrm{f}(\mathrm{x})+\mathrm{g} (\mathrm{x})u
\end{equation} with inputs $\mathrm{u}\in \mathbb{U}\subseteq \mathbb{R}^m$, structured and unstructured form, and $\mathrm{x}=\begin{bmatrix}
\mathrm{x}_1 &  \mathrm{x}_2& .. & \mathrm{x}_{\mathrm{n}} \\
\end{bmatrix}^\mathrm{T}\in \mathbb{X}\subseteq \mathbb{R}^{\mathrm{n}}$. 

The stabilization problem for a system (\ref{shshshasdasf}) intends to build a control system via feedback law $\mathrm{u_{P\&I}}=\mathrm{u}(\mathrm{x})$ that stabilizes the states of a closed-loop system around an equilibrium point $\mathrm{x}^*=0$. The control objective is to design a feedback control law $\mathrm{u_{P\&I}}=\mathrm{k}(\mathrm{x})$ that globally asymptotically converges the system states to the desired one i.e., $\mathrm{x}\rightarrow\mathrm{x_{d}}$. Numerous techniques are proposed in the literature for the stabilization of control of NLS. Some of them are briefly overviewed in the introduction section and above subsections

The issues and limitations of the recent stabilization procedures are outlined below:
\begin{itemize}
    \item Computing virtual derivatives and successive derivations of virtual inputs in BS lead to an explosion of complexity \cite{PANcomposite}, making computer involvement inevitable \cite{Issuesbackstepping}.
    \item The objective behind the CTCLF for control law design is to ensure the stability of the entire cascade system. The main concern is that even though each subsystem may be stable on its own, the interconnection of these subsystems might render the entire cascade unstable.
    \item There is no organized way for developing the feedback law in step ($S_4$) when utilizing the classical I\&I approach for practical applications \cite{WangCDC, WangCDC}.
    \item The step $(S_4)$ is replaced by a horizontal contraction-based design to ensure the attractivity of the manifold $\mathrm{M}$ in the I\&I HCP. But, the selection of $\Xi(\mathrm{x})=\Theta ^{\mathrm{T}}(\mathrm{x})\Theta (\mathrm{x})$  for the FLF is not defined properly and selected randomly. Moreover, solving a matrix inequality requires certain lengthy and tedious calculations. Therefore, the I\&I HCP \cite{WangCDC} necessitates various time-consuming and complex computations with inequality solutions.
    \item Deriving evaluation criteria and a systematic form of CCM for general NLS is complicated and challenging.
\end{itemize}

\begin{remark}
In the existing literature, different methodologies have been proposed for the stabilization and control of NLS. But, they are provided for a particular system structure. A systematic and universal method for developing a stabilizing feedback law for real-world systems is challenging due to:
\begin{enumerate}
    \item the varied and complex dynamics of systems and 
    \item the unavailability of system dynamics in a proper structured form.
\end{enumerate}
\end{remark}

Concerning the issues and challenges associated with the aforementioned recent stabilization procedures, a constructive and systematic strategy with more apparent degrees of freedom to achieve the stabilization and control of the NLS is proposed.  The proposed P\&I approach is based on the choice of an appropriate implicit manifold and the generation of a suitable passive output and a related storage function. Moreover, the proposed P\&I approach does not involve the SOS programming utilizing the solver Mosek and the parser YALMIP. The nature of $\mathrm{f}(\mathrm{x})$ and $\mathrm{g}(\mathrm{x})$ may not be in an appropriate structure. They can be in any form. Therefore, this paper attempts to sketch the outline of a proposed theory which provides a stepping stone to be general enough.

\section{The Proposed P\&I Approach}\label{sec3}
The proposed P\&I approach is based on
\begin{itemize}
    \item an immersion of a dynamical system into a stable system using a preserving mapping and 
    \item the generation of a suitable passive output and corresponding storage function.
\end{itemize}  
This involves a particular PR structure imposed on the tangent bundle associated with the control system.
\subsection{Main Result}\label{main esult}
Given a single input system
\begin{align}\label{feedbackoriginalsystem}
\begin{split}
\dt{\mathrm{x}}=\mathrm{f(\mathrm{x}, \lambda)}\hspace{1.5cm}
\dt{\lambda}=\mathrm{u} 
\end{split} 
\end{align}
with $(\mathrm{x}, \lambda)\in (\mathbb{R}^{n-1},\mathbb{R})$ and without any particular structure. 
\begin{remark}
The dynamics (\ref{feedbackoriginalsystem}) is chosen  for illustration of the proposed methodology. This method can be generalized for any system which is affine in $\mathrm{u}$. 
\end{remark}
 %The proposed P\&I approach ensures NLS stabilization by executing the four steps $(S_1-S_4)$ outlined below. 
The target system, the manifold invariance condition, and the implicit manifold condition—i.e., the three essential steps of the classical I\&I approach—are merged in step $(S_1)$. 
The notion of tangent space and passivity theory is invoked to propose the P\&I approach in the following three steps $(S_2-S_4)$ to ensure the \textit{manifold attractivity}.
 
$(S_1)$ \textbf{Construction of the implicit manifold:}
Assume that there exist 
$\beta:\mathbb{R}^{\mathrm{h}}\rightarrow \mathbb{R}^{\mathrm{h}}$, $\pi: \mathbb{R}^{\mathrm{h}}\rightarrow \mathbb{R}^n$,   $\Psi:\mathbb{R}^{\mathrm{n}}\rightarrow \mathbb{R}^{\mathrm{n-h}}$
with the $\mathrm{h < n}$ such that the following statement hold.
%To propose a systematic approach by enlarging and incorporating the idea of Immersion and Invariance (I\&I) via the concept of implicit manifold and its structure
%For a given system,  it is assumed that there exists a $\mathrm{C}^{\infty}$  mapping $\varphi(\mathrm{x}):\mathbb{R}^{\mathrm{n}}\rightarrow \mathbb{R}$ such that the subsystem $\dt{\mathrm{x}}=\mathrm{f(\mathrm{x}, \varphi (\mathrm{\mathrm{x}}))}$ has a GES/GAS equilibrium at the origin. 

The target dynamics $\dt{\eta}=\beta(\eta)$  with $\mathrm{x}=\eta$ is defined such that the subsystem $\dt{\mathrm{x}}=\mathrm{f(\mathrm{x}, \varphi (\mathrm{\mathrm{x}}))}$ for $\mathrm{C}^{\infty}$  mapping $\varphi(\mathrm{x}):\mathbb{R}^{\mathrm{n}}\rightarrow \mathbb{R}$ has a GES/GAS equilibrium at the origin by considering the relationship $\mathrm{\lambda}=\varphi (\mathrm{\mathrm{x}})$. This defines the implicit manifold  $\Psi(\mathrm{x}, \lambda )=\mathrm{\lambda}-\varphi (\mathrm{\mathrm{x}})=0$, the implicit manifold $\mathbb{M}=\left \{(\mathrm{x},\lambda) \in \mathbb{R}^{\mathrm{n-1}}\times\mathbb{R} | \Psi(\mathrm{x}, \lambda):=\lambda-\varphi(\mathrm{x})=0 \right \}$, and $\pi(\eta)=\mathrm{col}(\eta, \varphi(\mathrm{\eta}))$.
%\begin{itemize}
%    \item it is assumed that there exists a $\mathrm{C}^{\infty}$  mapping $\varphi(\mathrm{x}):\mathbb{R}^{\mathrm{n}}\rightarrow \mathbb{R}$ such that the subsystem $\dt{\mathrm{x}}=\mathrm{f(\mathrm{x}, \varphi (\mathrm{\mathrm{x}}))}$ has a GES/GAS equilibrium at the origin.
%    \item the target dynamics $\dt{\zeta}=\beta(\zeta)=\mathrm{f(\zeta, \varphi(\zeta))}$  with $\mathrm{x}=\zeta$ is defined by considering the relationship $\mathrm{\lambda}=\varphi (\mathrm{\mathrm{x}})$. This defines the implicit manifold (off-the-manifold) $\Psi(\mathrm{x}, \lambda )=\mathrm{\lambda}-\varphi (\mathrm{\mathrm{x}})=0$ and $\pi(\zeta)=\mathrm{col}(\zeta, \varphi(\mathrm{\zeta}))$.
%\end{itemize}
 %In order to ensure the \textit{manifold attractivity and trajectory boundedness} in a systematic way without any assumption and consideration, the amalgamation of concept of tangent space and passivity theory is invoked to propose P\&I approach in next three steps $(S_2-S_4)$.
\begin{lemma}
The implicit manifold $\mathbb{M}$ is invariant.
\end{lemma}
\textbf{Proof:} As the relationship $\mathrm{\lambda}=\varphi (\mathrm{\mathrm{x}})$ is considered above, the manifold $\varphi (\mathrm{\mathrm{x}})-\mathrm{\lambda}=0$ is written to get 
\begin{equation}\label{gradphi}
    \triangledown \Psi(\mathrm{x}, \lambda)=\begin{bmatrix}
\frac{\partial \varphi(\mathrm{x})}{\partial \mathrm{x}} & -1
\end{bmatrix}
\end{equation}
i.e., normal to the velocity vector field  
$\begin{bmatrix}
\dt{\mathrm{x}}\\ 
\dt{\lambda}
\end{bmatrix}$. Thus, the product of (\ref{gradphi}) and the velocity vector field gives $\frac{\partial \varphi(\mathrm{x})}{\partial \mathrm{x}} \dt{\mathrm{x}}-\dt{\lambda}=0\Rightarrow \dt{\varphi}({\mathrm{x}})-\dt{\varphi}({\mathrm{x}})=0$. It means the velocity vector field is always tangent to $\mathbb{M}$. Hence, $\mathbb{M}$ is  invariant.

{$(S^{\dagger\dagger}_2)$}  \textbf{Tangent space structure for control systems:}
Consider an n-dimensional manifold $\mathbb{M}$ with tangent bundle $\mathbb{T}_{\mathbb{M}}$, such that all $\mathrm{p}\in {\mathbb{M}}$, $\mathbb{T}_{\mathrm{p}}{\mathbb{M}}$ has the following structure
\begin{equation}
    \mathbb{T}_{\mathrm{p}}{\mathbb{M}} = \mathbb{H}_{\mathrm{p}} \oplus  \mathbb{V}_{\mathrm{p}}: \hspace{0.3cm}  \mathbb{H}_{\mathrm{p}} \cap   \mathbb{V}_{\mathrm{p}}=0
\end{equation}
where $\mathbb{H}_{\mathrm{p}}$ is the horizontal space and $\mathbb{V}_{\mathrm{p}}$ is the vertical space.  
\begin{equation}\label{TpMm}
  \mathrm{Then} \hspace{0.9cm} \mathbb{T}_{\mathrm{p}}{\mathbb{M}}= \mathbb{H}_{\mathrm{p}} \oplus  \mathbb{V}_{\mathrm{p}}=(\dt{\mathrm{x}}, 0)\oplus (0, \dt{\lambda})=(\dt{\mathrm{x}},\dt{\lambda})
\end{equation} 
is written for given system (\ref{feedbackoriginalsystem}) at any point $\mathrm{p}\in\mathbb{M}$. With the implicit manifold $\Psi (\mathrm{x} ,\lambda )$ obtained in $S_1$, the normal vector direction is given by $\triangledown \Psi (\mathrm{x} ,\lambda )$. Thus, a PR metric $\mathrm{R}$ on space $\mathbb{T}_{\mathrm{p}}{\mathbb{M}}$ can be defined as
\begin{small}
\begin{align}\label{metric}
    \begin{split}
        \mathrm{R}&=\triangledown \Psi (\mathrm{x} ,\lambda )^{\mathrm{T}}\triangledown \Psi (\mathrm{x} ,\lambda )\\&=\begin{bmatrix}
\left ( \frac{\partial \varphi}{\partial \mathrm{x}}\right )^{\mathrm{T}}\left ( \frac{\partial \varphi}{\partial \mathrm{x}}\right ) & \left ( -\frac{\partial \varphi}{\partial \mathrm{x}}\right )^{\mathrm{T}}\\ 
-\left ( \frac{\partial \varphi}{\partial \mathrm{x}}\right ) & \mathrm{I}
\end{bmatrix}=\begin{bmatrix}
\mathrm{\mathrm{m}}_{11} & \mathrm{\mathrm{m}}_{12}\\ 
 \mathrm{\mathrm{m}}_{21}&\mathrm{\mathrm{m}}_{22}
\end{bmatrix}.
    \end{split}
\end{align}
\end{small}
which is intuitively a natural choice. To proceed with the passive output, the obtained implicit manifold and a  PR  metric $\mathrm{R}$ are utilized in the  splitting of tangent space [further details in Appendix]. The off-the-manifold dynamics can now be viewed as evolving in the tangent space with metric as $\left ( \mathbb{T}_{\mathbb{M}},\mathrm{R} \right )$.
%\begin{remark}
%In order the get the passive output, the metric $\chi$ is replaced with PR metric $\mathrm{R}$=$\triangledown \Psi (\mathrm{x} ,\lambda )^{\mathrm{T}}\triangledown \Psi (\mathrm{x} ,\lambda )$ as a natural choice. 
%The metric $\mathrm{R}$ is obtained using $\triangledown \Psi (\mathrm{x} ,\lambda )^{\mathrm{T}}\triangledown \Psi (\mathrm{x} ,\lambda )$ as $\triangledown \Psi (\mathrm{x} ,\lambda )$ is full rank. 
%\end{remark}
\begin{remark}
To obtain the passive output, the metric $\chi$ is replaced with semi-Riemannian metric $\mathrm{R}$ as a natural choice. (Details and proof can be found in Appendix)
%The metric $\mathrm{R}$ is obtained using $\triangledown \Psi (\mathrm{x} ,\lambda )^{\mathrm{T}}\triangledown \Psi (\mathrm{x} ,\lambda)$. 
\end{remark}
For $(\mathbb{M}, \mathrm{R})$, the splitting is visualized as follows:
\begin{small}
\begin{equation}
     (\dt{\mathrm{x}},\dt{\lambda})=\left (\dt{\mathrm{x}}, -\mathrm{m}_{22}^{-1}\mathrm{m}_{21} \dt{\mathrm{x}} \right )\oplus \left (0, \dt{\lambda}+\mathrm{m}_{22}^{-1}\mathrm{m}_{21} \dt{\mathrm{x}} \right ) =\widetilde{\mathbb{H}}_{\mathrm{p}} \oplus \widetilde{\mathbb{V}}_{\mathrm{p}}
\end{equation}
\end{small}

As $\dt{\lambda}$  is along the vertical direction, the passive output is chosen as a component $\dt{\lambda}+\mathrm{m}_{22}^{-1}\mathrm{m}_{21} \dt{\mathrm{x}}$ which  is in the same direction or parallel to $\dt{\lambda}$ \cite{ShadabArxiv}.  Roughly speaking, the idea is to bring the component $\dt{\lambda}+\mathrm{m}_{22}^{-1}\mathrm{m}_{21}$ of $\widetilde{\mathbb{H}}_{\mathrm{p}}$ to the component $\mathrm{m}_{22}^{-1}\mathrm{m}_{21} \dt{\mathrm{x}}$ of $\widetilde{\mathbb{V}}_{\mathrm{p}}$.
\begin{figure}[ht!]
    \centering
    \includegraphics[width=\linewidth]{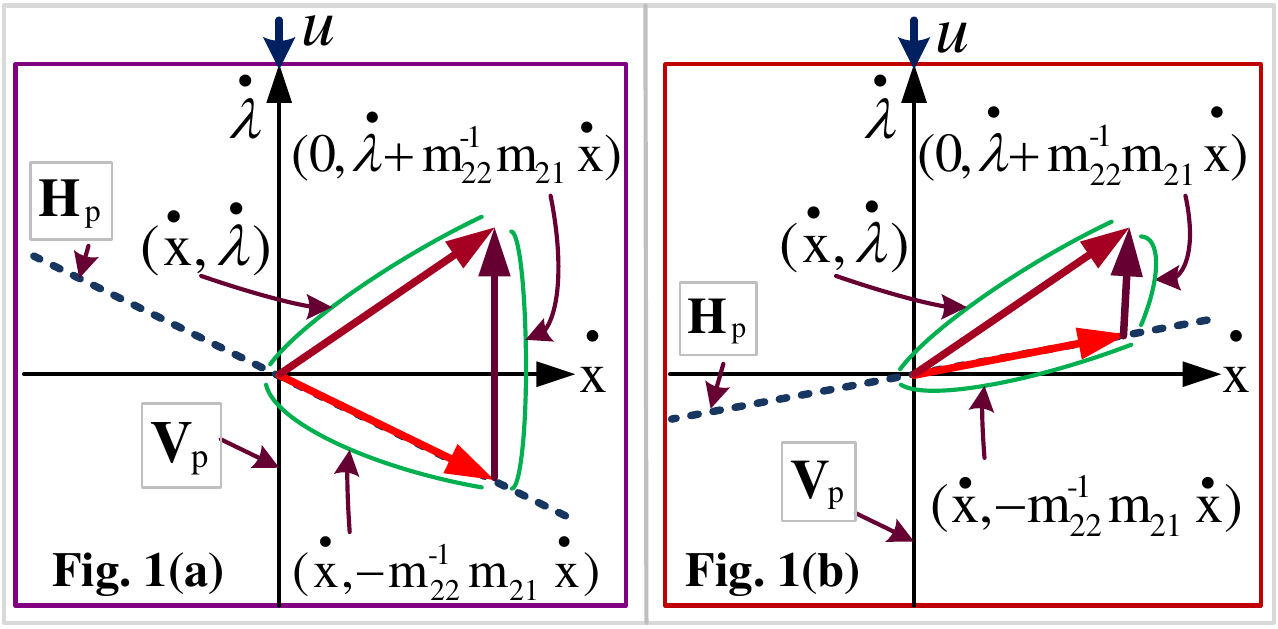}
    \caption{Geometrical interpretation: vertical vector $\mathbb{V}_{\mathrm{p}}$ is along the fiber direction and $\mathbb{H}_{\mathrm{p}} \oplus  \mathbb{V}_{\mathrm{p}}=\mathbb{T}_{\mathrm{p}}{\mathbb{M}}$. }
    \label{Geometrical interpretation}
\end{figure}
The geometrical interpretation for the splitting tangent vector is shown in Fig. \ref{Geometrical interpretation}. The different possibilities of splitting tangent vector are shown in Fig. \ref{Geometrical interpretation}(a) and Fig. \ref{Geometrical interpretation}(b). The discussion about the passive output and storage function is provided in Appendix.

\textbf{$(S^{\dagger\dagger}_3)$  Passive output:}
The component of $\mathrm{u}$ tangent vector along $\dt{\lambda}$  is used to define the \textit{passive output} 
%$\mathrm{y}_1={\int_{0}^{t}\dt{\lambda}\hspace{0.08cm}\mathrm{dt}}$ and $\mathrm{y}_2=
%{\int_{0}^{t}(\mathrm{m}_{22}^{-1}\mathrm{m}_{21}\dt{\mathrm{x}})\hspace{0.08cm}\mathrm{dt}}$
\begin{equation}\label{GammaPassive}
    \mathrm{y}=\mathrm{y}_1+\mathrm{y}_2=\int_{0}^{t}\dt{\lambda}\hspace{0.08cm}\mathrm{dt}
+{\int_{0}^{t}(\mathrm{m}_{22}^{-1}\mathrm{m}_{21}\dt{\mathrm{x}})\hspace{0.08cm}\mathrm{dt}}
\end{equation}
with the help of passivity theory. Here, $\mathrm{y}_1={\int_{0}^{t}\dt{\lambda}\hspace{0.08cm}\mathrm{dt}}$ and $\mathrm{y}_2={\int_{0}^{t}(\mathrm{m}_{22}^{-1}\mathrm{m}_{21}\dt{\mathrm{x}})\hspace{0.08cm}\mathrm{dt}}$ are defined. If $\mathrm{m}_{22}^{-1}\mathrm{m}_{21}$ is a constant then $\mathrm{y}=({\lambda}+\mathrm{m}_{22}^{-1}\mathrm{m}_{21}{\mathrm{x}})$ is defined. If $\mathrm{m}_{22}^{-1}\mathrm{m}_{21}$ is a function of $\mathrm{x}$ then, it can be written as the gradient of any function $\mathrm{q(x)}$ i.e.,  $\mathrm{m}_{22}^{-1}\mathrm{m}_{21}\mathrm{(x)}=\triangledown \mathrm{q(x)}$. Then
\begin{equation}\label{Gamma}
    \mathrm{y}=\int_{0}^{t}(\dt{\lambda}+\triangledown \mathrm{q(x)} \dt{\mathrm{x}})\mathrm{dt}=(\lambda+\mathrm{q(x)})
\end{equation}
\begin{remark}
\textbf{The condition  $\mathrm{m}_{22}^{-1}\mathrm{m}_{21}\mathrm{(x)}=\triangledown \mathrm{q(x)}$ is related to the condition of integrability and integrable connection in differential geometry.} 
\end{remark}
\textbf{In some cases, the systems will no longer have a closed-form expression. In such cases, finding the implicit manifold and passive output may be difficult or even impossible in different approaches including the P\&I method to stabilize the systems in upper triangular form.}

\begin{figure}[ht!]
    \centering
    \includegraphics[width=0.8\linewidth]{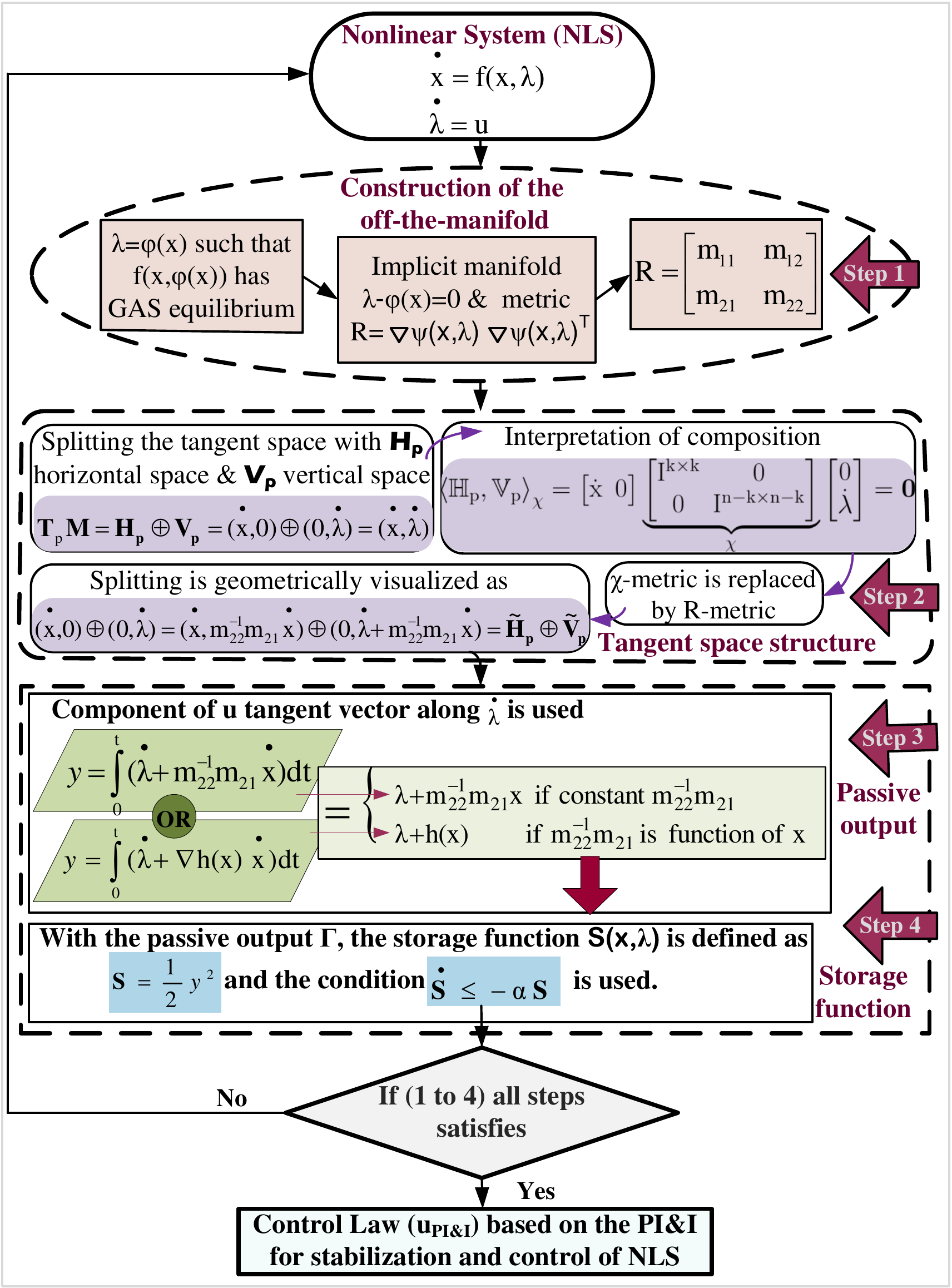}
    \caption{Flowchart depicting the proposed P\&I approach for stabilization and control of NLS in systematic and simplified steps.}
    \label{flowchart}
\end{figure}

$(S^{\dagger\dagger}_4)$ \textbf{Storage function:}
%\begin{remark}
%The interpretation is that the off-the-manifold dynamics converges to the implicit manifold exponentially where, the rate of convergence is decided by $\alpha$.
%\end{remark}
With $\mathrm{y}$, the candidate Lyapunov function $\mathbb{S}(\mathrm{x}, \lambda)$ (i.e., storage function) is defined as
\begin{equation}\label{Storage function}
  \mathbb{S}(\mathrm{x}, \lambda)=\frac{1}{2}{\mathrm{y}}^2=\frac{1}{2}(\lambda+\mathrm{q(x)})^2.
\end{equation}
The convergence of the off-the-manifold dynamics to the implicit manifold at an exponential rate $\alpha$ is accompanied by selecting the condition
\begin{equation}\label{exponential laypunov}
    \dt{\mathbb{S}}\leq -\alpha\mathbb{S}.
\end{equation}
One can use the condition (\ref{exponential laypunov}) along with the storage function (\ref{Storage function}) and passive output (\ref{Gamma}) to get
\begin{equation}\label{exponential laypunov solution}
    (\lambda+\mathrm{q(x)})(\dt{\lambda}+\frac{\partial \mathrm{q(x)}}{\partial \mathrm{x}}\dt{\mathrm{x}})\leq -\frac{\alpha}{2}(\lambda+\mathrm{q(x)})^2
\end{equation}
\begin{equation}\label{sjksdkc}
 \Rightarrow (\dt{\lambda}+\frac{\partial \mathrm{q(x)}}{\partial \mathrm{x}}\dt{\mathrm{x}})+\frac{\alpha}{2}(\lambda+\mathrm{q(x)})=0
\end{equation}
%\begin{equation}
 %   u+\frac{\alpha}{2}(\lambda+\mathrm{h(x)})+\frac{\partial \mathrm{h(x)}}{\partial \mathrm{x}}\dt{\mathrm{x}}=0
%\end{equation}
The equation (\ref{sjksdkc}) is modified by substituting (\ref{feedbackoriginalsystem}) in terms of the final control law as
\begin{equation}\label{final control law}
  \mathrm{u}=-\frac{\alpha}{2}\lambda-\frac{\alpha}{2}\mathrm{q(x)}-\frac{\partial\mathrm{q(x)}}{\partial \mathrm{x}}\mathrm{f(\mathrm{x}, \lambda)}.
\end{equation}
The above-defined control law ensures the GAS equilibrium point of the system to zero/origin.
\begin{theorem}
The proposed P\&I based control law (\ref{final control law}) ensures the GAS/GES equilibrium point of the system (\ref{feedbackoriginalsystem}) to zero/origin.
\end{theorem}
\textbf{Proof:}  The storage function obtained as (\ref{Storage function}) can be interpreted as the Lyapunov function
\begin{equation}
   \mathrm{\mathbb{S}(x)}=\frac{1}{2}\sum_{i=0}^{n-1}(\mathrm{x}_{i+1}+\mathrm{q_i(x)})^2=\frac{1}{2}\sum_{i=0}^{n-1}(\mathrm{x}_{i+1}-\varphi_i(\mathrm{x}))^2
\end{equation}
 with $\mathrm{q_i(x)}=-\varphi_i(\mathrm{x})$ mentioned in \cite{ZamaniTAC}. As $(\mathrm{x}, \lambda)\in (\mathbb{R}^{\mathrm{n}-1},\mathbb{R})$, it is noted that $\mathrm{x_n}=\lambda$ is used.  The condition $\dt{\mathbb{S}}\leq -\alpha\mathbb{S}$ of the exponential stability is utilized to obtain the proposed P\&I based control law. The obtained control law (\ref{final control law}) ensures the manifold attractivity of the invariant manifold $\Psi(\mathrm{x}, \lambda)=0$, and the dynamics  on the 
 said invariant manifold converges exponentially to the equilibrium point. Thus, one can say that the global attractivity of the manifold and exponential stability of the dynamics on the manifold ensures GAS/GES of the equilibrium point for the system (\ref{feedbackoriginalsystem}).
The complete stepwise procedure for stabilization of NLS via the proposed P\&I approach is shown in Fig. \ref{flowchart}.

\begin{remark}
In the rest of this paper, the
controller of Subsection \ref{main esult} will be referred to as the P\&I approach.
\end{remark}
\subsection{P\&I approach applicability  with Precursory Examples}
The last crucial stage of manifold attractivity in the classical I\&I method lacks a straightforward and systematic procedure, which hampers the implementation of the classical I\&I in practical applications. The matrix $\Xi(\mathrm{x})=\Theta ^{\mathrm{T}}(\mathrm{x})\Theta (\mathrm{x})$ (refer Ex. 3 in \cite{WangTAC})  is randomly chosen and not defined properly in I\&I HCP \cite{WangTAC}. Moreover, the I\&I HCP \cite{WangCDC} necessitates various time-consuming and complex computations with inequality solutions. Therefore, the control law is obtained in a systematic and methodical way by the proposed P\&I approach.
\subsubsection{P\&I approach for the systems with suitable structure}
\begin{example}\label{Example 1}
Consider a two-dimensional NLS  
\begin{align}\label{ex1_system}
    \begin{split}
         \dt{\mathrm{x}}_1=-\mathrm{x}_1+\theta_a \mathrm{x}_1^3\mathrm{x}_2\hspace{1.5cm}
          \dt{\mathrm{x}}_2=\mathrm{u}
    \end{split}
\end{align}
\end{example}
from \cite{Astolfi, WangTAC} with known $\theta_a>0$, $\mathrm{x}:=\mathrm{col}(\mathrm{x}_1, \mathrm{x}_2)\in \mathbb{R}^2$ , and $\mathrm{u} \in \mathbb{R}$.  
To start with $S_1$,  $\varphi (\mathrm{x}_1)=\mathrm{x}_2=-\mathrm{x}_1^2$ is selected to get stable subsystem
\begin{equation}
    \dt{\mathrm{x}}_1=-\mathrm{x}_1-\theta_a \mathrm{x}_1^5
\end{equation}
and it leads to the selection of target dynamics $\dt{\eta}=\beta(\eta)=-\eta-\theta_a\eta^5$ with $\mathrm{x}_1=\eta$. The target dynamics is defined on the implicit manifold 
 \begin{equation}
   \Psi (\mathrm{x}, \lambda)= \Psi (\mathrm{x}_1, \mathrm{x}_2)= \mathrm{x}_2+\mathrm{x}_1^2=0.
\end{equation}
As per $(S^{\dagger\dagger}_2)$, $\mathrm{R}(\mathrm{x})=  \triangledown \Psi(\mathrm{x}_1, \mathrm{x}_2)^{\mathrm{T}} \triangledown \Psi(\mathrm{x}_1, \mathrm{x}_2) $
\begin{small}
\begin{equation}
   =\begin{bmatrix}
2\mathrm{x}_1\\ 
1
\end{bmatrix}\begin{bmatrix}
2\mathrm{x}_1&
1
\end{bmatrix}=\begin{bmatrix}
4\mathrm{x}_1^2 &2\mathrm{x}_1 \\ 
 2\mathrm{x}_1& 1
\end{bmatrix}
\end{equation}
\end{small}
is defined. Here $\mathrm{m}_{21}=2\mathrm{x}_1$ with $\mathrm{m}_{22}=1$ gives $\triangledown \mathrm{q(x)}=2\mathrm{x}_1$ and $\mathrm{q(x)}=\mathrm{x}_1^2$. As $\dt{\mathrm{x}}_2$ is along the vertical direction, the passive output $\dt{\mathrm{x}}_2+\mathrm{m}_{22}^{-1}\mathrm{m}_{21}\dt{\mathrm{x}}_1=\dt{\mathrm{x}}_2+2\mathrm{x}_1\dt{\mathrm{x}}_1$ is
chosen as a component  which is in the same direction or parallel to $\dt{\mathrm{x}}_2$. Then by the preliminary definition of passive output in  $(S^{\dagger\dagger}_3)$,  $\mathrm{y}={\mathrm{y}}_1+{\mathrm{y}}_2$ with ${\mathrm{y}}_1=\int_{0}^{t}(\triangledown_{ \mathrm{x}_1}\Psi(\mathrm{x})\dt{\mathrm{x}}_1)\mathrm{dt}=\mathrm{x}_1^2$ and ${\mathrm{y}}_2=\int_{0}^{t}(\triangledown_{\mathrm{x}_2}\Psi(\mathrm{x}) \dt{\mathrm{x}}_2\mathrm{dt})=\mathrm{x}_2$ is chosen using (\ref{Gamma}).

With the passive output ${\mathrm{y}}$, the storage function  
\begin{equation}
    \mathbb{S}(\mathrm{\mathrm{x}_1}, \mathrm{\mathrm{x}_2})=\frac{1}{2}(\mathrm{x}_1^2+\mathrm{x}_2)^2
\end{equation}
from $(S^{\dagger\dagger}_4)$ and (\ref{Storage function}) is defined. The obtained P\&I control law from (\ref{final control law}) lead to final control law \begin{equation}\label{secondp&Ilaw}
   \mathrm{u_{P\&I}}=-\frac{1}{2}(\alpha-4)\mathrm{x_1}^2-\frac{1}{2}\alpha\mathrm{x_2}-2\theta_a\mathrm{x_1}^4\mathrm{x_2}.
\end{equation}
\begin{remark}\label{eg1remark}
The $\mathrm{u_{P\&I}}$ (\ref{secondp&Ilaw}) and the I\&I HCP-based control law (refer Ex. 4 of \cite{WangTAC}) obtained for Example \ref{Example 1} are exactly same. %Similarly, the $\mathrm{u_{P\&I}}$ (\ref{wanglabel}) is exactly same as the control law obtained via I\&I HCP approach for $\alpha=2$ (refer Ex. 3 of \cite{WangTAC}). The trajectory boundedness and GAS equilibrium of the CLS (system (\ref{ex1_system}) with (\ref{secondp&Ilaw}) and system (\ref{Wangsystem3}) with (\ref{wanglabel})) are already proved in \cite{WangTAC}.
%For the same control law, the trajectory boundedness and GAS of origin of the system (\ref{Wangsystem3}) in closed-loop with (\ref{wanglabel}) is already proved. Hence, the system  (\ref{Wangsystem3})  with the proposed P\&I control law (\ref{wanglabel}) has GAS equilibrium.
\end{remark}
\begin{example}\label{example W2}
Consider an academic example
\begin{align}\label{Wangsystem2}
    \begin{split}
        \dt{\mathrm{x}}_1 &=-\mathrm{x}_1+\mathrm{x}_1^2+\mathrm{x}_1\mathrm{x}_2 \\
\dt{\mathrm{x}}_2 &=\mathrm{x}_3\\
\dt{\mathrm{x}}_3& =\mathrm{u}
    \end{split}
\end{align}
\end{example}
with ${\mathrm{x}}^*=0$ as an equilibrium. Here, the stabilization is easily achieved by considering the first-order target dynamics as well as second-order dynamics.  The sequential approach ensures the GES of an equilibrium by using first-order dynamics. However, the use of second-order dynamics facilitates the system's equilibrium point to be GAS.

To start with $S_1$, the first-order target dynamics $\dt{\eta}=\beta(\eta)=-\eta$ with $\mathrm{x}_1=\eta$ is selected to get globally exponentially stable subsystem
\begin{equation}
    \dt{\mathrm{x}}_1=-\mathrm{x}_1.
\end{equation}
under certain condition.
%This provides the relation/virtual control law $\varphi_1 (\mathrm{x}_1)=\mathrm{x}_2=-\mathrm{x}_1$. 
The target dynamics is defined on the implicit manifold 
 \begin{equation}\label{om1}
    \Psi_1(\mathrm{x_1}, \mathrm{x_2})= \mathrm{x}_2+\mathrm{x}_1=0
\end{equation}
\begin{equation}\label{om2}
   \Psi_1(\mathrm{x_1}, \mathrm{x_2})= \mathrm{x}_1(\mathrm{x}_2+\mathrm{x}_1)=0.
\end{equation}
Based on these two possible implicit manifolds (\ref{om1}) and (\ref{om2}), two different cases (Case 1 and Case 2) are discussed below. These cases are presented to get clear insight about the uniqueness of implicit manifold, passive output, and storage function.

\textbf{Case 1:} In this case, the implicit manifold $ \Psi_1(\mathrm{x_1}, \mathrm{x_2})= \mathrm{x}_2+\mathrm{x}_1=0$ is considered.
As per $(S^{\dagger\dagger}_2)$, $\mathrm{R}(\mathrm{x})=  \triangledown \Psi_1(\mathrm{x}_1, \mathrm{x}_2)^{\mathrm{T}} \triangledown \Psi_1(\mathrm{x}_1, \mathrm{x}_2) $
\begin{small}
\begin{equation}
   =\begin{bmatrix}
1\\ 
1
\end{bmatrix}\begin{bmatrix}
1&
1
\end{bmatrix}=\begin{bmatrix}
1 &1 \\ 
1& 1
\end{bmatrix}
\end{equation}
\end{small}
is defined. As the system is available in proper system structure, the P\&I approach is applied recursively/sequentially to get the final control law. 

Here $\mathrm{m}_{21}=1$ and $\mathrm{m}_{22}=1$.   Then by the preliminary definition of passive output in  $(S^{\dagger\dagger}_3)$,  $\mathrm{y_1}=\int_{0}^{t}(\dt{\mathrm{x}}_2+\dt{\mathrm{x}}_1)\mathrm{dt}$  is chosen using (\ref{Gamma}).

With the passive output ${\mathrm{y_1}}$, the storage function  
\begin{equation}
    \mathbb{S}_1(\mathrm{\mathrm{x}_1}, \mathrm{\mathrm{x}_2})=\frac{1}{2}(\mathrm{x}_1+\mathrm{x}_2)^2
\end{equation}
from $(S^{\dagger\dagger}_4)$ and (\ref{Storage function}) is defined. The obtained control law from (\ref{final control law}) lead to virtual control law 
\begin{equation}\label{x3virtualcontrollaw}
   \mathrm{x_3}= \mathrm{x_1}- \mathrm{x^2_1}- \mathrm{x_1}\mathrm{x_2}-\frac{\alpha_2}{2}(\mathrm{x_1}+\mathrm{x_2}).
\end{equation}

\begin{remark}
The subsystem $\dt{\mathrm{x}}_1$ is globally exponentially stabilized  by the selection of the virtual control law $\mathrm{x}_2=-\mathrm{x}_1$ via the notion of target dynamics. This helps in \begin{itemize}
    \item defining the implicit manifold $ \Psi_1(\mathrm{x_1}, \mathrm{x_2})$ and
    \item deriving the virtual control law $\mathrm{x_3}= \mathrm{x_1}- \mathrm{x^2_1}- \mathrm{x_1}\mathrm{x_2}-\frac{\alpha_2}{2}(\mathrm{x_1}+\mathrm{x_2})$.
\end{itemize}  
\end{remark}

\begin{figure}
    \centering
    \includegraphics[width=\linewidth]{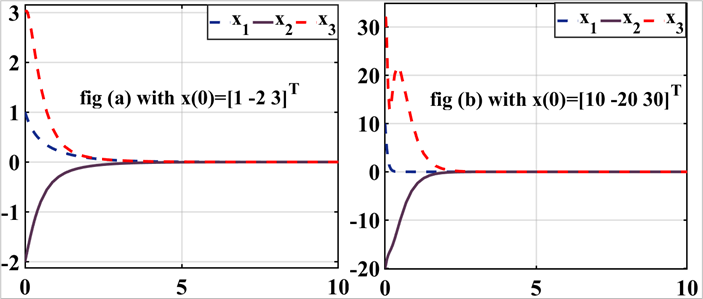}
    \caption{Figure associated with Example \ref{example W2}: Time evolution of state trajectories of system (\ref{Wangsystem2}) for different initial conditions}
    \label{Wangsyst2fig1}
\end{figure}

\begin{remark}
The obtained virtual control law $\mathrm{x}_3$ in (\ref{x3virtualcontrollaw}) renders the subsystem
\begin{align}
    \begin{split}
         \dt{\mathrm{x}}_1 &=-\mathrm{x}_1+\mathrm{x}_1^2+\mathrm{x}_1\mathrm{x}_2 \\
\dt{\mathrm{x}}_2 &=\mathrm{x}_3
    \end{split}
\end{align}
GES to an equilibrium point i.e., zero. Therefore, the goal is to get the final control law $\mathrm{u}$ that globally asymptotically stabilizes the complete system (\ref{Wangsystem2}).
\end{remark} 
This is achieved by defining the implicit manifold
\begin{equation}\label{impliManifoldaxqqs}
  \Psi_2(\mathrm{x})= \Psi_2(\mathrm{x_1}, \mathrm{x_2}, \mathrm{x_3})=\mathrm{x_3}-\mathrm{x_1}+ \mathrm{x^2_1}+ \mathrm{x_1}\mathrm{x_2}+\frac{\alpha_2}{2}(\mathrm{x_1}+\mathrm{x_2})=0
\end{equation}
using the virtual control law (\ref{x3virtualcontrollaw}).
After getting the implicit manifold $\Psi_2(\mathrm{x})$,  a PR metric $\mathrm{R}=\triangledown \Psi_2(\mathrm{x})^{\mathrm{T}}\triangledown \Psi_2(\mathrm{x})$ is defined  as
\begin{small}
\begin{equation}\label{reimanmetric}
    =\begin{bmatrix}
\mathrm{m}_{11} & \mathrm{m}_{12} &\mathrm{m}_{13} \\ 
 \mathrm{m}_{21}& \mathrm{m}_{22} & \mathrm{m}_{23}\\ 
\mathrm{m}_{31} &\mathrm{m}_{32}  & \mathrm{m}_{33}
\end{bmatrix}=\begin{bmatrix}
(-1+2\mathrm{x_1}+\mathrm{x_2}+\frac{\alpha_2}{2})^2 &(-1+2\mathrm{x_1}+\mathrm{x_2}+\frac{\alpha_2}{2})(\mathrm{x_1}+\frac{\alpha_2}{2})  &-1+2\mathrm{x_1}+\mathrm{x_2}+\frac{\alpha_2}{2}\\ 
(-1+2\mathrm{x_1}+\mathrm{x_2}+\frac{\alpha_2}{2})(\mathrm{x_1}+\frac{\alpha_2}{2}) & (\mathrm{x_1}+\frac{\alpha_2}{2})^2 &(\mathrm{x_1}+\frac{\alpha_2}{2})\\ 
 (-1+2\mathrm{x_1}+\mathrm{x_2}+\frac{\alpha_2}{2})&(\mathrm{x_1}+\frac{\alpha_2}{2})& 1
\end{bmatrix}
\end{equation}
\end{small}
to obtain $\mathrm{m}_{13}=\mathrm{m}_{31}=(-1+2\mathrm{x_1}+\mathrm{x_2}+\frac{\alpha_2}{2})$, $\mathrm{m}_{23}=\mathrm{m}_{32}=(\mathrm{x_1}+\frac{\alpha_2}{2})$, and $\mathrm{m}_{33}=1$.
The definition of passive output (\ref{GammaPassive}) 
\begin{small}
\begin{equation}
    \mathrm{y_2}=\int_{0}^{t}(\dt{\mathrm{x}}_3+\mathrm{m}_{33}^{-1}\mathrm{m}_{23}\dt{\mathrm{x}}_2+\mathrm{m}_{33}^{-1}\mathrm{m}_{31}\dt{\mathrm{x}}_1)\mathrm{dt}=\mathrm{x_3}-\mathrm{x_1}+ \mathrm{x^2_1}+ \mathrm{x_1}\mathrm{x_2}+\frac{\alpha_2}{2}(\mathrm{x_1}+\mathrm{x_2})
\end{equation}
\end{small}
from step $(S^{\dagger\dagger}_3)$ is fulfilled.
 \begin{remark}\label{remark333}
As the input $\mathrm{u}$ is available in $\dt{\mathrm{x}}_3$, the elements in metric $\mathrm{R}$ having connection with subscript $3$ is taken to define the passive output (see Fig. \ref{Geometrical interpretation}).
 \end{remark}
With the passive output $\mathrm{y_2}$, the storage function 
\begin{align}
    \begin{split}
        \mathbb{S}_2(\mathrm{x}_1,\mathrm{x}_2, \mathrm{x}_3)=&\frac{1}{2}{\mathrm{y^2_2}}=\frac{1}{2}(\mathrm{x_3}-\mathrm{x_1}+ \mathrm{x^2_1}+ \mathrm{x_1}\mathrm{x_2}+\frac{\alpha_2}{2}(\mathrm{x_1}+\mathrm{x_2}))^2
    \end{split}
\end{align}
is defined using $(S^{\dagger\dagger}_4)$ and (\ref{Storage function}). The condition $\dt{\mathbb{S}}\leq - \alpha \mathbb{S}$ from (\ref{exponential laypunov}) is utilized and extended to get the final control law. Upon substitution, the final control law
\begin{small}
\begin{equation}\label{wangsys2control lwa}
    \mathrm{u}= -\left ( (-\mathrm{x}_1+\mathrm{x}_1^2+\mathrm{x}_1\mathrm{x}_2)\left (-1+2\mathrm{x_1}+\mathrm{x_2}+\frac{\alpha_2}{2} \right)+\mathrm{x_3}\left ( \mathrm{x_1}+\frac{\alpha_2}{2} \right )+\frac{\alpha}{2}\left ( \mathrm{x_3}-\mathrm{x_1}+\mathrm{x_1}^2+\mathrm{x_1}\mathrm{x_2}+ \frac{\alpha_2}{2}(\mathrm{x_1}+\mathrm{x_2})\right )\right ).
\end{equation}
\end{small}
is obtained that ensures the convergence of system trajectory $\mathrm{x(t)}$ to an equilibrium point. The time evolution of state trajectories of system (\ref{Wangsystem2}) for different initial conditions with $\alpha_1=2$, $\alpha_2=8$, and $\alpha=12$ are shown in Fig. \ref{Wangsyst2fig1}.
Consider a coordinate transformation
\begin{align}
    \begin{split}
        \zeta=\mathbb{T}(\mathrm{x})=\begin{bmatrix}
\zeta_1\\ 
\zeta_2\\ 
\zeta_3
\end{bmatrix}=\begin{bmatrix}
\mathbb{T}_1(\mathrm{x})\\ 
\mathbb{T}_2(\mathrm{x})\\ 
\mathbb{T}_3(\mathrm{x})
\end{bmatrix} =\begin{bmatrix}
\mathrm{x_1}\\
 \mathrm{x_2}+\mathrm{x_1}\\
\mathrm{x_3}-\mathrm{x_1}+ \mathrm{x^2_1}+ \mathrm{x_1}\mathrm{x_2}+\frac{\alpha_2}{2}(\mathrm{x_1}+\mathrm{x_2})
\end{bmatrix}
    \end{split}
\end{align}
obtained with the help of the defined target dynamics and implicit manifolds. The system dynamics (\ref{Wangsystem2}) is transformed to
\begin{align}\label{sgfgiue}
    \begin{split}
        \dt{\zeta}_1=&-\zeta_1+\zeta_1\zeta_2\\
        \dt{\zeta}_2=&-\frac{\alpha_2}{2}\zeta_2+\zeta_3\\
\dt{\zeta}_3=&-\frac{\alpha}{2}\zeta_3
    \end{split}
\end{align}
using the global diffeomorphism mapping  $\zeta=\mathbb{T}(\mathrm{x})$ and the obtained P\&I based control law (\ref{wangsys2control lwa}). 
%As the linear system exhibits the GES behaviour, the system (\ref{sgfgiue}) can be made linear by setting $\zeta_1\zeta_2\approx 0$ to ensure the GES behaviour. The approximation $\zeta_1\zeta_2\approx 0$ can be done by properly selecting the values of the manipulating variables $\alpha_2$ and $\alpha$.
The system (\ref{sgfgiue}) can be made linear by setting $\zeta_1\zeta_2\approx 0$ to ensure the GES behavior since the linear system shows the GES behavior. By carefully choosing the values of the manipulating variables $\alpha_2$ and $\alpha$, it is possible to achieve $\zeta_1\zeta_2\approx 0$.
\begin{remark}
%By setting the values of the manipulating variables $\alpha_2$ and $\alpha$ properly, the exponential convergence is achieved. Even for the random values of  $\alpha_2$ and $\alpha$, the closed-loop system exhibit GAS behaviour.
The exponential convergence is achieved by appropriately adjusting the manipulating variables $\alpha_2$ and $\alpha$. Even with random $\alpha_2$ and $\alpha$ values, the closed-loop system exhibits GAS behavior.
\end{remark}

\textbf{Case 2:} In this case, the implicit manifold $ \Psi_1(\mathrm{x_1}, \mathrm{x_2})= \mathrm{x}_1(\mathrm{x}_2+\mathrm{x}_1)=0$ is considered. Based on the steps  $(S^{\dagger\dagger}_2)$,  $(S^{\dagger\dagger}_3)$, and  $(S^{\dagger\dagger}_4)$, the virtual control law
\begin{equation}\label{virtualx333}
    \mathrm{x_3}=\frac{-1}{\mathrm{x_1}}\left ( (2\mathrm{x_1}+\mathrm{x_2})(-\mathrm{x}_1+\mathrm{x}_1^2+\mathrm{x}_1\mathrm{x}_2)+\frac{\alpha_2}{2}(\mathrm{x}^2_1+\mathrm{x}_1\mathrm{x}_2) \right )
\end{equation}
is obtained. In next successive iterations, the implicit manifold  
\begin{equation}
   \Psi_2(\mathrm{x_1}, \mathrm{x_2}, \mathrm{x_3})= \mathrm{x_1}\mathrm{x_3}+\left ( (2\mathrm{x_1}+\mathrm{x_2})(-\mathrm{x}_1+\mathrm{x}_1^2+\mathrm{x}_1\mathrm{x}_2)+\frac{\alpha_2}{2}(\mathrm{x}^2_1+\mathrm{x}_1\mathrm{x}_2) \right )=0
\end{equation}
is obtained with the help of the virtual control law (\ref{virtualx333}) in order to get the final control law
\begin{small}
\begin{align}\label{cl_local}
    \begin{split}
        \mathrm{u}=&\frac{-1}{\mathrm{x}_1}\left ( \left ( -4\mathrm{x}_1+6\mathrm{x}^2_1+4\mathrm{x}_1\mathrm{x}_2-\mathrm{x}_2+2\mathrm{x}_1\mathrm{x}_2+\mathrm{x}_3+\frac{\alpha_2}{2}(2\mathrm{x}_1+\mathrm{x}_3) \right ) (-\mathrm{x}_1+\mathrm{x}_1^2+\mathrm{x}_1\mathrm{x}_2)+\left (2\mathrm{x}^2_1-\mathrm{x}_1+\mathrm{x}^2_1+\left ( \frac{\alpha_2}{2}\mathrm{x}_1  \right )   \right )\mathrm{x}_3\right)\\&-\frac{1}{\mathrm{x}_1}\left ( -2\mathrm{x}^2_1+2\mathrm{x}^3_1+2\mathrm{x}^2_1\mathrm{x}_2-\mathrm{x}_2\mathrm{x}_1+\mathrm{x}_2\mathrm{x}^2_1+\mathrm{x}^2_2\mathrm{x}_1+\mathrm{x}_1\mathrm{x}_3+\frac{\alpha}{2}(\mathrm{x}^2_1+\mathrm{x}_1\mathrm{x}_2) \right )
    \end{split}
\end{align}
\end{small}
via the systematic four step procedure of P\&I approach based on the steps $(S_1)$, $(S^{\dagger\dagger}_2)$,  $(S^{\dagger\dagger}_3)$, and  $(S^{\dagger\dagger}_4)$.

\begin{theorem}
For the system (\ref{Wangsystem2}), the obtained control law (\ref{cl_local}) with $\alpha > 0$ $\alpha_{2} > 0$ via the proposed P\&I approach renders the complete closed-loop control system (locally) exponentially stable.
\end{theorem}

\textbf{Proof:} 
Consider a coordinate transformation
\begin{align}
    \begin{split}
        \zeta=\mathbb{T}(\mathrm{x})=\begin{bmatrix}
\zeta_1\\ 
\zeta_2\\ 
\zeta_3
\end{bmatrix}=\begin{bmatrix}
\mathbb{T}_1(\mathrm{x})\\ 
\mathbb{T}_2(\mathrm{x})\\ 
\mathbb{T}_3(\mathrm{x})
\end{bmatrix} =\begin{bmatrix}
\mathrm{x_1}\\
\mathrm{x_1}(\mathrm{x_2}+\mathrm{x_1})\\
\mathrm{x_1}\mathrm{x_3}+\left ( (2\mathrm{x_1}+\mathrm{x_2})(-\mathrm{x}_1+\mathrm{x}_1^2+\mathrm{x}_1\mathrm{x}_2)+\frac{\alpha_2}{2}(\mathrm{x}^2_1+\mathrm{x}_1\mathrm{x}_2) \right )
\end{bmatrix}
    \end{split}
\end{align}
obtained with the help of the defined target dynamics and implicit manifolds. The system dynamics (\ref{Wangsystem2}) is transformed to
\begin{align}\label{sgfgiue1}
    \begin{split}
        \dt{\zeta}_1=&-\zeta_1+\zeta_2\\
        \dt{\zeta}_2=&-\frac{\alpha_2}{2}\zeta_2+\zeta_3\\
\dt{\zeta}_3=&-\frac{\alpha}{2}\zeta_3
    \end{split}
\end{align}
using the global diffeomorphism mapping  $\zeta=\mathbb{T}(\mathrm{x})$ and the obtained P\&I based control law (\ref{cl_local}). 
Being a linear system, it is obvious that the system (\ref{sgfgiue1}) is  stable with $\alpha_{2} > 0$ and $\alpha > 0$ and it assures the exponential convergence of $\zeta_{\mathrm{m}}$ for $ \mathrm{m}=1,2,3$ to zero.
\begin{remark}
The control law (\ref{cl_local}) ensures the locally exponential convergence. The controller shoots up the trajectory and causes the CLS to become unstable at $\mathrm{x_1}=0$.
\end{remark}

\textbf{An alternate approach facilitating the second-order target dynamics:}
Two cases (Case 1 and Case 2) are discussed based on the selection of the first-order target dynamics. It is possible to achieve $\zeta_1\zeta_2\approx 0$  by carefully choosing the values of the manipulating variables $\alpha_2$ and $\alpha$. Even if the proper combinations of  $\alpha_2$ and $\alpha$ are unable to achieve $\zeta_1\zeta_2\approx 0$, the complete CLS still exhibits the GAS behaviour. The control law (\ref{cl_local}) ensures the locally exponential convergence in Case 2. The controller (\ref{cl_local}) shoots up the trajectory and causes the CLS to become unstable at $\mathrm{x_1}=0$. Therefore, an alternate approach facilitating the second-order target dynamics is considered.  

For a given system (\ref{Wangsystem2}), the second-order target dynamics
\begin{align}
    \begin{split}
        &\dt{\eta}_1=-\eta_1\\
&\dt{\eta}_2=-\frac{\alpha_2}{2}\eta_2
    \end{split}
\end{align}
with ${\eta}_1=\mathrm{x_1}$ and ${\eta}_2=\mathrm{x_2}$ is considered. The target dynamics is defined on the implicit manifolds
\begin{align}
    \begin{split}
        &\Psi_1(\mathrm{x_1}, \mathrm{x_2})=\mathrm{x_2}+\mathrm{x_1}=0\\
&\Psi_2(\mathrm{x_2}, \mathrm{x_3})=\mathrm{x_3}+\frac{\alpha_2}{2}\mathrm{x_2}=0
    \end{split}
\end{align}
and it leads to implicit manifold vector $\Psi (\mathrm{x})=\mathrm{col}\left ( \underset{\Psi_1}{\underbrace{\mathrm{x_2}+\mathrm{x_1}}}, \hspace{0.09cm}\underset{\Psi_2}{\underbrace{\mathrm{x}_3+\frac{\alpha_2}{2}\mathrm{x}_2}} \right )$. The combined implicit manifold 
\begin{equation}\label{systemwangT2}
 \Phi(\mathrm{x}_1,\mathrm{x}_2,\mathrm{x}_3)= \Phi(\mathrm{x})=\Psi_1+\Psi_2=\mathrm{x}_1+\left ( 1+\frac{\alpha_2}{2} \right )\mathrm{x}_2+\mathrm{x}_3=0
\end{equation}
is obtained by adding the elements of  implicit manifold vector $\Psi (\mathrm{x})$. After getting the implicit manifold $\Psi(\mathrm{x})$,  a PR metric $\mathrm{R}=\triangledown \Phi(\mathrm{x})^{\mathrm{T}}\triangledown \Phi(\mathrm{x})$ is defined  as
\begin{equation}
    \mathrm{R}=\begin{bmatrix}
1\\ 
1+\alpha_2\\ 
1
\end{bmatrix}\begin{bmatrix}
1& 
1+\alpha_2& 
1
\end{bmatrix}=\begin{bmatrix}
1& 
1+\alpha_2& 
1 \\ 
1+\alpha_2 &(1+\alpha_2)^2  &1+\alpha_2 \\ 
1& 
1+\alpha_2& 
1
\end{bmatrix}
\end{equation}
to obtain $\mathrm{m}_{13}=\mathrm{m}_{31}=1$, $\mathrm{m}_{23}=\mathrm{m}_{32}=(1+\frac{\alpha_2}{2})$, and $\mathrm{m}_{33}=1$.
The definition of passive output (\ref{GammaPassive}) 
\begin{small}
\begin{equation}
    \mathrm{y}=\int_{0}^{t}(\dt{\mathrm{x}}_3+\mathrm{m}_{33}^{-1}\mathrm{m}_{23}\dt{\mathrm{x}}_2+\mathrm{m}_{33}^{-1}\mathrm{m}_{31}\dt{\mathrm{x}}_1)\mathrm{dt}=\Psi_1+\Psi_2=\mathrm{x}_1+\left ( 1+\frac{\alpha_2}{2} \right )\mathrm{x}_2+\mathrm{x}_3
\end{equation}
\end{small}
from step $(S^{\dagger\dagger}_3)$ is fulfilled. With the passive output $\mathrm{y_2}$, the storage function 
\begin{align}
    \begin{split}
        \mathbb{S}(\mathrm{x}_1,\mathrm{x}_2, \mathrm{x}_3)=&\frac{1}{2}{\mathrm{y^2}}=\frac{1}{2}\left ( \mathrm{x}_1+\left ( 1+\frac{\alpha_2}{2} \right )\mathrm{x}_2+\mathrm{x}_3 \right )^2
    \end{split}
\end{align}
is defined using $(S^{\dagger\dagger}_4)$ and (\ref{Storage function}). The condition $\dt{\mathbb{S}}\leq - \alpha \mathbb{S}$ from (\ref{exponential laypunov}) is utilized and extended to get the final control law. Upon substitution, the final control law
\begin{small}
\begin{equation}\label{wangsys2controlawww}
    \mathrm{u}= \mathrm{x}_1-\mathrm{x}^2_1-\mathrm{x}_1\mathrm{x}_2-\left ( 1+\frac{\alpha_2}{2} \right )\mathrm{x}_3-\frac{\alpha}{2}\left ( \mathrm{x}_1+\left ( 1+\frac{\alpha_2}{2} \right )\mathrm{x}_2+\mathrm{x}_3 \right )
\end{equation}
\end{small}
is obtained that ensures the convergence of system trajectory $\mathrm{x(t)}$ to an equilibrium point.
 The time evolution of state trajectories of system (\ref{Wangsystem2}) for different initial conditions with the control law (\ref{wangsys2controlawww}), $\alpha_2=20$, and $\alpha=10$.
In accordance with the explanation above (in Case1, Case 2, and An alternate approach facilitating second-order target dynamics), stabilization can be easily achieved by carefully choosing the target dynamics.
\begin{remark}
In general, the storage function is far from unique. Nonuniqueness itself arises from the fact that there might be various implicit manifolds depending on the choice of target dynamics.
\end{remark}
\textbf{Discussion:} The stabilization of system (\ref{Wangsystem2}) 
\begin{equation}
     \dt{\mathrm{x}}_1 =-\mathrm{x}_1+\mathrm{x}_1^2+\mathrm{x}_1\mathrm{x}_2 \hspace{1.5cm}
\dt{\mathrm{x}}_2=\mathrm{x}_3\hspace{1.5cm}
\dt{\mathrm{x}}_3 =\mathrm{u}
\end{equation}
is also achieved via the I\&I Virtual Contraction Procedure (I\&I VCP) in \cite{WangTAC} (Example 2 in \cite{WangTAC}). In I\&I VCP, the \textit{manifold attractivity and trajectory boundedness} i.e., the last crucial step of I\&I approach is ensured via virtual contraction. 

As per the approach described in \cite{WangTAC}, the second-order target dynamics 
\begin{align}
    \begin{split}
        &\dt{\eta}_1=-\eta_1\\
&\dt{\eta}_2=-\eta_2
    \end{split}
\end{align}
is specified to get the implicit manifolds
\begin{equation}
    \Psi_1(\mathrm{x_1, x_2})=\mathrm{x_2+x_1}=0 \hspace{2cm}\Psi_2(\mathrm{x_2, x_3})=\mathrm{x_2+x_3}=0
\end{equation}
and its related vector $\Psi (\mathrm{x})=\mathrm{col}(\underset{\Psi_1}{\underbrace{\mathrm{x}_1 +\mathrm{x_2}}}, \hspace{0.09cm}\underset{\Psi_2}{\underbrace{\mathrm{x}_2+\mathrm{x}_3}})$. The dynamics of the off-the-manifold
\begin{align}
    \begin{split}
        &\dt{\Psi}_1=-{\Psi}_1+\mathrm{x_1}{\Psi}_1+{\Psi}_2\\
&\dt{\Psi}_2={\Psi}_2-{\Psi}_1+\mathrm{x_1}+\mathrm{u}
    \end{split}
\end{align}
coordinate is given by 
\begin{itemize}
    \item defining each element of implicit manifold vector as a off-the-manifold coordinate and
    \item differentiating the these elements.
\end{itemize}
After defining the  dynamics of the off-the-manifold coordinate, the prolongation system 
\begin{equation}\label{inivcopcm}
    \frac{\mathrm{d} \delta \Psi}{\mathrm{dt}}=\begin{bmatrix}
-1+\mathrm{x_1} & 1\\ 
-1+\triangledown_{ \Psi_1}\mathrm{u} & 1+\triangledown_{ \Psi_2}\mathrm{u}
\end{bmatrix}\delta \Psi
\end{equation}
is computed by a concept of variational dynamics. To get the final control law, the Finsler-Lyapunov Function 
\begin{equation}
    \mathrm{V}(\mathrm{x}, \Psi, \delta \Psi):=\delta \Psi^{\mathrm{T}}\mathrm{P(\mathrm{x},\Psi)}\delta \Psi
\end{equation}
with $\mathrm{P}:\mathbb{R}^{n}\times\mathbb{R}^{n-h}\rightarrow \mathbb{R}_{>0}^{(n-h)\times (n-h)}$ is defined and the condition $\dt{\mathrm{V}}(\mathrm{x}, \Psi, \delta \Psi)\leq -\alpha \mathrm{V}(\mathrm{x}, \Psi, \delta \Psi)$ is verified. In this example, the positive definiteness matrix 
\begin{equation}
    \mathrm{P(x_1)}=\begin{bmatrix}
1+2\mathrm{x_1}^2 &\mathrm{x_1}  \\ 
\mathrm{x_1} & 1
\end{bmatrix} 
\end{equation}
is selected. The time derivative of $\mathrm{V}$, the integration of obtained term, some lengthy and tedious calculations, and addition of $\mathrm{-x_1}$ provide the final control law 
\begin{equation}
    \mathrm{u}=\frac{5}{2}\mathrm{x_1}^3-3\mathrm{x_1}^2\mathrm{x_2}-\frac{1}{2}\mathrm{x_1}\mathrm{x_2}^2+\mathrm{x_1}^2-\mathrm{x_1}\mathrm{x_3}-\mathrm{x_1}-2\mathrm{x_2}-2\mathrm{x_3}
\end{equation}
by satisfying the manifold attractivity and trajectory boundedness condition. The I\&I VCP method can be viewed as an application of the theory of generation of Control Contraction Metric (CCM). According to contraction theory, if a system has the Incremental Lyapunov Function (ILF) and Contraction Metric (CM) of the differential state, all NLS trajectories converge incrementally and exponentially to one single trajectory. In I\&I VCP, the FLF is used as a ILF and the matrix in (\ref{inivcopcm}) is treated as CM. Here, the contraction of off-the-manifold dynamics is discussed instead of system dynamics. Therefore, this method is labeled as I\&I VCP \cite{WangTAC, WangCDC}. In order to get the final control law, the LMI and YALMIP toolboxes are required.

The similar problem is solved via the proposed P\&I method for comparison purpose.

On other hand, the selection of $\Psi_2=-\mathrm{x_1}{\Psi}_1$ renders the subsystem $\dt{\Psi}_1=-{\Psi}_1$. The implicit manifold is defined as $\Phi(\mathrm{x}, \Psi_1, \Psi_2)=\Psi_2+\mathrm{x_1}{\Psi}_1=0$.  To fulfill the step $(S_2)$ of the proposed P\&I approach, a PR metric $\mathrm{R}$ is defined  as
\begin{equation}
    \mathrm{R}=\triangledown \Phi(\mathrm{x}, \Psi_1, \Psi_2)^{\mathrm{T}}\triangledown \Phi(\mathrm{x}, \Psi_1, \Psi_2)=\begin{bmatrix}
\mathrm{x^2_1} &\mathrm{x_1} \\ 
 \mathrm{x_1}& 1
\end{bmatrix}
\end{equation}
The definition of passive output (\ref{GammaPassive}) 
\begin{small}
\begin{equation}
    \mathrm{y}=\int_{0}^{t}(\dt{\Psi}_2+\mathrm{m}_{22}^{-1}\mathrm{m}_{21}\dt{\mathrm{x}}_1)\mathrm{dt}={\Psi}_2+\mathrm{x}_1{\Psi}_1
\end{equation}
\end{small}
from step $(S_3)$ is fulfilled. With the obtained passive output $\mathrm{y}$, the candidate Lyapunov Function (Storage function)
\begin{equation}
    \mathbb{S}(\Psi_1,\Psi_2)=\frac{1}{2}{\mathrm{y}}^2=\frac{1}{2}(\Psi_2+\mathrm{x}_1\Psi_1)^2
\end{equation}
from step $(S_4)$ is defined. The condition $\dt{\mathbb{S}}\leq - \alpha \mathbb{S}$ from (\ref{exponential laypunov}) is utilized to get the final control law
\begin{equation}
    \mathrm{u}=-\left ( \Psi_2-\Psi_1-2\mathrm{x_1}\Psi_1+2\mathrm{x^2_1} \Psi_1+\mathrm{x_1}\Psi_2+\mathrm{x_1}\mathrm{x_2}\Psi_1+\frac{\alpha}{2}(\Psi_2+\mathrm{x_1}\Psi_1)\right ).
\end{equation}
%\begin{remark}
%The condition of positive definiteness matrix $\mathrm{P}$ is relaxed and replaced by the pseudo-Riemannian metric $\mathrm{R}$. Upon comparison, the off-the-diagonal elements of both $\mathrm{P}$ and $\mathrm{R}$ are same i.e., $\mathrm{x_1}$. This is a generalization of a Riemannian manifold in which the requirement of positive-definiteness is relaxed.
%\end{remark}
\begin{remark}
In contrast to the I\&I VCP, the proposed P\&I approach is systematic, methodical, and does not required  any positive-definiteness matrix $\mathrm{P}$, the integration of obtained term, solution of matrix inequality,  some lengthy and tedious calculations, and addition of $\mathrm{-x_1}$ to get the final control law.
\end{remark}

\subsubsection{P\&I approach for the systems without any suitable structure}
\begin{example}\label{example W3}
A specific system 
\begin{align}\label{Wangsystem3}
    \begin{split}
        \dt{\mathrm{x}}_1&=-\mathrm{x}_1+\mathrm{x}_1^2+\mathrm{x}_1\mathrm{x}_2+\mathrm{x}_1\mathrm{x}_3\\
\dt{\mathrm{x}}_2& =\mathrm{x}_3\\
\dt{\mathrm{x}}_3& =-\mathrm{x}_3+\mathrm{u}
    \end{split}
\end{align}
\end{example}
from \cite{WangTAC} with $\mathrm{x}^*=0$ and $\mathrm{x}=(\mathrm{x}_1, \mathrm{x}_2, \mathrm{x}_3)$ is highlighted in order to validate the performance of the P\&I approach for the system without suitable structure. 
The subsystem $\dt{\mathrm{x}}_1=0$ exhibits the bounded nature of $\mathrm{x_1(t)}$ for ${\mathrm{x}}_1=0$.  The target dynamics $\dt{\eta}=\beta(\eta)=- \eta$ with $\mathrm{x}_2=\eta$ is chosen to get $\pi(\eta)=\mathrm{col}(0, \eta, -\eta)$ and implicit manifold vector $\Psi (\mathrm{x})=\mathrm{col}(\underset{\Psi_1}{\underbrace{\mathrm{x}_1}}, \hspace{0.09cm}\underset{\Psi_2}{\underbrace{\mathrm{x}_2+\mathrm{x}_3}})$. 
The combined implicit manifold 
\begin{equation}\label{systemT2}
 \Phi(\mathrm{x}_1,\mathrm{x}_2,\mathrm{x}_3)= \Phi(\mathrm{x})=\Psi_1+\Psi_2=\mathrm{x}_1+\mathrm{x}_2+\mathrm{x}_3=0
\end{equation}
to satisfy the step $(S_1)$ for the system (\ref{Wangsystem3}) is obtained by adding the elements of the $\Psi (\mathrm{x})$.
\begin{remark}
Specifically, It can be observed that the selection of ${\mathrm{x}}_2+{\mathrm{x}}_3=-\mathrm{x}_1$ renders the subsystem $\dt{\mathrm{x}}_1=-\mathrm{x}_1$.
Thus, the  implicit manifold 
\begin{equation}\label{simpleoff-manifold}
 \Phi(\mathrm{x}_1,\mathrm{x}_2,\mathrm{x}_3)= \Phi(\mathrm{x})=\mathrm{x}_1+\mathrm{x}_2+\mathrm{x}_3=0
\end{equation}
is obtained easily just by inspection.
\end{remark}
To fulfill the step $(S^{\dagger\dagger}_2)$, a PR metric $\mathrm{R}$ is defined  as
\begin{small}
\begin{equation}\label{reimanmetric1}
    \triangledown \Phi(\mathrm{x})^{\mathrm{T}}\triangledown \Phi(\mathrm{x})=\begin{bmatrix}
\mathrm{m}_{11} & \mathrm{m}_{12} &\mathrm{m}_{13} \\ 
 \mathrm{m}_{21}& \mathrm{m}_{22} & \mathrm{m}_{23}\\ 
\mathrm{m}_{31} &\mathrm{m}_{32}  & \mathrm{m}_{33}
\end{bmatrix}=\begin{bmatrix}
1 &1  &1 \\ 
1 & 1 & 1\\ 
 1&  1& 1
\end{bmatrix}
\end{equation}
\end{small}
to obtain $\mathrm{m}_{13}=\mathrm{m}_{31}=1$, $\mathrm{m}_{23}=\mathrm{m}_{32}=1$, and $\mathrm{m}_{33}=1$.
The definition of passive output (\ref{GammaPassive}) 
\begin{small}
\begin{equation}
    \mathrm{y}=\int_{0}^{t}(\dt{\mathrm{x}}_3+\mathrm{m}_{33}^{-1}\mathrm{m}_{23}\dt{\mathrm{x}}_2+\mathrm{m}_{33}^{-1}\mathrm{m}_{31}\dt{\mathrm{x}}_1)\mathrm{dt}=\mathrm{x}_3+\mathrm{x}_2+\mathrm{x}_1
\end{equation}
\end{small}
from step $(S^{\dagger\dagger}_3)$ is fulfilled.
 \begin{remark}
As the input $\mathrm{u}$ is available in $\dt{\mathrm{x}}_3$, the elements in metric $\mathrm{R}$ having connection with subscript $3$ is taken to define the passive output (see Fig. \ref{Geometrical interpretation}).
 \end{remark}
With the passive output $\mathrm{y}$, the storage function 
\begin{align}
    \begin{split}
        \mathbb{S}(\mathrm{x}_1,\mathrm{x}_2, \mathrm{x}_3)=&\frac{1}{2}{\mathrm{y}}^2=\frac{1}{2}(\underset{\lambda}{\underbrace{\mathrm{x}_3}}+\underset{\mathrm{q(\mathrm{x}_1,\mathrm{x}_2)}}{\underbrace{\mathrm{x}_2+\mathrm{x}_1}})^2\\=&\frac{1}{2}(\lambda+\mathrm{q(\mathrm{x}_1,\mathrm{x}_2)})^2
    \end{split}
\end{align}
is defined using $(S^{\dagger\dagger}_4)$ and (\ref{Storage function}). The condition $\dt{\mathbb{S}}\leq - \alpha \mathbb{S}$ from (\ref{exponential laypunov}) is extended as
\begin{small}
\begin{equation}
    (\underset{\dt{\lambda}}{\underbrace{\dt{\mathrm{x}}_3}}+\underset{\frac{\partial \mathrm{q(\mathrm{x}_1,\mathrm{x}_2)}}{\partial \mathrm{x}_2}\dt{\mathrm{x}}_2+\frac{\partial \mathrm{q}(\mathrm{x}_1,\mathrm{x}_2) }{\partial \mathrm{x}_1}\dt{\mathrm{x}}_1}{\underbrace{ \dt{\mathrm{x}}_2+\dt{\mathrm{x}}_1}})+\frac{\alpha}{2}\underset{\lambda+\mathrm{q(\mathrm{x}_1,\mathrm{x}_2)}}{\underbrace{(\mathrm{x}_3+\mathrm{x}_2+\mathrm{x}_1)}}=0.
\end{equation}
\end{small}
Upon substitution, the final control law
\begin{small}
   \begin{equation}\label{wanglabel}
   \mathrm{u_{P\&I}}=\mathrm{x}_3-\frac{\alpha}{2}(\mathrm{x}_1+\mathrm{x}_3+\mathrm{x}_2)+\mathrm{x}_1-\mathrm{x}_1^2-\mathrm{x}_1\mathrm{x}_2-\mathrm{x}_1\mathrm{x}_3
\end{equation}
\end{small}
is obtained that ensure the convergence of system trajectory $\mathrm{x(t)}$ to an equilibrium point.

\textbf{Discussion:} 
The stabilization of system (\ref{Wangsystem3}) 
\begin{equation}
     \dt{\mathrm{x}}_1 =-\mathrm{x}_1+\mathrm{x}_1^2+\mathrm{x}_1\mathrm{x}_2+\mathrm{x}_1\mathrm{x}_3 \hspace{1.5cm}
\dt{\mathrm{x}}_2=\mathrm{x}_3\hspace{1.5cm}
\dt{\mathrm{x}}_3 =\mathrm{u}
\end{equation}
is also achieved via the I\&I Horizontal Contraction Procedure (I\&I HCP) in \cite{WangTAC} (Example 3 in \cite{WangTAC}). In I\&I HCP, the \textit{manifold attractivity and trajectory boundedness} i.e., the last crucial step of I\&I approach is ensured via horizontal contraction. The subsystem $\dt{\mathrm{x}}_1=0$ exhibits the bounded nature of $\mathrm{x_1(t)}$ for ${\mathrm{x}}_1=0$.  The target dynamics $\dt{\eta}=\beta(\eta)=- \eta$ with $\mathrm{x}_2=\eta$ is chosen to get $\pi(\eta)=\mathrm{col}(0, \eta, -\eta)$ and implicit manifold vector $\Psi (\mathrm{x})=\mathrm{col}(\underset{\Psi_1}{\underbrace{\mathrm{x}_1}}, \hspace{0.09cm}\underset{\Psi_2}{\underbrace{\mathrm{x}_2+\mathrm{x}_3}})$.
The off-the-manifold dynamics
\begin{align}
    \begin{split}
        \dt{\Psi}_{\mathrm{a}}&=-{\Psi}_{\mathrm{a}}+{\Psi}_{\mathrm{a}}^2+{\Psi}_{\mathrm{a}}{\Psi}_{\mathrm{b}}\\
\dt{\Psi}_{\mathrm{b}}&={\mathrm{u}} 
    \end{split}
\end{align}
is defined for the implicit manifold description ($S_3$ of the proposition \ref{prop1}). The variational dynamics (\ref{prol22}) is selected as
\begin{small}
\begin{equation}\label{djfdkljn}
    \sigma(\mathrm{x})=\begin{bmatrix}
-1+2\mathrm{x_1}+\mathrm{x_2}+\mathrm{x_3} &\mathrm{x_1}  &\mathrm{x_1}  \\
 0& 0 &1  \\
\triangledown_{\mathrm{x_1}}\Upsilon & \triangledown_{\mathrm{x_2}}\Upsilon  & -1+\triangledown_{\mathrm{x_3}} \Upsilon \\
\end{bmatrix}
\end{equation}
\end{small}
and the 
\begin{small}
\begin{equation}\label{sjhjdsdjf}
    \Theta(\mathrm{x})=\begin{bmatrix}
(\mathrm{x_1}+1)e^{\mathrm{x_1}+\mathrm{x_2}+\mathrm{x_3}} & \mathrm{x_1}e^{\mathrm{x_1}+\mathrm{x_2}+\mathrm{x_3}}  \\
1 & 1 \\
\end{bmatrix}
\end{equation}
for the matrix $\Xi(\mathrm{x})=\Theta ^{\mathrm{T}}(\mathrm{x})\Theta (\mathrm{x})$ is randomly chosen with 
\end{small}
\begin{equation}
    \Xi(\mathrm{x})=\triangledown \Psi=\begin{bmatrix}
1 &0  &0  \\
0 & 1 &  1\\
\end{bmatrix}
\end{equation}
as a natural choice for the FLF (\ref{flfwang}). The derivative of the FLF (\ref{prol1}) is solved to get
\begin{small}
\begin{equation}\label{hggfghhchfc}
    \mathrm{u_{I\&IHCP}}=\Upsilon(\mathrm{x})= -\mathrm{x_2}-\mathrm{x_1}\mathrm{x_2}-\mathrm{x_1}\mathrm{x_3}-\mathrm{x_3}-\mathrm{x_1}^2
\end{equation}
\end{small}
by considering (\ref{djfdkljn}) and (\ref{sjhjdsdjf}). 
 
The $ \mathrm{u_{P\&I}}$ is exactly same as the control law (\ref{hggfghhchfc}) in I\&I HCP approach for  $\alpha_2=2$. For the same control law, the trajectory boundedness and GAS of origin of the system (\ref{Wangsystem3}) in closed-loop with (\ref{wanglabel}) is already proved. Hence, the system  (\ref{Wangsystem3})  with the proposed P\&I control law (\ref{wanglabel}) has GAS equilibrium.

\begin{remark}
In I\&I HCP, the selection of $\Xi(\mathrm{x})=\Theta ^{\mathrm{T}}(\mathrm{x})\Theta (\mathrm{x})$ (refer (\ref{sjhjdsdjf})) for the FLF is not defined properly and selected randomly. Moreover, solving a matrix inequality requires certain lengthy and tedious calculations. In the proposed P\&I approach, the pseudo-Riemannian metric is properly defined in (\ref{reimanmetric}) and related calculations are systematically carried out to obtained the control law for GAS equilibrium point of the system.
\end{remark}
%\begin{remark}\label{eg1remark1}
%The $\mathrm{u_{P\&I}}$ (\ref{secondp&Ilaw}) and the I\&I HCP-based control law (refer Ex. 4 of \cite{WangTAC}) obtained for Example \ref{Example 1} are exactly same. Similarly, the $\mathrm{u_{P\&I}}$ (\ref{wanglabel}) is exactly same as the control law obtained via I\&I HCP approach for $\alpha=2$ (refer Ex. 3 of \cite{WangTAC}). The trajectory boundedness and GAS equilibrium of the CLS (system (\ref{ex1_system}) with (\ref{secondp&Ilaw}) and system (\ref{Wangsystem3}) with (\ref{wanglabel})) are already proved in \cite{WangTAC}.
%For the same control law, the trajectory boundedness and GAS of origin of the system (\ref{Wangsystem3}) in closed-loop with (\ref{wanglabel}) is already proved. Hence, the system  (\ref{Wangsystem3})  with the proposed P\&I control law (\ref{wanglabel}) has GAS equilibrium.
%\end{remark}
\begin{remark}\label{eg1remark11}
In contrast to the I\&I HCP,  the proposed P\&I approach constructs a control law in a systematic way without any inequality calculation.
\end{remark}
\subsubsection{P\&I approach for the nontriangular nonlinear systems affine in $(\mathrm{x_1},\mathrm{x_2})$}
The main goal is to stabilize the normal form \cite{olfati2001nonlinear}
\begin{align}
    \begin{split}
       &\dt{\mathrm{z}}=\mathrm{f}(\mathrm{z},\mathrm{x_1},\mathrm{x_2})\\
       &\dt{\mathrm{x}}_1={\mathrm{x_2}}\\
       &\dt{\mathrm{x}}_2={\mathrm{u}}
    \end{split}
\end{align} in its simplest possible form \cite{olfati2001nonlinear} given as
\begin{align}\label{huh}
    \begin{split}
     &\dt{\mathrm{z}}=\mathrm{f}(\mathrm{z})+\mathrm{g}_1(\mathrm{z})\mathrm{x_1}+\mathrm{g}_2(\mathrm{z})\mathrm{x_2}\\
    &\dt{\mathrm{x}}_1={\mathrm{x_2}}\\
    &\dt{\mathrm{x}}_2={\mathrm{u}}
    \end{split}
\end{align}
where the vector fields $\mathrm{f}(\mathrm{z})$, $\mathrm{g}_1(\mathrm{z})$, $\mathrm{g}_2(\mathrm{z})$ are all smooth and $\mathrm{f}(0)=0$. The goal is to stabilize the $\mathrm{z(t)}\rightarrow 0$. It is clearly seen that the cross term $\mathrm{g}_1(\mathrm{z})\mathrm{x_1}+\mathrm{g}_2(\mathrm{z})\mathrm{x_2}$ is available in $\dt{\mathrm{z}}$. Hence, the control law $\mathrm{u}$ obtained via direct calculation results a local solution. To start with, the term $\mathrm{g}_2(\mathrm{z})\mathrm{x_2}=0$ is considered. This results into
\begin{equation}
    \dt{\mathrm{z}}=\mathrm{f}(\mathrm{z})+\mathrm{g}_1(\mathrm{z})\mathrm{x_1}=\mathrm{f}(\mathrm{z})+\mathrm{g}_1(\mathrm{z})\sigma(\mathrm{z}).
\end{equation}
Here, we have to calculate $\sigma(\mathrm{z})$ such that the dynamics of $\mathrm{z}$-subsystem is GAS. As $\mathrm{x_1}=\sigma_1(\mathrm{z})$, the implicit manifold $\Psi_1$ can be obtained as:
\begin{equation}
    \Psi_1=\mathrm{x_1}-\sigma_1(\mathrm{z})=0.
\end{equation}
Once the implicit manifold $\Psi_1$ is obtained, the four step procedure is easily applied to get the virtual control law $\sigma_2(\mathrm{z, x_1})$. Thus, the four step calculation provides
\begin{align}
    \begin{split}
         \Psi_2=\dt{\Psi}_1+\frac{\alpha_1}{2}\Psi_1&=\dt{\mathrm{x}}_1-\frac{\partial \sigma_1}{\partial \mathrm{z}}\dt{\mathrm{z}}+\frac{\alpha_1}{2}(\mathrm{x_1}-\sigma_1(\mathrm{z}))\\
&={\mathrm{x}}_2-\frac{\partial \sigma}{\partial \mathrm{z}}\left ( \mathrm{f}(\mathrm{z})+\mathrm{g}_1(\mathrm{z})\mathrm{x_1}+\mathrm{g}_2(\mathrm{z})\mathrm{x_2} \right )+\frac{\alpha_1}{2}(\mathrm{x_1}-\sigma_1(\mathrm{z}))\\
&={\mathrm{x}}_2-\left ( \mathrm{L_f\sigma(\mathrm{z})}+\mathrm{x_1} \mathrm{L_{g1}\sigma(\mathrm{z}}) \right )-\mathrm{x_2}\mathrm{L_{g2}\sigma(\mathrm{z}})+\frac{\alpha_1}{2}(\mathrm{x_1}-\sigma_1(\mathrm{z}))\\
&\Rightarrow \mathrm{x_2}\left ( 1- \mathrm{L_{g2}\sigma(\mathrm{z}}) \right )-\left ( \mathrm{L_f\sigma(\mathrm{z})}+\mathrm{x_1} \mathrm{L_{g1}\sigma(\mathrm{z}}) \right )+\frac{\alpha_1}{2}(\mathrm{x_1}-\sigma_1(\mathrm{z})).
    \end{split}
\end{align}
with $M(\mathrm{z})=\frac{1}{\left ( 1- \mathrm{L_{g2}\sigma(\mathrm{z}}) \right )}$, the above equation is modified as:
\begin{align}
    \begin{split}
        \Psi_2 &= M(\mathrm{z})\left ( \mathrm{x_2}- \frac{\left ( \mathrm{L_f\sigma(\mathrm{z})}+\mathrm{x_1} \mathrm{L_{g1}\sigma(\mathrm{z}}) -\frac{\alpha_1}{2}(\mathrm{x_1}-\sigma_1(\mathrm{z}))\right )}{M(\mathrm{z})}\right ) \\
 \Psi_2&=M(\mathrm{z})\left (\mathrm{x_2}-\sigma_2(\mathrm{z, x_1})  \right )
    \end{split}
\end{align}
where $\sigma_2(\mathrm{z, x_1})=\frac{\left ( \mathrm{L_f\sigma(\mathrm{z})}+\mathrm{x_1} \mathrm{L_{g1}\sigma(\mathrm{z}}) -\frac{\alpha_1}{2}(\mathrm{x_1}-\sigma_1(\mathrm{z}))\right )}{M(\mathrm{z})}$. Here, we got the second implicit manifold $\Psi_2=M(\mathrm{z})\left (\mathrm{x_2}-\sigma_2(\mathrm{z, x_1})  \right )$. Now, again applying the four step P\&I procedure to newly obtained implicit manifold $\Psi_2=M(\mathrm{z})\left (\mathrm{x_2}-\sigma_2(\mathrm{z, x_1})  \right )$. This will render
\begin{align}\label{jhh}
    \begin{split}
        \dt{\Psi}_2=\omega&=M(\mathrm{z})\left ( \dt{\mathrm{x}}_2-\dt{\sigma}_2(\mathrm{z,x_1}) \right )+\dt{M}\left (\mathrm{x_2}-\sigma_2(\mathrm{z, x_1})  \right )+\frac{\alpha_2}{2}\\
        &=M(\mathrm{z})\left (u-\sigma_3(\mathrm{z, x_1,x_2})  \right )
    \end{split}
\end{align}
Substituting the system dynamics into the (\ref{jhh}) to get the final control law
\begin{equation}\label{hgjg}
    \mathrm{u}=\frac{1}{M(\mathrm{z})}\left ( \left (M(\mathrm{z})\dt{\sigma}_2(\mathrm{z,x_1} \right )-\dt{M}\left (\mathrm{x_2}-\sigma_2(\mathrm{z, x_1})  \right )-\frac{\alpha_2}{2}(M(\mathrm{z})\left (\mathrm{x_2}-\sigma_2(\mathrm{z, x_1})  \right )) \right ).
\end{equation}
The obtained smooth state feedback control law (\ref{hgjg}) via P\&I approach with some positive gain $\alpha_i$ for $i=1,2$ globally asymptotically stabilizes $(\mathrm{z, x_1, x_2})$ for the nonlinear system (\ref{huh}). 
\section{The proposed P\&I Control for Systems in parametric strict feedback (PSF) form}\label{sec4}
The identification of the CM and variational dynamics are the most important aspects of the Backstepping design for incremental stability.  As the trajectories approach each other in the incremental stability concept, the idea of infinitesimal distance comes into the picture, forcing the system dynamics to be prolonged. As a result, the focus of this methodology is mostly based on evaluating variational dynamics and identifying the CM. In order to reduce the mathematical complexity and to provide a systematic stabilization procedure, the proposed P\&I approach for the system in parametric strict feedback form is presented.
\subsection{Applicability of the P\&I approach to the systems in PSF form}
Consider a system is  PSF form
\begin{small}
\begin{align}\label{AP21}
\begin{split}
&\dt{\mathrm{x}}_1= \Omega_1(\mathrm{x},\mathrm{u})=\gamma_1 (\mathrm{x_1})+\vartheta_1 \mathrm{x_2}\\
&\dt{\mathrm{x}}_2= \Omega_2(\mathrm{x},\mathrm{u})=\gamma_2 (\mathrm{x_1,x_2})+\vartheta_2\mathrm{x_3}\\
&....\\
&\dt{\mathrm{x}}_{\mathrm{n-1}}= \Omega_{\mathrm{n-1}}(\mathrm{x},\mathrm{u})=\gamma_{\mathrm{n-1}} (\mathrm{x}_1, \mathrm{x}_2,..,\mathrm{x}_{\mathrm{n-1}})+\vartheta_{\mathrm{n-1}}\mathrm{x_{\mathrm{n}}}\\
 & \dt{\mathrm{x}}_{\mathrm{n}}= \Omega_{\mathrm{n}}(\mathrm{x},\mathrm{u})=\gamma_{\mathrm{n}} (\mathrm{x})+\widetilde{\lambda} (\mathrm{x})\mathrm{u}\\
\end{split}
\end{align}
\end{small}
with the parametric strict-feedback form $\Omega$, the system state $\mathrm{x} \in \mathbb{R}^{\mathrm{n}}$, control input $\mathrm{u} \in \mathrm{U}\subseteq  \mathbb{R}$, the parameters $\vartheta_{\mathrm{i}} \in \mathbb{R}$ for $\mathrm{i}=1,...,\mathrm{n}$,   the smooth function $\gamma_{\mathrm{i}}:\mathbb{R}^{\mathrm{i}}\to \mathbb{R}$ for $\mathrm{i}=1,...,\mathrm{n}$,  and  $\widetilde{\lambda} (\mathrm{x}) \neq 0$ over the domain of interest.
The first step (S1) in the proposed P\&I approach is the selection of target dynamics $\dt{\eta}=-\frac{\alpha_1}{2}\eta$ with ${\mathrm{x}} _1=\eta$ for construction of the implicit manifold in order to get an exponentially stable solution of $\dt{\mathrm{x}}_1$. It renders the off-the-manifold 
\begin{equation}\label{91}
\gamma_1 (\mathrm{x_1})+\vartheta_1 \mathrm{x_2}+\frac{\alpha_1}{2}\mathrm{x_1}=0   \Rightarrow  \mathrm{x_2}-\varphi_1(\mathrm{x_1})=0 
\end{equation}
which ultimately leads to the virtual control law
\begin{equation}
 \mathrm{x_2}=\underset{\varphi_1(\mathrm{x_1})}{\underbrace{\frac{-1}{\vartheta_1 }\left (  \gamma_1 (\mathrm{x_1})+\frac{\alpha_1}{2}\mathrm{x_1} \right )}}
\end{equation}
to \textit{exponentially stabilize the solution of  $\dt{\mathrm{x}}_1$}.As per the $S2^{\dagger\dagger}$, the semi-Riemannian metric $\mathrm{R}$ for off-the-manifold $\mathrm{x_2}-\varphi_1(\mathrm{x_1})=0$ is defined as
\begin{small}
\begin{align}\label{93}
    \begin{split}
        \mathrm{R_1}&=\triangledown \Psi (\mathrm{x_1} ,\mathrm{x_2} )^{\mathrm{T}}\triangledown \Psi (\mathrm{x_1} ,\mathrm{x_2})\\&=\begin{bmatrix}
\left ( \frac{\partial \varphi_1}{\partial \mathrm{x_1}}\right )^{\mathrm{T}}\left ( \frac{\partial \varphi_1}{\partial \mathrm{x_1}}\right )& \left ( -\frac{\partial \varphi_1}{\partial \mathrm{x_1}}\right )^{\mathrm{T}}\\ 
-\left ( \frac{\partial \varphi_1}{\partial \mathrm{x_1}}\right ) & \mathrm{1}
\end{bmatrix}=\begin{bmatrix}
\mathrm{\mathrm{m}}_{11(1)} & \mathrm{\mathrm{m}}_{12(1)}\\ 
 \mathrm{\mathrm{m}}_{21(1)} & \mathrm{\mathrm{m}}_{22(1)}
\end{bmatrix}
    \end{split}
\end{align}
\end{small}
from (\ref{metric}). In step $S3^{\dagger\dagger}$, the component $\left (0, \dt{\mathrm{x}}_2+\mathrm{\mathrm{m}}_{22(1)}^{-1}\mathrm{\mathrm{m}}_{21(1)} \dt{\mathrm{x}}_1 \right )$ of virtual law $\mathrm{x_2}$ tangent vector along $\dt{\mathrm{x}}_2$  is utilized to define the passive output ${\mathrm{y}}_1$
\begin{align}\label{94}
    \begin{split}
    {\mathrm{y}}_1&=\int_{0}^{t}(\dt{\mathrm{x}}_2+\mathrm{m}_{22(1)}^{-1}\mathrm{m}_{21(1)}\dt{\mathrm{x}}_1)\mathrm{dt}\\
        &=\int_{0}^{t}(\dt{\mathrm{x}}_2-\left ( \frac{\partial \varphi_1}{\partial \mathrm{x_1}}\right )\dt{\mathrm{x}}_1)\mathrm{dt} =\mathrm{x_2}-\varphi_1(\mathrm{x_1}).
    \end{split}
\end{align}
In line with third step of selecting the passive output, the storage function 
\begin{equation}\label{95}
    \mathbb{S}_1(\mathrm{x_1}, \mathrm{x_2})=\frac{1}{2}{\mathrm{y}}^2=\frac{1}{2}(\mathrm{x_2}-\varphi_1(\mathrm{x_1}))^2
\end{equation}
from step $S4^{\dagger\dagger}$ is selected and the condition
\begin{equation}\label{96}
    \dt{\mathbb{S}}_1\leq -\alpha_2\mathbb{S}_1
\end{equation}
is utilized to get virtual control law 
\begin{small}
\begin{equation}\label{x2forx2d}
    \mathrm{x_3}=\underset{\varphi_2(\mathrm{x_1, x_2})}{\underbrace{\frac{-1}{\vartheta_2 }\left ( \gamma_2 (\mathrm{x_1}, \mathrm{x_2})-\left ( \frac{\partial \varphi_1}{\partial \mathrm{x_1}}\dt{\mathrm{x}}_1\right )+\frac{\alpha_2}{2}(\mathrm{x_2}-\varphi_2(\mathrm{x_1,x_2})) \right )}}.
\end{equation}
\end{small}
The virtual control law $\mathrm{x_3}$ derived via the proposed P\&I approach in (\ref{x2forx2d}) is responsible for \textit{the exponential convergence of the solution of $\mathrm{x_2}$-dynamics}. This virtual control law $\mathrm{x_3}$ from (\ref{x2forx2d}) has allowed to get the implicit manifold 
\begin{equation}\label{dvhj}
    \mathrm{x_3}-\varphi_2(\mathrm{x_1, x_2})=0
\end{equation}
is obtained. After getting the implicit manifold, the steps in the proposed P\&I approach are applied to derive the next virtual control law $\mathrm{x_4}$. The semi-Riemannian metric 
\begin{small}
\begin{equation}
    \mathrm{R}_2=\begin{bmatrix}
\begin{bmatrix}
 ( \frac{\partial \varphi_2}{\partial \mathrm{x_1}}) \\ ( \frac{\partial \varphi_2}{\partial \mathrm{x_2}})
\end{bmatrix}\begin{bmatrix}
\left ( \frac{\partial \varphi_2}{\partial \mathrm{x_1}}\right ) & \left ( \frac{\partial \varphi_2}{\partial \mathrm{x_2}}\right )
\end{bmatrix}& -\begin{bmatrix}
 ( \frac{\partial \varphi_2}{\partial \mathrm{x_1}}) \\ ( \frac{\partial \varphi_2}{\partial \mathrm{x_2}})
\end{bmatrix}\\ 
\begin{bmatrix}
 (- \frac{\partial \varphi_2}{\partial \mathrm{x_1}}) &  (- \frac{\partial \varphi_2}{\partial \mathrm{x_2}} )
\end{bmatrix} & \mathrm{1}
\end{bmatrix}
\end{equation}
\end{small}
using (\ref{dvhj})
\begin{small}
\begin{equation}
    =\begin{bmatrix}
\mathrm{\mathrm{m}}_{11(2)} & \mathrm{\mathrm{m}}_{12(2)}  & \mathrm{\mathrm{m}}_{13(2)}\\ 
 \mathrm{\mathrm{m}}_{21(2)}&\mathrm{\mathrm{m}}_{22(2)}   &\mathrm{\mathrm{m}}_{23(2)} \\
\mathrm{\mathrm{m}}_{31(2)} & \mathrm{\mathrm{m}}_{32(2)}  & \mathrm{\mathrm{m}}_{33(2)}\\
\end{bmatrix}
\end{equation}
\end{small}
 and the passive output 
\begin{small}
\begin{equation}
    \mathrm{y}=\int_{0}^{t}(\dt{\mathrm{x}}_3+\mathrm{m}_{33}^{-1}\mathrm{m}_{23}\dt{\mathrm{x}}_2+\mathrm{m}_{33}^{-1}\mathrm{m}_{31}\dt{\mathrm{x}}_1)\mathrm{dt}
\end{equation}
\end{small}
is defined via systematic way in order to get the virtual control law $\mathrm{x_4}$
\begin{small}
\begin{equation}
    \mathrm{x_4}=\underset{\varphi_3(\mathrm{x_1, x_2, x_3})}{\underbrace{\frac{-1}{\vartheta_3 }\left ( \gamma_3 (\mathrm{x_1}, \mathrm{x_2})-\left ( \frac{\partial \varphi_2}{\partial \mathrm{x}}\dt{\mathrm{x}}\right )+\frac{\alpha_3}{2}(\mathrm{x_2}-\varphi_3(\mathrm{x_1,x_2,x_3})) \right )}}.
\end{equation}
\end{small}
The application of the four steps procedure in a repeatative mode provides the final control law 
\begin{small}
\begin{equation}\label{shc}
    \mathrm{u_{P\&I}}={\frac{-1}{{\widetilde{\lambda} (\mathrm{x})}}\left ( \gamma_{\mathrm{n}}(\mathrm{x})-\left ( \frac{\partial \varphi_{\mathrm{n-1}}}{\partial \mathrm{x}}\dt{\mathrm{x}}\right )+\frac{\alpha_{\mathrm{n}}}{2}(\mathrm{x_{\mathrm{n}}}-\varphi_{\mathrm{n-1}}(\mathrm{x})) \right )}
\end{equation}
\end{small}
at the end that exponentially stabilizes the system (\ref{AP21}) to equilibrium point.
\begin{theorem}\label{theo1}
For any system of the form (\ref{AP21}), the obtained control law via the proposed P\&I approach
\begin{equation}\label{Psfflaw}
    \mathrm{u_{P\&I}}=\frac{-1}{\widetilde{\lambda} (\mathrm{x})}\left (\gamma_{\mathrm{n}} (\mathrm{x})+\mathrm{b_n(x)} \right )
\end{equation}
where $\alpha_{\mathrm{m}} > 0 \hspace{0.1cm} \mathrm{for} \hspace{0.1cm} \mathrm{m}=1,..,\mathrm{n}$,
\begin{small}
\begin{equation*}
    \mathrm{b_m(x)} =\frac{\alpha_{\mathrm{m}}}{2}(\mathrm{x_m}-{\varphi}_{\mathrm{m-1}}(\mathrm{x}))-\frac{\partial \varphi_{\mathrm{m-1}}}{\partial \mathrm{x}}\Omega(\mathrm{x, u}), \hspace{0.1cm} \mathrm{for} \hspace{0.1cm} \mathrm{m}=1,..,\mathrm{n},
\end{equation*}
\end{small}
\begin{equation*}
{\varphi}_{\mathrm{m}}(\mathrm{x})=\frac{-1}{\vartheta_{\mathrm{m}}}\left ( \gamma_{\mathrm{m}}+\mathrm{b_m} \right ), \hspace{0.1cm} \mathrm{for} \hspace{0.1cm} \mathrm{m}=1,..,\mathrm{n-1},
\end{equation*}
\begin{equation*}
    {\varphi}_{-1}(\mathrm{x})={\varphi}_{0}(\mathrm{x})=0 \hspace{0.1cm} \forall \mathrm{x} \in\mathbb{R}^{\mathrm{n}}, \vartheta_0=0, \hspace{0.1cm} \mathrm{and}\hspace{0.1cm} \mathrm{x}_0=0,
\end{equation*}
renders the complete closed-loop control system Globally Exponentially Stable (GES).
\end{theorem}

\textbf{Proof:} The coordinate transformation is defined as
\begin{small}
\begin{equation}\label{transformastion}
     \zeta=\mathbb{T}(\mathrm{x}) =\begin{bmatrix}
\mathrm{x_1}-\varphi_0(\mathrm{x})  \\
 \mathrm{x_2}-\varphi_1(\mathrm{x})\\
 \mathrm{x_3}-\varphi_2(\mathrm{x})\\
\vdots \\
\mathrm{x_n}-\varphi_{\mathrm{n}-1}(\mathrm{x})
\end{bmatrix}
\end{equation}
\end{small}
 with a global diffeomorphism mapping $\varphi:\mathbb{R}^{\mathrm{n}}\to \mathbb{R}^{\mathrm{n}}$. The system dynamics (\ref{AP21}) is transformed to
\begin{small}
\begin{align}
    \begin{split}
        \dt{\zeta}_{\mathrm{m}}&=-\frac{\alpha_{\mathrm{m}}}{2}{\zeta}_{\mathrm{m}}+\vartheta_{\mathrm{m}}\zeta_{\mathrm{m}+1} \hspace{0.5cm}\mathrm{m}=1,...,\mathrm{n}\\
\dt{\zeta}_{\mathrm{n}}&= -\frac{\alpha_{\mathrm{n}}}{2}{\zeta}_{\mathrm{n}}+\gamma_{\mathrm{n}} (\mathrm{x})+\widetilde{\lambda} (\mathrm{x})\mathrm{u}-\frac{\partial \varphi_{\mathrm{n-1}}}{\partial \mathrm{x}}\Omega(\mathrm{x, u})+\frac{\alpha_{\mathrm{n}}}{2}{\zeta}_{\mathrm{n}}
    \end{split}
\end{align}
\end{small}
via (\ref{transformastion}). The proposed P\&I control law (\ref{Psfflaw}) or (\ref{shc}) provides the closed-loop system
\begin{small}
\begin{equation}\label{cltfseg}
    \begin{bmatrix}
\dt{\zeta}_1 \\
\dt{\zeta}_2 \\
\vdots \\
\dt{\zeta}_{\mathrm{n-1}}\\
\dt{\zeta}_{\mathrm{n}}\\
\end{bmatrix}=\begin{bmatrix}
-\frac{\alpha_{1}}{2} &\vartheta_1  & \cdots  & 0 &0  \\
0 & -\frac{\alpha_{2}}{2}&   \cdots &0 & 0  \\
\vdots &\vdots  & \cdots  & \vdots &\vdots  \\
0 &0  & \cdots  & -\frac{\alpha_{\mathrm{n-1}}}{2} & \vartheta_{\mathrm{n-1}}  \\
0 &0  & \cdots  & 0 &-\frac{\alpha_{\mathrm{n}}}{2}  \\
\end{bmatrix}\begin{bmatrix}
{\zeta}_1 \\
{\zeta}_2 \\
\vdots \\
{\zeta}_{\mathrm{n-1}}\\
{\zeta}_{\mathrm{n}}\\
\end{bmatrix}
\end{equation}
\end{small}
It is obvious that the system (\ref{cltfseg}) is globally stable with $\alpha_{\mathrm{m}} > 0 \hspace{0.1cm} \mathrm{for} \hspace{0.1cm} \mathrm{m}=1,..,\mathrm{n}$ and it assures the exponential convergence of $\zeta_{\mathrm{m}}$ for $ \mathrm{m}=1,..,\mathrm{n}$ to zero.
\subsection{Illustration of results of the P\&I approach  and the theorem \ref{theo1} for systems in PSF form}
\begin{figure}[ht!]
    \centering
    \includegraphics[width=\linewidth]{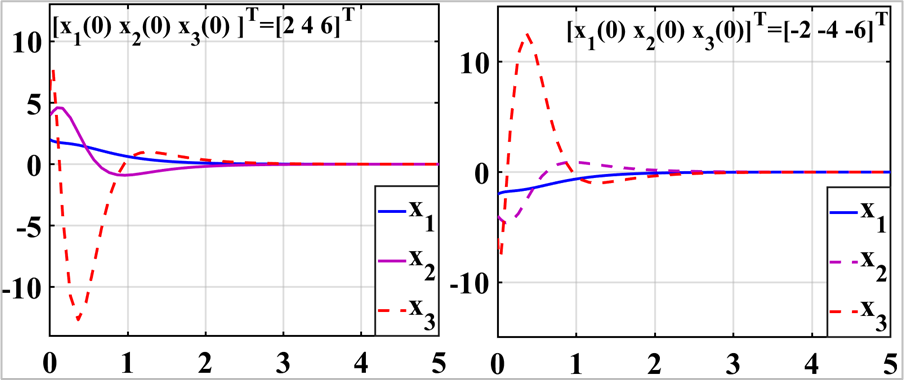}
    \caption{Figure associated with Example \ref{examples2}: Time evolution of state trajectories of system (\ref{exa2}) for $\begin{bmatrix}
\alpha_1 & \alpha_2 & \alpha_3 \\
\end{bmatrix}^{\mathrm{T}}=\begin{bmatrix}
12 & 12 & 4 \\
\end{bmatrix}^{\mathrm{T}}$.}
    \label{example2}
\end{figure}

\begin{example}\label{examples2}
Consider a system from \cite{jouffryo2002integrator} augmented by a chain of 2 integrators
\begin{align}\label{exa2}
    \begin{split}
        \dt{\mathrm{x}}_{1}=-\mathrm{x}_{1}^3+\mathrm{x}_{2}\hspace{0.8cm}
\dt{\mathrm{x}}_{2}=\mathrm{x}_{3}\hspace{0.8cm}
\dt{\mathrm{x}}_{3}=u.
    \end{split}
\end{align}
Being a third order system, the above mentioned theorem \ref{theo1} is utilized to obtain the final control law $\mathrm{u_{P\&I}}=-\mathrm{b_3}$ with
\begin{small}
\begin{align*}
    \begin{split}
\mathrm{b_1}=&\frac{\alpha_1}{2}\mathrm{x_1},\hspace{0.8cm}
\varphi_1(\mathrm{x})=\mathrm{x_1}^3-\frac{\alpha_1}{2}\mathrm{x_1},\\
\mathrm{b_2}=&\left ( \frac{\alpha_1}{2}-3 \mathrm{x_1}^2\right )(-\mathrm{x}_{1}^3+\mathrm{x}_{2})+\frac{\alpha_2}{2}(\mathrm{x_2}-\mathrm{x_1}^3+\frac{\alpha_1}{2}\mathrm{x_1}),\\ 
 \varphi_2(\mathrm{x})=&-\left ( \frac{\alpha_1}{2}-3 \mathrm{x_1}^2\right )(-\mathrm{x}_{1}^3+\mathrm{x}_{2})-\frac{\alpha_2}{2}(\mathrm{x_2}-\mathrm{x_1}^3+\frac{\alpha_1}{2}\mathrm{x_1}),\\
    \end{split}
\end{align*}
\end{small}
and
\begin{align*}
    \begin{split}
     \mathrm{b_3}& = \frac{\alpha_3}{2} (\mathrm{x_3}+ ( \frac{\alpha_1}{2}-3 \mathrm{x_1}^2 )(-\mathrm{x}_{1}^3+\mathrm{x}_{2})\\&
     +\frac{\alpha_2}{2}(\mathrm{x_2}-\mathrm{x_1}^3+\frac{\alpha_1}{2}\mathrm{x_1})) +\left ( \frac{\alpha_1}{2}-3 \mathrm{x_1}^2\right )\\& 
     \left ( -3\mathrm{x}_{1}^2(-\mathrm{x}_{1}^3+\mathrm{x}_{2})+ \mathrm{x}_{3} \right ) +(-\mathrm{x}_{1}^3+\mathrm{x}_{2})\\& (-6\mathrm{x}_{1}(-\mathrm{x}_{1}^3+\mathrm{x}_{2})) + \frac{\alpha_2}{2}(\mathrm{x_3}+ ( \frac{\alpha_1}{2}-3 \mathrm{x_1}^2)(-\mathrm{x}_{1}^3+\mathrm{x}_{2}))
    \end{split}
\end{align*}
 that stabilizes the system (\ref{exa2}) exponentially to equilibrium point $(0,0,0)$. The stabilization result for different initial conditions is shown in Fig. \ref{example2}.
\end{example}
\begin{example}
Consider a magnetic levitation system
\begin{align}\label{MLevit}
    \begin{split}
        \dt{\mathrm{x}}_1&=\frac{1}{\mathrm{m_b}}\mathrm{x}_2\\
\dt{\mathrm{x}}_2&=\frac{1}{2\mathrm{c_c}}\mathrm{x}_3-\mathrm{m_b}\mathrm{g}\\
\dt{\mathrm{x}}_3&=\frac{-2\mathrm{R_c}}{\mathrm{c_c}}(1-\mathrm{x_1})\mathrm{x_3}+2\sqrt{\mathrm{x_3}}\mathrm{u}
    \end{split}
\end{align}
\end{example}
with ${\mathrm{x}}_1$ as a ball displacement, momentum of ball ${\mathrm{x}}_2$, square of the flux linkage associated with the electromagnet ${\mathrm{x}}_3$, mass of the ball $\mathrm{m_b}$, gravity $\mathrm{g}$, resistance of the coil $\mathrm{R_c}$, and a positive constant that depends on the number of coil turns $\mathrm{c_c}$ \cite{Jeltsema_Magnetic, zamaniCDC}. By using the Theorem \ref{theo1} and the $\alpha_{\mathrm{m}}=2$ for $\mathrm{m}=1,...,\mathrm{n}$, the results
\begin{figure}[ht!]
    \centering
    \includegraphics[width=\linewidth]{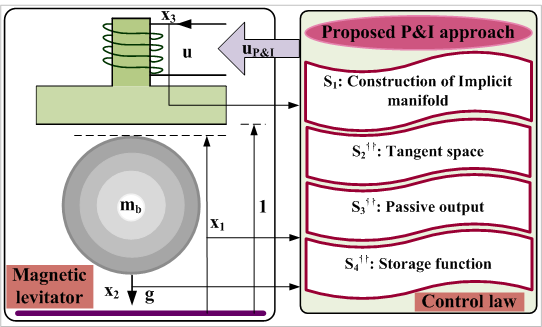}
    \caption{A magnetic levitated ball system with proposed P\&I controller}
    \label{magnetic levitator}
\end{figure}
\begin{equation}
\varphi_1(\mathrm{x_1})=-\mathrm{m_b}\mathrm{x_1}\Rightarrow\mathrm{x_2}+ \mathrm{m_b}\mathrm{x_1}=0,
\end{equation}
\begin{equation}
    \varphi_2(\mathrm{x_1, x_2})=2\mathrm{c_cm_bg}-2\mathrm{c_cx_2}-2\mathrm{c_c}(\mathrm{x_2+m_bx_1})\Rightarrow \mathrm{x_3}-2\mathrm{c_cm_bg}+2\mathrm{c_cx_2}+2\mathrm{c_c}(\mathrm{x_2+m_bx_1})=0,
\end{equation}
and
\begin{align}
    \begin{split}
        \mathrm{b_1}&=\mathrm{x_1}\hspace{1.25cm}\mathrm{b_2}=\mathrm{x_2+(x_2+m_bx_1)}\\
        \mathrm{b_3}&=\mathrm{2c_cx_2+4c_c\left ( \frac{x_3}{2c_c}-m_bg \right )+(x_3-2c_cm_bg+2c_cx_2+2c_c(x_2+m_bx_1))}
    \end{split}
\end{align}
are obtained. The final control law
\begin{small}\label{u_ML}
\begin{equation}
   \mathrm{u}=\frac{-1}{2\sqrt{\mathrm{x_3}}}\left [\frac{-2\mathrm{R_c}}{c_c}(1-\mathrm{x_1})\mathrm{x_3}+ \mathrm{2c_cx_2+4c_c\left ( \frac{x_3}{2c_c}-m_bg \right )+(x_3-2c_cm_bg+2c_cx_2+2c_c(x_2+m_bx_1))} \right ]
\end{equation}
\end{small}
is obtained using the proposed P\&I approach that globally exponentially stabilizes the magnetic levitation system (\ref{MLevit}) to an equilibrium point.
\begin{figure}[ht!]
    \centering
    \includegraphics[width=\linewidth]{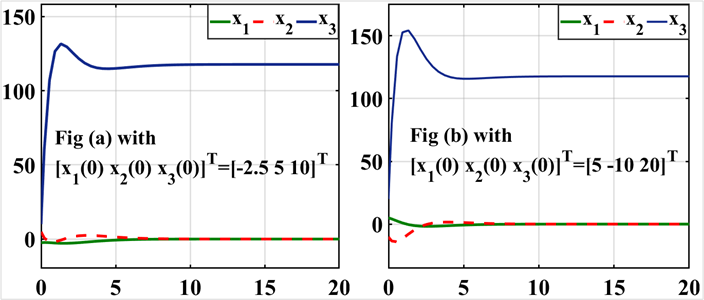}
    \caption{Time evolution of states of magnetic levitation system for different initial conditions}
    \label{ML_fig1}
\end{figure}
The system parameters $\mathrm{m_b}=3$, $\mathrm{g}=9.81$,  $\mathrm{R_c}=10$, and $\mathrm{c_c}=2$ are considered for the simulation results. Fig. \ref{ML_fig1} demonstrates the convergence of system states to equilibrium points for different initial conditions.
 
\begin{figure}[ht!]
    \centering
    \includegraphics[width=\linewidth]{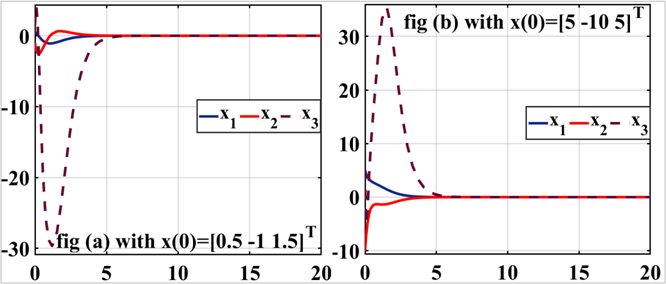}
    \caption{Time evolution of states of DC motor which drives a manipulator arm with a load for different initial conditions}
    \label{Dc fig1}
\end{figure}

\begin{figure}[ht!]
    \centering
    \includegraphics[width=\linewidth]{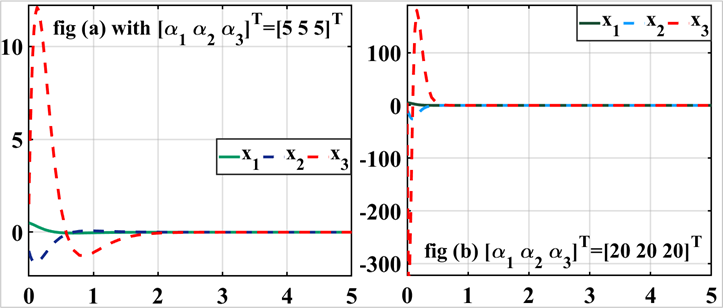}
    \caption{Time evolution of states of DC motor which drives a manipulator arm with a load for different values of the flexible variables $\alpha$}
    \label{Dc fig2}
\end{figure}

\begin{example}\label{DCwithMani}
The obtained results in the Theorem \ref{theo1} are applied to ensure the stabilization and control of an electromechanical system
\begin{align}\label{DCwithMani1}
    \begin{split}
        \dt{\mathrm{x}}_1&=\mathrm{x}_2\\
\dt{\mathrm{x}}_2&=\mathrm{x}_3-\theta_a\mathrm{sin(x}_1)-\theta_b\mathrm{x}_2\\
\dt{\mathrm{x}}_3&=-\theta_c\mathrm{x}_2-\theta_d\mathrm{x}_3+\mathrm{u}\\
    \end{split}
\end{align}
\end{example}
composed by a DC motor which drives a manipulator arm with a load \cite{rabai2013adaptiveInIDcmot, tong2010directDCeex4}. By using the Theorem \ref{theo1},
\begin{align}
    \begin{split}
        \mathrm{b_1}&=\frac{\alpha_1}{2}\mathrm{x_1},\hspace{2.5cm}\mathrm{b_2}=\frac{\alpha_2}{2}\left (\mathrm{x_2} +\frac{\alpha_1}{2}\mathrm{x_1}\right)+\left (\frac{\alpha_1}{2}\mathrm{x_1}\right),\\
\mathrm{b_3}&=\frac{\alpha_3}{2}\left ( \mathrm{x_3}-\theta_a \mathrm{sin(x_1)-\theta_b\mathrm{x_2}+\frac{\alpha_1}{2}}\mathrm{x_2}+\frac{\alpha_2}{2}\left(\mathrm{x_2}+ \frac{\alpha_1}{2}\mathrm{x_1}\right) \right)+\mathrm{x_2}\left ( -\theta_a \mathrm{cos(x_1)+\frac{\alpha_1\alpha_2}{4}} \right )\\&+(\mathrm{x}_3-\theta_a\mathrm{sin(x}_1)-\theta_b\mathrm{x}_2)\left (-\theta_b+ \frac{\alpha_1}{2}+\frac{\alpha_2}{2} \right ),
    \end{split}
\end{align}
and the final control law
\begin{equation}
    \mathrm{u}=-\left (-\theta_c\mathrm{x}_2-\theta_d\mathrm{x}_3+\mathrm{b_3} \right )
\end{equation}
that globally exponentially stabilizes the equilibrium of system (\ref{DCwithMani1}) are directly obtained. According to the parameters listed in \cite{rabai2013adaptiveInIDcmot}, the closed-loop controlled system's results have been simulated. As per the \cite{rabai2013adaptiveInIDcmot}, the obtained parameters are $\theta_a=35.5391$, $\theta_b=0.2821$, $\theta_c=2.3112$, and $\theta_d=3.11\times 10^3$.  The time evolution of states of the system with different initial conditions are depicted in Fig. \ref{Dc fig1}. The overall closed-loop performance is entirely stable and converges to zero for small as well as large initial conditions of system states. 
 The time evolution of states of the system with different values of the flexible variables $\alpha_i$ are depicted in Fig. \ref{Dc fig2}. The convergence time can be adjusted by increasing the values of $\alpha_{\mathrm{i}}$ for $\mathrm{i}=1,2,3$.  The CLS exhibits overshoot during the transient period and settles in finite time as the value of flexible variables increases.

\section{P\&I approach and Techniques based on the Generation of Control Contraction Metric}\label{sec5}
In \cite{IanManchester}, the concept of a control contraction metric (CCM) for NLS is proposed, and it is proved that the existence of a CCM is sufficient to guarantee universal exponential stabilizability. The derived criteria for the existence of CCM in \cite{IanManchester} can be framed as a convex feasibility problem, and the CCM stabilizability condition can be thought of as a differential version of the CLF condition \cite{IanManchester}. However, it is a time-consuming and drawn-out procedure to derive evaluation criteria and a systematic form of CM for general NLS. Moreover, the CCM controller requires the optimization parser YALMIP and the solver Mosek with a semidefinite program to provide a dual metric for controller design. The adaptive control method in \cite{slotinelopez2019contraction} based on the CCM approach has formulated the parameter-dependent dual CM condition and utilizes the adaptive law for parameter estimation \cite{nayyer2022passivitylcss}.
%\end{itemize}

The control design methods based on the generation of CCM can be unified in the proposed  P\&I methodology. Additionally, the target dynamics are set to achieve the parameter-independent implicit manifolds, passive output, and associated storage function to mitigate the impacts of the parametric changes in some systems (wherever possible).
The examples mentioned in this section are solved in a systematic way by the P\&I approach,
\begin{itemize}
    \item without any assumption in the selection of matrix, unlike the I\&I HCP \cite{WangTAC},
    \item without  SOS optimization, solver mosek, and YALMIP,
    \item without any construction of parameter-dependent dual CM and parameter estimation.
\end{itemize}

\begin{example}\label{example 3}
A specific system 
\begin{align}\label{systemT1}
    \begin{split}
       &\dt{\mathrm{x}}_1=-\mathrm{x}_1+\mathrm{x}_3\hspace{1cm}
\dt{\mathrm{x}}_2 =\mathrm{x}_1^2-\mathrm{x}_2-2\mathrm{x}_1\mathrm{x}_3+\mathrm{x}_3\\
&\dt{\mathrm{x}}_3=-\mathrm{x}_2+u
    \end{split}
\end{align}
\end{example}
with $\mathrm{x}^*=0$ from \cite{IanManchester, Andrieuthirdorder} is used to highlight the interest of the proposed P\&I controller. Unbounded nonlinearity is observed in the dynamics of $\mathrm{x_2}$. Since the three-dimensional state $\mathrm{x}$=$[\mathrm{x_1},\mathrm{x_2}, \mathrm{x_3}]$=1/3 fails to meet the rank condition, it is evident that this system is not feedback linearizable.
%The LQR problem necessitates the solution of the algebraic Riccati equation and a linearized version of the system at the origin. The CCM controller requires the optimization parser YALMIP and the solver Mosek with a semidefinite program to provide a dual metric for controller design.

\begin{figure}[ht!]
    \centering
    \includegraphics[width=\linewidth]{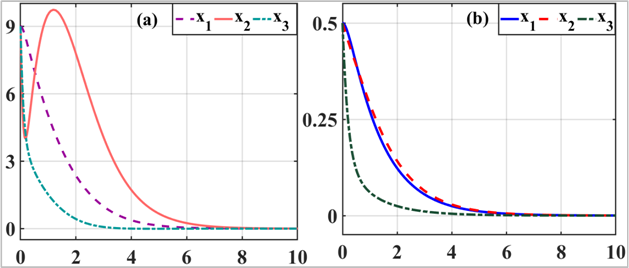}
    \caption{Time evolution of system behavior for different initial conditions}
    \label{Ianexample}
\end{figure}

The convergence of the solution of $\dt{\mathrm{x}}_1$ to zero is simply obtained by substituting $\mathrm{x_3}=-\alpha_1 \mathrm{x_1}$. With ${\mathrm{u}}=0$, $\mathrm{x}_2=\alpha_2\mathrm{x}_3$  the exponential convergence of the solution trajectory of the subsystem $\dt{\mathrm{x}}_3$ to zero is ensured. These general observations provided a implicit manifold vector $\Psi (\mathrm{x})=\mathrm{col}(\alpha_1 \mathrm{x}_1+\mathrm{x}_3, \hspace{0.09cm}\mathrm{x}_2-\alpha_2\mathrm{x}_3)$.
The combined implicit manifold (\ref{systemT2}) for the system (\ref{systemT1}) is obtained by adding the elements of the $\Phi (\mathrm{x})$.
\begin{equation}\label{systemT21}
   \Phi(\mathrm{x}_1,\mathrm{x}_2, \mathrm{x}_3)=\underset{\Psi_a}{\underbrace{ \alpha_1 \mathrm{x}_1+\mathrm{x}_3}}+\underset{\Psi_b}{\underbrace{\mathrm{x}_2-\alpha_2\mathrm{x}_3}}=0
\end{equation}
The solution of (\ref{exponential laypunov}) and (\ref{final control law}) with steps $(S_2), (S_3)$, and $(S_4)$ gives
\begin{small}
\begin{align}\label{systemT7}
    \begin{split}
       \mathrm{u_{P\&I}}&=\left ( \frac{-1}{1-\alpha_2} \right ) ( \alpha_1(-\mathrm{x}_1+\mathrm{x}_3)  +(\mathrm{x}_1^2-\mathrm{x}_2-2\mathrm{x}_1\mathrm{x}_3\\& +\mathrm{x}_3) - \mathrm{x}_2 +\frac{\alpha}{2}(\alpha_1 \mathrm{x}_1+\mathrm{x}_2+(1-\alpha_2) \mathrm{x}_3)).
    \end{split}
\end{align}
\end{small}
The solution of system (\ref{systemT1}) with control law (\ref{systemT7}) to initial state $\mathrm{x}_0=[0.5, 0.5, 0.5]^{\mathrm{T}}$ in Fig. \ref{Ianexample}(b) and $\mathrm{x}_0=[9, 9, 9]^{\mathrm{T}}$ in Fig. \ref{Ianexample}(a). Fig. \ref{Ianexample} with $\alpha_1=2, \alpha_2=10$, and $\alpha=20$ exhibits the  GAS behaviour of the proposed controller. 
For the small as well as larger initial conditions, the P\&I approach exhibits the convergence of system trajectories to zero with improved performance over the CCM controller (refer \cite{IanManchester} for comparison purposes).
\begin{example}
Consider a nonlinear contracting system dynamics \cite{slotinelopez2019contraction} 
\begin{align}\label{sle1}
    \begin{split}
        \dt{\mathrm{x}}_1&=\mathrm{x}_3-\theta_{\mathrm{a}}\mathrm{x}_1\\
\dt{\mathrm{x}}_2&= \mathrm{x}_1^2-\mathrm{x}_2\\
 \dt{\mathrm{x}}_3&=\mathrm{tanh(\mathrm{x_2})}-\theta_{\mathrm{b}}\mathrm{x_3}-\theta_{\mathrm{c}}\mathrm{x}_1^2+\mathrm{u}
    \end{split}
\end{align}
\end{example}
with the known parameters $\theta_{\mathrm{a}}$, $\theta_{\mathrm{b}}$, and $\theta_{\mathrm{c}}$ that has no appropriate structure since it is neither feedback linearizable nor in strict-feedback form.  The dual contraction metric takes a non-flat form when the dual  control contraction metric (CCM) condition is formulated using the YALMIP toolbox as a sum-of-squares optimization in \cite{slotinelopez2019contraction}. Furthermore,  it is demonstrated in \cite{slotinelopez2019contraction} that the system cannot be stabilized using a quadratic Lyapunov function. A simple and systematic controller is built in the proposed P\&I approach through the construction of the off-the-manifold by selecting the second-order target dynamics and utilizing the corresponding passive output and storage function.

A straightforward and systematic controller is built to perform regulation as well as tracking problems via the proposed P\&I approach.

\textbf{Regulation problem:}
A common practice to get an implicit manifold is to equate the right-hand side $\mathrm{x}_3-\theta_{\mathrm{a}}\mathrm{x}_1$ with $-\alpha_1\mathrm{x}_1$ and $-\theta_{\mathrm{b}} \mathrm{x}_1^2-\mathrm{x}_2$ with $-\alpha_2\mathrm{x}_2$. This ultimately leads to a combined off-the-manifold
\begin{equation}\label{sdkjndn}
     \mathrm{x}_3-\theta_{\mathrm{a}}\mathrm{x}_1+\alpha_1\mathrm{x}_1-\theta_{\mathrm{b}} \mathrm{x}_1^2-\mathrm{x}_2+\alpha_2\mathrm{x}_2=0.
\end{equation}
\begin{remark}
Implementing the four-step P\&I approach culminates in the \textit{parameter-dependent pseudo-Riemannian metric, passive output, and storage function} for the aforementioned parameter-based implicit manifold (\ref{sdkjndn}). The solution obtained from this implicit manifold requires the parameter estimations \cite{shadab2022persistence, shadab2023finite, shadab2022gaussian, ShadabDREM}.
%The implementation of four step procedure of the P\&I approach to above-mentioned parameter-based off-the-manifold (\ref{sdkjndn}) provides the parameter-dependent pseudo-Riemannian metric, passive output, and storage function.
\end{remark}
It is clearly seen that the solutions of the
$\dt{\mathrm{x}}_1=\mathrm{x}_3-\theta_{\mathrm{a}}\mathrm{x}_1$  and $\dt{\mathrm{x}}_2= -\theta_{\mathrm{b}}\mathrm{x}_1^2-\mathrm{x}_2$ of (\ref{sle1}) can be  globally exponentially stabilize by selecting $\mathrm{x_3}=-\alpha_1\mathrm{x_1}$ and $\mathrm{x}_1^2=\alpha_2\mathrm{x_2}$ respectively. 
The selection of such virtual control laws leads to a subsystem  
\begin{align}\label{seconb}
    \begin{split}
        \dt{\mathrm{x}}_1=-(\alpha_1+\theta_{\mathrm{a}})\mathrm{x}_1\\
\dt{\mathrm{x}}_2=-(1+\alpha_2\theta_{\mathrm{b}}) \mathrm{x}_2.
    \end{split}
\end{align}
For exponential convergence of the solutions of $\dt{\mathrm{x}}_1$ and $\dt{\mathrm{x}}_2$, the conditions  drawn from (\ref{subtrac}) are as follows:
\begin{align}\label{regccondition}
    \begin{split}
        &(\theta_{\mathrm{a}}+\alpha_1)>>0\\
        &\theta_{\mathrm{b}}\alpha_2\approx 0
    \end{split}
\end{align}
Based on the above conditions (\ref{regccondition}), the exponential convergence of the subsystem (\ref{seconb}) is guaranteed by selecting the proper values of flexible variables $(\alpha_1$ and $\alpha_2)$.  It means the selected flexible variables $(\alpha_1$ and $\alpha_2)$ in the final control law robustify the performance of the closed-loop system and make it parameter insensitive.
By setting $\alpha_1>>0$ and $\alpha_2\approx 0$, it is observed that the solution trajectory of the subsystem (\ref{seconb}) converges exponentially to zero irrespective of the values of $\theta_{\mathrm{a}}$ and $\theta_{\mathrm{b}}$.
The set of allowable parameter variations, for instance, is
$\theta_{\mathrm{a}}\in(-10, 10)$ for $\alpha_1=10$.  Similarly, $\theta_{\mathrm{2}}\alpha_2\approx 0$ must be satisfied for exponential convergence. Therefore,  the value of $\alpha_2$ should be close to zero or set to zero.
\begin{remark}
By adjusting the values of the flexible variables $\alpha_1$ and $\alpha_2$, the impact of the parametric changes in the system can be mitigated.
\end{remark}
The implicit manifolds are derived as:
\begin{align}\label{implimanif}
    \begin{split}
        &\mathrm{x_3}+\alpha_1\mathrm{x_1}=0\\
        &\mathrm{x}_1^2-\alpha_2\mathrm{x_2}=0
    \end{split}
\end{align}
The derived implicit manifolds (\ref{implimanif}) are used to obtain the \textit{parameter-independent combined implicit manifold}
\begin{small}
\begin{equation}\label{slot2}
   \Psi(\mathrm{x}_1,\mathrm{x}_2, \mathrm{x}_3)= \underset{\Psi_a}{\underbrace{ \alpha_1 \mathrm{x}_1+\mathrm{x}_3}}+\underset{\Psi_b}{\underbrace{\mathrm{x}_1^2-\alpha_2\mathrm{x}_2}}=0.
\end{equation}
\end{small}

\begin{figure}[ht!]
    \centering
    \includegraphics[width=\linewidth]{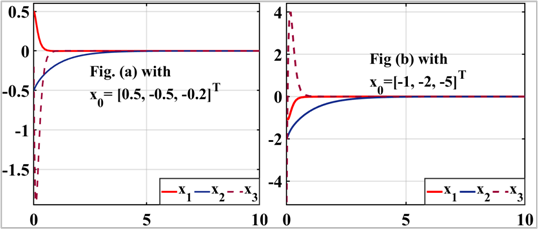}
    \caption{Time evolution (in second) of system behavior for different initial conditions with $\alpha_1=10, \alpha_2=0.1$, and $\alpha=20$}
    \label{slotinenewmatchedexam}
\end{figure}

\begin{figure}[ht!]
    \centering
    \includegraphics[width=\linewidth]{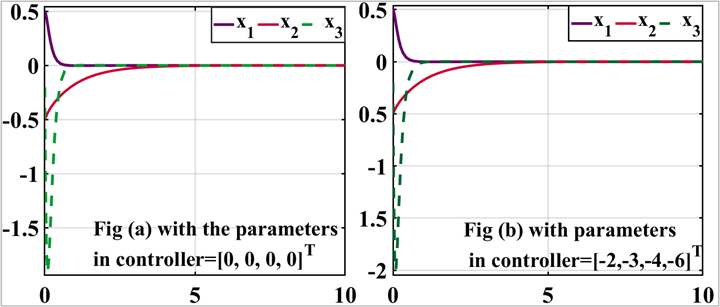}
    \caption{Time evolution (in second) of system behavior for different parameters in the controller $\alpha_1=10, \alpha_2=0.1$, and $\alpha_3=20$}
    \label{slotinenewmatchedexamf2}
\end{figure}

All the three states $\mathrm{x}_1$, $\mathrm{x}_2$, and $\mathrm{x}_3$  are appeared in the combined off-the-manifold (\ref{slot2}). Therefore, the P\&I based control law
\begin{small}
\begin{align}\label{kjdbkjjkdf}
    \begin{split}
        &\mathrm{u_{P\&I}}=-(\mathrm{tanh(\mathrm{x_2})}-\theta_{\mathrm{c}}\mathrm{x_3}-\theta_{\mathrm{d}}\mathrm{x}_1^2+(\alpha_1+2\mathrm{x}_1)(\mathrm{x}_3-\theta_{\mathrm{a}}\mathrm{x}_1)\\&-\alpha_2(-\theta_{\mathrm{b}} \mathrm{x}_1^2-\mathrm{x}_2)+\frac{\alpha}{2}({\alpha_1 \mathrm{x}_1+\mathrm{x}_3}+{\mathrm{x}_1^2-\alpha_2\mathrm{x}_2}))
    \end{split}
\end{align}
\end{small}
is obtained via the proposed P\&I approach in four steps $(\textsl{S}_1-\textsl{S}_4^{\dagger\dagger}$) by defining $\mathrm{R}=\triangledown \Psi(\mathrm{x})^{\mathrm{T}}\triangledown \Psi(\mathrm{x})$
\begin{equation}\label{pseudir}
   = \begin{bmatrix}
(\alpha_1+2\mathrm{x_1})^2 &-\alpha_2(\alpha_1+2\mathrm{x_1})  & (\alpha_1+2\mathrm{x_1})\\ 
 -\alpha_2(\alpha_1+2\mathrm{x_1})& \alpha_2^2 &-\alpha_2 \\ 
 (\alpha_1+2\mathrm{x_1})& -\alpha_2 & 1
\end{bmatrix}
\end{equation}
and $\mathrm{y}=\int_{0}^{t}(\dt{\mathrm{x}}_3+\mathrm{m}_{33}^{-1}\mathrm{m}_{23}\dt{\mathrm{x}}_2+\mathrm{m}_{33}^{-1}\mathrm{m}_{31}\dt{\mathrm{x}}_1)\mathrm{dt}$ with 
\begin{small}
\begin{align}
    \begin{split}
        \mathbb{S}(\mathrm{x}_1,\mathrm{x}_2, \mathrm{x}_3)=\frac{1}{2}({\alpha_1 \mathrm{x}_1+\mathrm{x}_3}+{\mathrm{x}_1^2-\alpha_2\mathrm{x}_2})^2
    \end{split}
\end{align}
\end{small}
where, 
$\mathrm{m}_{33}^{-1}\mathrm{m}_{23}=\mathrm{m}_{33}^{-1}\mathrm{m}_{32}=-\alpha_2$ and $\mathrm{m}_{33}^{-1}\mathrm{m}_{31}=\mathrm{m}_{33}^{-1}\mathrm{m}_{13}=(\alpha_1+2\mathrm{x_1})$. 
Solving the condition $\dt{\mathbb{S}}\leq - \alpha \mathbb{S}$ ensures the overall GAS behaviour of the system trajectories to zero.
Based on the selection of the combined off-the-manifold, the solution trajectories of $\dt{\mathrm{x}}_1$ and  $\dt{\mathrm{x}}_2$ exhibit the GES behaviour. As the second-order target dynamics is considered, the solution of system (\ref{sle1}) with the nominal parameters $\begin{bmatrix}
\theta_{\mathrm{a}}&\theta_{\mathrm{b}}
&\theta_{\mathrm{c}}&\theta_{\mathrm{d}} 
\end{bmatrix}^{\mathrm{T}}=\begin{bmatrix}
-0.3 & -0.8 & -0.25 & -0.75
\end{bmatrix}^{\mathrm{T}}$, and control law (\ref{kjdbkjjkdf}) to the system with initial state $\mathrm{x}_0=[0.5, -0.5, -0.2]^{\mathrm{T}}$ in  left side and $\mathrm{x}_0=\begin{bmatrix}
-0.3 & -0.8 & -0.25 & -0.75
\end{bmatrix}^{\mathrm{T}}$ in  right side of Fig. \ref{slotinenewmatchedexam} exhibits the  GAS behaviour of the proposed controller. 
The parameters considered in the proposed controller are different from the actual system parameters in order to evaluate the robustness of the controller. The closed-loop performance of the system (\ref{sle1}) with the nominal parameters and the controller (\ref{kjdbkjjkdf})  with $\begin{bmatrix}
0 & 0 & 0 & 0
\end{bmatrix}^{\mathrm{T}}$ parameters are shown in Fig. \ref{slotinenewmatchedexamf2}(a). Fig. \ref{slotinenewmatchedexamf2}(b) represents the overall performance with  $\begin{bmatrix}
-2 & -3 & -4 & -6 
\end{bmatrix}^{\mathrm{T}}$ and initial condition $\mathrm{x_0}=\begin{bmatrix}
0.5 & -0.5 & -0.2
\end{bmatrix}^{\mathrm{T}}$
\begin{remark}\label{remksdvkjdsbj}
The obtained off-the-manifold (\ref{slot2}) and associated pseudo-Riemannian metric (\ref{pseudir}) are free from the matched and extended matched parametric uncertainties. As a result, the performance of the closed-loop system with the proposed P\&I controller remains unaffected even after setting all the parameter values in the controller to zero.
\end{remark}
\textbf{Tracking problem}
The aim of this tracking problem is to follow a desired trajectory $(\mathrm{x_d}, \mathrm{u_d})$ governed by a reference $\mathrm{x}_{1d}$. The tracking error is defined as:
\begin{equation}
    \mathrm{e_1}=\mathrm{x_1}-\mathrm{x}_{1d}\Rightarrow \mathrm{x_1}=\mathrm{e_1}+\mathrm{x}_{1d}
\end{equation}
Then, its derivative and modified dynamics are as follows:
\begin{align}\label{tr1}
    \begin{split}
        \dt{\mathrm{e}}_1&=\mathrm{x}_3-\theta_{\mathrm{a}}(\mathrm{e}_1+\mathrm{x}_{1d})-\dt{\mathrm{x}}_{1d} \\
\dt{\mathrm{x}}_2&=-\theta_{\mathrm{b}} (\mathrm{e}_1+\mathrm{x}_{1d})^2-\mathrm{x}_2\\
 \dt{\mathrm{x}}_3&=\mathrm{tanh(\mathrm{x_2})}-\theta_{\mathrm{c}}\mathrm{x_3}-\theta_{\mathrm{d}}(\mathrm{e}_1+\mathrm{x}_{1d})^2+\mathrm{u}.
    \end{split}
    \end{align}
The aim is to design a control law  that drives $ \mathrm{x}_1$ to $ \mathrm{x}_{1d}$. The objective is formulated so that the error reaches zero irrespective of parameter variations. Therefore, the  virtual control laws
\begin{align}\label{troffm}
    \begin{split}
       & \mathrm{x}_3=\theta_{\mathrm{a}}\mathrm{x}_{1d}+\dt{\mathrm{x}}_{1d}-\alpha_1\mathrm{e}_{1}\Rightarrow \mathrm{x}_3-\theta_{\mathrm{a}}\mathrm{x}_{1d}+\dt{\mathrm{x}}_{1d}-\alpha_1\mathrm{e}_{1}=0 \\ &(\mathrm{e}_1+\mathrm{x}_{1d})^2=\alpha_2\mathrm{x}_2\Rightarrow (\mathrm{e}_1+\mathrm{x}_{1d})^2-\alpha_2\mathrm{x}_2=0
    \end{split}
\end{align}
are defined in such way that the externally manipulating variables $(\alpha_1)$ and $(\alpha_2)$ dominate the system parameters. 
It leads to a subsystem
\begin{align}\label{subtrac}
\begin{split}
       &\dt{\mathrm{e}}_1=-(\theta_{\mathrm{a}}+\alpha_1)\mathrm{e}_1\\ 
       &\dt{\mathrm{x}}_2=-(1+\theta_{\mathrm{b}}\alpha_2)\mathrm{x}_2
    \end{split}
\end{align}
For exponential convergence of the solutions of $\dt{\mathrm{e}}_1$ and $\dt{\mathrm{x}}_2$, the conditions  drawn from (\ref{subtrac}) are as follows:
\begin{align}\label{traccondition}
    \begin{split}
        &(\theta_{\mathrm{a}}+\alpha_1)>>0\\
        &\theta_{\mathrm{b}}\alpha_2\approx 0
    \end{split}
\end{align}
\begin{remark}
In both the regulation and tracking problems, the criteria/conditions obtained for exponential convergence of state trajectories are identical. Therefore, the Remark \ref{remksdvkjdsbj} is also applicable for tracking problem.
\end{remark}
In order to get the final control law, the combined off-the-manifold 
\begin{equation}\label{tr2}
    \mathrm{x}_3-\theta_{\mathrm{a}}\mathrm{x}_{1d}-\dt{\mathrm{x}}_{1d}+\alpha_1\mathrm{e}_{1} +(\mathrm{e}_1+\mathrm{x}_{1d})^2-\alpha_2\mathrm{x}_2=0
\end{equation}
is obtained by adding the two implicit manifolds derived in (\ref{troffm}). The  control law
\begin{align}\label{trackinglaw}
    \begin{split}
        &\mathrm{u_{P\&I}}=-(\mathrm{tanh(\mathrm{x_2})}-\theta_{\mathrm{c}}\mathrm{x_3}-\theta_{\mathrm{d}})(\mathrm{e}_1+\mathrm{x}_{1d})^2+(\alpha_1+\\&2(\mathrm{e}_1+\mathrm{x}_{1d}))(\mathrm{x}_3-\theta_{\mathrm{a}}(\mathrm{e}_1+\mathrm{x}_{1d}))-\alpha_2(-\theta_{\mathrm{b}} \mathrm{x}_1^2-\mathrm{x}_2)\\&+\frac{\alpha}{2}({\alpha_1 (\mathrm{e}_1+\mathrm{x}_{1d})+\mathrm{x}_3}+{(\mathrm{e}_1+\mathrm{x}_{1d})^2-\alpha_2\mathrm{x}_2}))
    \end{split}
\end{align}
is obtained by utilizing the four steps of the proposed P\&I approach.

\begin{figure}[ht!]
    \centering
    \includegraphics[width=\linewidth]{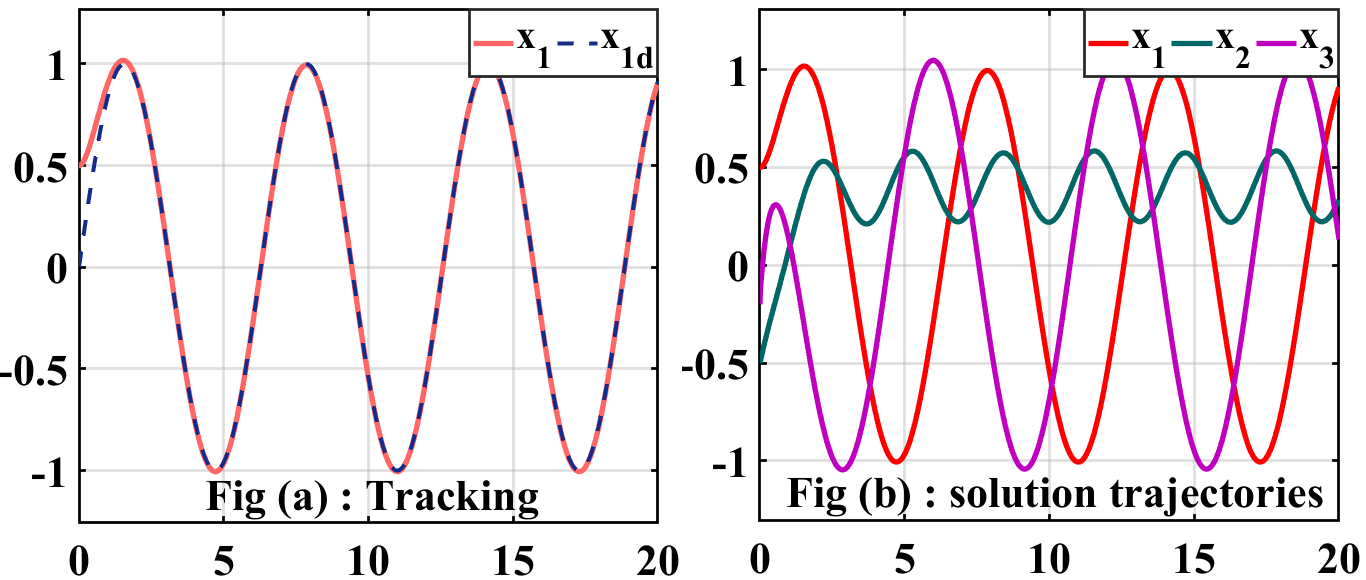}
    \caption{Time evolution (in second) of system behavior for different parameters in the controller $\alpha_1=200, \alpha_2=0.1$, and $\alpha=2$}
    \label{tracking1}
\end{figure}

The virtual control laws are selected to achieve the least number of system parameters in the combined implicit manifold. The selection of target dynamics ensures the exponential convergence of $\mathrm{e}_1(t)$ and ${\mathrm{x}}_2(t)$ by satisfying the condition (\ref{traccondition}). The tracking performance of the CLS with the nominal parameters $\begin{bmatrix}
\theta_{\mathrm{a}}&\theta_{\mathrm{b}}
&\theta_{\mathrm{c}}&\theta_{\mathrm{d}} 
\end{bmatrix}^{\mathrm{T}}=\begin{bmatrix}
-0.3 & -0.8 & -0.25 & -0.75
\end{bmatrix}^{\mathrm{T}}$ and control law (\ref{trackinglaw}) to the system with initial state $\mathrm{x}_0=[0.5, -0.5, -0.2]^{\mathrm{T}}$  is shown in Fig. \ref{tracking1}. The overall CLS exhibits the  GAS behaviour of the proposed controller.  The P\&I guarantees the exponential convergence of $\mathrm{e}_1(t)$ and satisfies the objective of designing a control law that drives $\mathrm{x}_1$ to $\mathrm{x}_{1d}$ as depicted in Fig. \ref{tracking1}. 
The convergence rate is decided by selecting the proper values of $\alpha_1$ and $\alpha_2$. The convergence can be made faster by increasing the values of $\alpha_1$ and $\alpha_3$ and by choosing $\alpha_2 \approx 0$. 
The CLS exhibits overshoot during the transient period and settles in finite time as the value of flexible variables increases.

\begin{figure}[ht!]
    \centering
    \includegraphics[width=\linewidth]{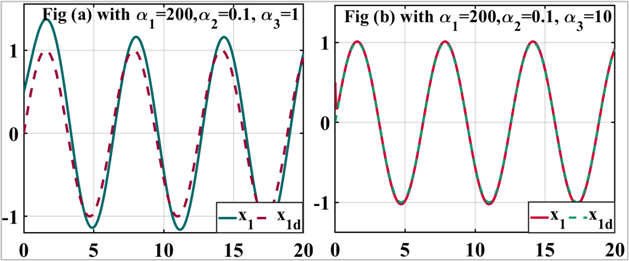}
    \caption{Time evolution (in second) of system behavior for different parameters in the controller $\alpha_1=200, \alpha_2=0.1$, and $\alpha=20$}
    \label{tracking2}
\end{figure}

\begin{figure}[ht!]
    \centering
    \includegraphics[width=\linewidth]{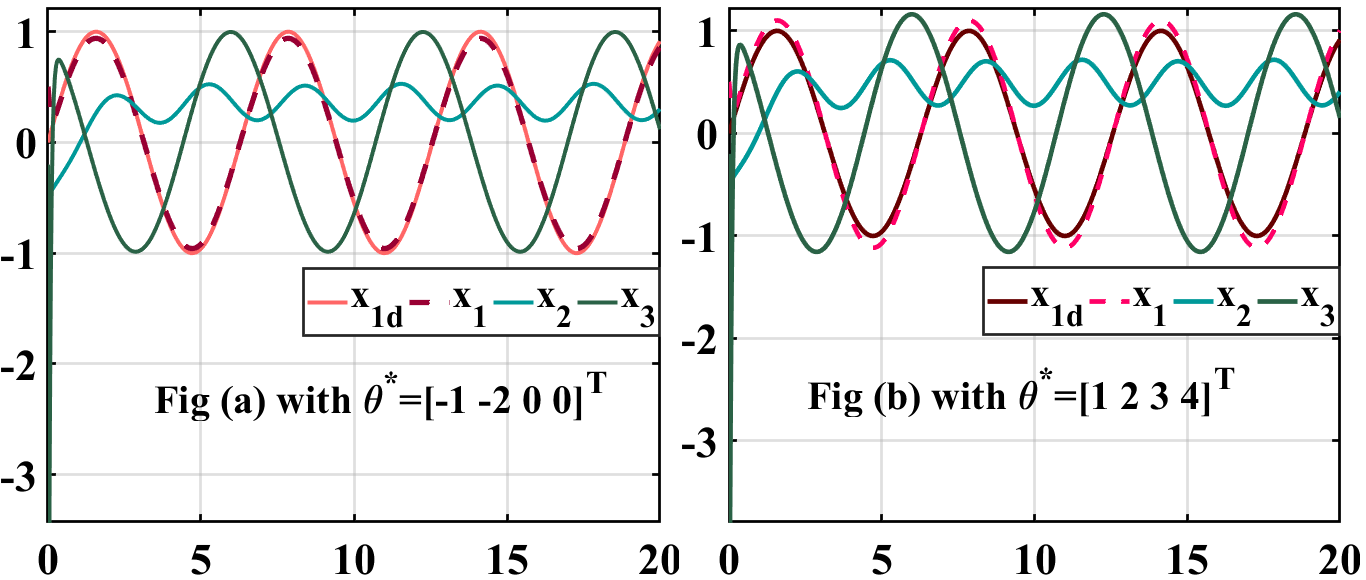}
    \caption{Time evolution (in second) of system behavior for different parameters in the controller $\alpha_1=100, \alpha_2=0.00005$, and $\alpha=30$}
    \label{tracking3}
\end{figure}

To assess and verify the performance of the proposed controller, many scenarios are taken into account with various parametric values. As per the Fig. \ref{tracking2} and Fig. \ref{tracking3}, the defined objective of tracking is also met by setting the parameters $\begin{bmatrix}
0 & 0 &0  &0 
\end{bmatrix}^T$, $\begin{bmatrix}
-1 & -2 &0  &0 
\end{bmatrix}^T$, $\begin{bmatrix}
1 & 2 &3  &4 
\end{bmatrix}^T$ in the controller. Moreover, the superior performance is obtained over the universal adaptive control proposed in \cite{LopezUniversal} for the $\theta \in [-0.4, 0.5]\times [-1, 0.6]\times [-0.6, 0.75]\times[-1.75, 0.4]$. Therefore, the proposed P\&I demonstrates system trajectories convergence with improved performance for both small and large parameter values in the controller while preserving the system parameters to nominal values without any parameter estimation.
\section{Role of the implicit manifold and P\&I approach in Systems with specific structure}\label{sec6}

\subsection{Feedforward system}
The stabilization of a class of systems in feedforward form is addressed in this subsection. This type of system can be stabilized using \textit{the classical forwarding approach}, as explained in \cite{Krstic, Sepulchre2011ConstructiveNC, Mazenc}. Forwarding generally applies to the systems in upper-triangular form. Forwarding as a complement to backstepping begins from the input and proceeds forward.  The proposed P\&I method is used to construct a new class of control laws for the feedforward systems.

 Consider a system
\begin{align}
    \begin{split}
        \dt{\mathrm{x}}_1&=\mathrm{f}_1(\mathrm{x}_2)\\
         \dt{\mathrm{x}}_2&=\mathrm{f}_2(\mathrm{x}_2)+\mathrm{g}(\mathrm{x}_2)u
    \end{split}
\end{align}
with $\mathrm{x}=\textrm{col}(\mathrm{x}_1, \mathrm{x}_2)\in \mathbb{R}\times \mathbb{R}^n$, $u \in \mathbb{R}$, $\mathrm{f}_1(0)=0$, $\mathrm{f}_2(0)=0$.

To begin with the P\&I for stabilization of systems in feedforward form, the concept of forwarding i.e., "begins from the input and proceeds forward" is adopted.

For convergence of $\mathrm{x}\mathrm{(t)}$ to equilibrium, the off-manifold $\mathrm{x}_1-\mu(\mathrm{x}_2)=0$ with some smooth function $\mu(\mathrm{x}_2)$ is obtained such that the condition (S2) holds.

The off-the-manifold $\mathrm{x}_1-\mu(\mathrm{x}_2)$ provide

\begin{equation}\label{feefr1}
    \dt{\mathrm{x}}_1=\frac{\partial\mu}{\partial \mathrm{x}_2}\dt{\mathrm{x}}_2\Rightarrow \mathrm{f}_1(\mathrm{x}_2)=\frac{\partial \mu(\mathrm{x}_2)}{\partial \mathrm{x}_2}
\end{equation}
with the set
\begin{equation}\label{feefr2}
    \mathbb{S}=\left \{ \mathrm{x}_2 \in \mathbb{R}^n | \frac{\partial \mu(\mathrm{x}_2)}{\partial \mathrm{x}_2}\mathrm{g}(\mathrm{x}_2) =0\right \}.
\end{equation}
The solution of (\ref{feefr1}) and (\ref{feefr2}) provide the function $\mu(\mathrm{x}_2)$. Hence, a direct application of the proposed P\&I procedure outlined from (\ref{GammaPassive}) to (\ref{final control law}) that gives the control law
\begin{small}
\begin{equation}\label{lawfeedf}
  u=\frac{1}{\left ( \frac{\partial \mu}{\partial \mathrm{x}_2} \right )\mathrm{g}(\mathrm{x}_2)}\left ( \mathrm{f}_1(\mathrm{x}_2)-  \frac{\partial \mu}{\partial \mathrm{x}_2} \mathrm{f}_2(\mathrm{x}_2)+ \alpha(\mathrm{x}_1-\mu(\mathrm{x}_2))\right )
\end{equation}
\end{small}
achieves global stabilization of the closed-loop system.

\begin{example}
Consider a system
\begin{align}\label{feedfordwardexample2}
    \begin{split}
        \dt{\mathrm{x}}_1&=\mathrm{x}_2^3\\
\dt{\mathrm{x}}_2&=-\mathrm{x}_2^3+(1+\mathrm{x}_2^2)u.
    \end{split}
\end{align}
If $u=0$ is considered, then the solution of $\dt{\mathrm{x}}_2=-\mathrm{x}_2^3$ converge to equilibrium point. It gives 
\begin{align}
    \begin{split}
        \dt{\mathrm{x}}_1&=\mathrm{f}_1(\mathrm{x}_2)=\mathrm{x}_2^3\\
\dt{\mathrm{x}}_2&=\mathrm{f}_2(\mathrm{x}_2)=-\mathrm{x}_2^3.
    \end{split}
\end{align}
The off-the-manifold $\mathrm{x}_1-\mu(\mathrm{x}_2)$ provide
\begin{equation}
    \dt{\mathrm{x}}_1=\frac{\partial\mu}{\partial \mathrm{x}_2}\dt{\mathrm{x}}_2\Rightarrow \mathrm{x}_2^3=\frac{\partial\mu}{\partial \mathrm{x}_2}(-\mathrm{x}_2^3)
\end{equation}
The solution of above equation gives $\mu(\mathrm{x}_2)=-\mathrm{x}_2$ and off-the-manifold $\mathrm{x}_1+\mathrm{x}_2=0$. Hence, a direct application of (\ref{lawfeedf}) that gives the control law
\begin{equation}
    u=\frac{-1}{(1+\mathrm{x}_2^2)}\left ( \alpha(\mathrm{x}_1+\mathrm{x}_2) \right )
\end{equation}
achieves global stabilization of the closed-loop system.
\end{example}

\begin{example}\label{sepulchreexamplefor}
Consider a strict feedforward system \cite{Sepulchre2011ConstructiveNC, Krstic}
\begin{align}\label{strictfeedforwardsysSepulvhre}
    \begin{split}
        \dt{\mathrm{x}}_1&=\mathrm{x}_2+\mathrm{g}_1(\mathrm{x}_2)u\\
\dt{\mathrm{x}}_2&=u
    \end{split}
\end{align}
with $\mathrm{g}_1(\mathrm{x}_2)=-\mathrm{x}_2^2$ is continuous and $\mathrm{g}_1(0)=0$. For the subsystem $\dt{\mathrm{x}}_2=u$, the storage function $\mathbb{S}_2=\frac{1}{2}\mathrm{x}_2^2$  provides a stabilizing feedback law $u=-\mathrm{x}_2$. With $u=-\mathrm{x}_2$, the complete system augmented by $\dt{\mathrm{x}}_2=-\mathrm{x}_2$ is written as
\begin{align}\label{augemneted dyn}
    \begin{split}
         \dt{\mathrm{x}}_1&=\mathrm{x}_2+\mathrm{x}_2^3\\
\dt{\mathrm{x}}_2&=-\mathrm{x}_2
    \end{split}
\end{align}
with the interconnection term $\mathrm{x}_2+\mathrm{x}_2^3$.
The off-the-manifold $\mathrm{x}_1+\mathrm{x}_2+\mathrm{x}_2^3=0$ can be obtained as $\dt{\mathrm{x}}_1=-\mathrm{x}_1$ is responsible for exponential stabilization of $\mathrm{x}_1$.

\begin{equation}\label{storaS1}
    \mathbb{S}_1=\underset{\mathrm{previous}}{\underbrace{\mathbb{S}_2}}+\frac{1}{2}\underset{\mathrm{off-the-manifold}}{\underbrace{(\mathrm{x}_1+\mathrm{x}_2+\mathrm{x}_2^3})}^2
\end{equation}
\begin{remark}
It is demonstrated in the above examples that the interconnection terms for CTCLF are determined directly from the off-the-manifold rather than by solving integral terms and system equations (Refer Ch. 6 of \cite{Sepulchre2011ConstructiveNC}).
\end{remark}
The time-derivative of (\ref{storaS1}) satisfies
\begin{equation}\label{S1dot1}
    \dt{\mathbb{S}}_1|_{u=-\mathrm{x}_3}=-\mathrm{x}_2^2+(\mathrm{x}_1+\mathrm{x}_2+\mathrm{x}_2^3)(-2\mathrm{x}_2^3).
\end{equation}
If $\mathrm{x}_1+\mathrm{x}_2+\frac{1}{3}\mathrm{x}_2^3=0$  as a modified version of the off-the-manifold is considered for ease of calculation, then the time-derivative of modified version of (\ref{storaS1}) is obtained as
\begin{equation}\label{S1dot2}
    \dt{\mathbb{S}}_1|_{u=-\mathrm{x}_3}=-\mathrm{x}_2^2.
\end{equation}
A conclusion “the control law $u=-\mathrm{x}_2$ has not guaranteed the asymptotic stability of augmented system (\ref{augemneted dyn})" can be drawn from (\ref{S1dot1}) and (\ref{S1dot2}). A new virtual control $v_1$ is added to get a control law $u=-\mathrm{x}_2+v_1$ that stabilizes the complete system. The time-derivative of (\ref{storaS1}) with a modified version of the off-the-manifold 
\begin{equation}
        \dt{\mathbb{S}}_1=-\mathrm{x}_2^2+\mathrm{x}_2v_1+(\mathrm{x}_1+\mathrm{x}_2+\frac{1}{3}\mathrm{x}_2^3)(v_1)
  \end{equation}
for the system
\begin{align}
    \begin{split}
            \dt{\mathrm{x}}_1&=\mathrm{x}_2+\mathrm{x}_2^3-\mathrm{x}_2v_1\\
\dt{\mathrm{x}}_2&=-\mathrm{x}_2+v_1
    \end{split}
\end{align}
can be rendered negative with 
  \begin{equation}
      v_1=-\mathrm{x}_2-(\mathrm{x}_1+\mathrm{x}_2+\frac{1}{3}\mathrm{x}_2^3).
       \end{equation}
The control law
       \begin{equation}
           u=-\mathrm{x}_2-\mathrm{x}_2-(\mathrm{x}_1+\mathrm{x}_2+\frac{1}{3}\mathrm{x}_2^3)
       \end{equation}
       guarantees GAS of the equilibrium $(\mathrm{x}_1, \mathrm{x}_2)=(0,0)$ of (\ref{strictfeedforwardsysSepulvhre}).
\end{example}

\begin{example}
    Consider a strict feedforward system
\begin{align}\label{f1}
    \begin{split}
       & \mathrm{\dt{x}_1=x_2+x^2_3+x_2\hspace{0.05cm}u}\\
&\mathrm{\dt{x}_2=x_3-x^2_3\hspace{0.05cm}u}\\
&\mathrm{\dt{x}_3=u}.
    \end{split}
\end{align}
\end{example}
As $\mathrm{u}$ is available all the three equations, (i.e., dynamics of $\mathrm{\dt{x}_1}, \mathrm{\dt{x}_2}$, and $\mathrm{\dt{x}_3}$) it is difficult to get implicit manifold directly. As the system is available in upper triangular form, therefore, a “bottom-up” recursive procedure can be applied. Therefore, considering a subsystem
\begin{align}\label{f2}
    \begin{split}
       &\mathrm{\dt{x}_2=x_3-x^2_3\hspace{0.05cm}u}\\
&\mathrm{\dt{x}_3=u}.
    \end{split}
\end{align}
In order to stabilize the solution of $\mathrm{\dt{x}_3}$  to zero exponentially, $\mathrm{u}=-\mathrm{x_3}$ should be chosen. With $\mathrm{u}=-\mathrm{x_3}$, the above subsystem (\ref{f2}) is modified as
\begin{align}\label{f3}
    \begin{split}
       &\mathrm{\dt{x}_2=x_3+x^3_3}\\
&\mathrm{\dt{x}_3=-x_3}.
    \end{split}
\end{align}
With the help of the storage function (Lyapunov function) $\mathbb{S}_3=\frac{1}{2}\mathrm{x^2_3}$, it is seen that the selection of $\mathrm{u}=-\mathrm{x_3}$ only stabilizes the solution of $\mathrm{\dt{x}_3}$ but not the solution of $\mathrm{\dt{x}_2}$. Therefore, the above subsystem is modified as
\begin{align}\label{f4}
    \begin{split}
       &\mathrm{\dt{x}_2=x_3+x^3_3}\\
    &\mathrm{\dt{x}_3=-x_3+v_2}.
    \end{split}
\end{align}
In order to stabilize the solution of $\mathrm{\dt{x}_2}$, the target dynamics is chosen such that
\begin{equation}\label{f5}
    \mathrm{\dt{x}_2}=-\mathrm{x_2}=\mathrm{x_3+x^3_3}.
\end{equation}
This selection of target dynamics $\mathrm{\dt{x}_2}=-\mathrm{x_2}$ provides the implicit manifold $\mathrm{x_2}+\mathrm{x_3+x^3_3}=0$. For simplification purposes, the obtained implicit manifold is written as:
\begin{equation}\label{f6}
    \Psi_1(\mathrm{x_2, x_3})= \mathrm{x_2}+\mathrm{x_3+\frac{1}{3}x^3_3}=0.
\end{equation}
By applying the proposed PI\&I approach, the storage function
$\frac{1}{2}\left ( \mathrm{x_2}+\mathrm{x_3+\frac{1}{3}x^3_3} \right )^2$ is combined with $\mathbb{S}_3$ for the augmented subsystem. This provides the combined storage function 
\begin{equation}\label{f7}
    \mathbb{S}_2=\frac{1}{2}\mathrm{x^2_3}+\frac{1}{2}\left ( \mathrm{x_2}+\mathrm{x_3+\frac{1}{3}x^3_3} \right )^2
\end{equation}
to get $\mathrm{v_2}$ for the subsystem (\ref{f4}).
The time derivative
\begin{equation}\label{f8}
    \dt{\mathbb{S}}_2=\mathrm{-x^2_3+x_3v_2+\left ( \mathrm{x_2}+\mathrm{x_3+\frac{1}{3}x^3_3} \right )\left ( 1+x^2_3 \right )v_2}
\end{equation}
of (\ref{f7}) helps in getting the
\begin{equation}\label{f9}
    \mathrm{v_2}=\mathrm{-\left ( \mathrm{x_2}+\mathrm{x_3+\frac{1}{3}x^3_3} \right )\left ( 1+x^2_3 \right )}.
\end{equation}
The equation $\mathrm{v_2}$ (\ref{f9}) is obtained by satisfying $\dt{\mathbb{S}}_2 \leq 0$ i.e., the Lyapunov condition of stabilty.
The control law 
\begin{equation}\label{f10}
    \mathrm{u=-x_3-\left ( \mathrm{x_2}+\mathrm{x_3+\frac{1}{3}x^3_3} \right )\left ( 1+x^2_3 \right )}
\end{equation}
from (\ref{f4}) and (\ref{f9}) is obtained and stabilizes the subsystem (\ref{f4}). The dynamics of $\dt{\mathrm{x}}_2$ and $\dt{\mathrm{x}}_3$ is stabilized using (\ref{f10}).
In order to stabilize the complete system, the value of $\mathrm{u}$ (\ref{f10}) in the original system dynamics (\ref{f1}) is substituted. Upon substitution,  dynamics of $\dt{\mathrm{x}}_1$ is written as:
\begin{equation}\label{f11}
    \dt{\mathrm{x}}_1=\mathrm{x_2+x^2_3-x_2x_3-x_2\left ( \mathrm{x_2}+\mathrm{x_3+\frac{1}{3}x^3_3} \right )(1+x^2_3)}
\end{equation}
In order to stabilize the solution of $\mathrm{\dt{x}_1}$, the target dynamics is chosen such that
\begin{equation}\label{f12}
    \mathrm{\dt{x}_1}=-\mathrm{x_1}=\mathrm{x_2+x^2_3-x_2x_3-x_2\left ( \mathrm{x_2}+\mathrm{x_3+\frac{1}{3}x^3_3} \right )(1+x^2_3)}.
\end{equation}
This selection of target dynamics $\mathrm{\dt{x}_1}=-\mathrm{x_1}$ provides the implicit manifold 
\begin{equation}\label{f13}
    \Psi_2(\mathrm{x_1, x_2, x_3})=\mathrm{x_1}+\mathrm{x_2+x^2_3-x_2x_3-x_2\left ( \mathrm{x_2}+\mathrm{x_3+\frac{1}{3}x^3_3} \right )(1+x^2_3)}=0.
\end{equation}
To get the complete storage function $\mathbb{S}=\mathbb{S}_2+\mathbb{S}_1$, the storage function $\mathbb{S}_1$ is required.
In order to get $\mathbb{S}_1$ using (\ref{f13}), a Pseudo-Riemannian metric
\begin{equation}
    \triangledown \Psi_2(\mathrm{x_1, x_2, x_3})^{\mathrm{T}}\triangledown \Psi_2(\mathrm{x_1, x_2, x_3})=\begin{bmatrix}
\mathrm{m}_{11} & \mathrm{m}_{12} &\mathrm{m}_{13} \\ 
 \mathrm{m}_{21}& \mathrm{m}_{22} & \mathrm{m}_{23}\\ 
\mathrm{m}_{31} &\mathrm{m}_{32}  & \mathrm{m}_{33}
\end{bmatrix}
\end{equation}
is needed to define. \textbf{ The definition of passive output for the feedforward system is given as: }

\begin{equation}
    \mathrm{y}=\int_{0}^{t}\left ( \dt{\mathrm{x}}_1+\begin{bmatrix}
\frac{\partial \mathrm{q(x_2, x_3)}}{\partial \mathrm{x_2}}  & 
\frac{\partial \mathrm{q(x_2, x_3)}}{\partial \mathrm{x_3}} 
\end{bmatrix}\begin{bmatrix}
\dt{\mathrm{x}}_2\\ 
\dt{\mathrm{x}}_3
\end{bmatrix} \right )\mathrm{dt}.
\end{equation}

\begin{equation}
\mathrm{The \hspace{0.1cm}term} \hspace{0.9cm}   \int_{0}^{t}\left (\frac{\partial \mathrm{q(x_2, x_3)}}{\partial \mathrm{x_2}} \dt{\mathrm{x}}_2 +
\frac{\partial \mathrm{q(x_2, x_3)}}{\partial \mathrm{x_3}} \dt{\mathrm{x}}_3 \right )\mathrm{dt}
\end{equation}
\textbf{can not be obtained in the separable form.} It is clearly seen that the passive output contains nonlinear multiplicative terms of $\mathrm{ x_2, x_3}$. Finding an explicit solution to a partial differential equation (PDE), which may be difficult or
even impossible to solve is a crucial stage in the different constructive approaches to stabilizing the systems in upper
triangular form. This function must be numerically evaluated or analytically approximated.

\textit{Discussion:}  
After completing the construction of the second-order stabilizing control law in Example \ref{sepulchreexamplefor} by considering the cross-term as implicit manifold, the stabilization of many third and higher-order systems through CTCLF will no longer have a closed-form expression, i.e., the CTCLF will contain nonlinear multiplicative terms of $\mathrm{x_1}, \mathrm{x_2},..., \mathrm{x_n}$ in the off-the-manifold.  \textbf{Finding an explicit solution to a partial differential equation (PDE), which may be difficult or even impossible to solve, is a crucial stage in the different constructive approaches to stabilizing the systems in upper triangular form. This function must be numerically evaluated or analytically approximated.} Several nested manifolds are enforced by increasing gains leading to multiple time scales in recursive designs. Because of the loss of robustness to high-frequency unmodeled dynamics, the high-gain nature of the recursive designs may lead to instability. Therefore, the saturation design  for upper-triangular system stabilization \cite{teel1, teel2} is based on the system's Jacobian linearization and utilizes low-gain and saturated feedback.

\subsection{Interlaced System}
Backstepping and forwarding are utilized to develop sequentially designed feedback control laws for the global stabilization of nonlinear feedback and feedforward systems. Interlaced systems extend the class of strict feedback and strict feedforward as they can be described as a scalar subsystem with many top-down and/or bottom-up augmentations. In an interlaced system, the lower triangular or upper triangular configurations cannot be exploited. The fundamental challenge in constructing or developing a Lyapunov function is that it only globally stabilize the closed-loop system, not making a system globally asymptotically stable. Therefore, simple composite Lyapunov functions are inadequate and are not suitable. A solution via the cross-term for a more general Lyapunov function rather than a simple composite Lyapunov function is presented in \cite{Sepulchre2011ConstructiveNC} to ensure the GAS.  A blend of backstepping, forwarding, and P\&I is used to achieve the global stabilization of a wider class of interlaced systems. Furthermore, instead of calculating the integral form of the system, the off-the-manifold is obtained directly in terms of the interconnection term and then used for the Lyapunov function.

\begin{example}\label{interexample}
To begin with, a third-order interlaced system 
\begin{align}\label{interlaced system sel}
    \begin{split}
        \dt{\mathrm{x}}_1&=\mathrm{x}_2+\mathrm{x}_2\mathrm{x}_3\\
\dt{\mathrm{x}}_2&=\mathrm{x}_3+\mathrm{x}_2^2\\
\dt{\mathrm{x}}_3&=\mathrm{x}_1\mathrm{x}_2\mathrm{x}_3+u
    \end{split}
\end{align}
from \cite{Sepulchre2011ConstructiveNC} is considered. The presented system is not in feedback form because of $\mathrm{x}_2 \mathrm{x}_3$, nor is it in feedforward form, due to the nonlinear terms $\mathrm{x}_1\mathrm{x}_2\mathrm{x}_3$ and $\mathrm{x}_2^2$. Because $\mathrm{x}_3$ is available as a virtual input in $\dt{\mathrm{x}}_2$, the middle equation is chosen instead of the top equation. The storage function $\mathbb{S}_1=\frac{1}{2}\mathrm{x}_2^2$ for scalar subsystem $\dt{\mathrm{x}}_2=\mathrm{x}_3+\mathrm{x}_2^2$ with $\dt{\mathbb{S}}_1=\mathrm{x}_2(\mathrm{x}_3+\mathrm{x}_2^2)\Rightarrow \mathrm{x}_3=-\mathrm{x}_2-\mathrm{x}_2^2$ for $\dt{\mathbb{S}}_1\leq 0$ renders a virtual control law $\mathrm{x}_3=-\mathrm{x}_2-\mathrm{x}_2^2$ . The substitution of virtual law $\mathrm{x}_3$ modifies the top equation as
\begin{equation}
    \dt{\mathrm{x}}_1=\mathrm{x}_2-\mathrm{x}_2^2-\mathrm{x}_2^3
\end{equation}
The goal is to stabilize $\dt{\mathrm{x}}_1$ as $\dt{\mathrm{x}}_1=-\mathrm{x}_1$
and it gives the off-the manifold $\mathrm{x}_1+\mathrm{x}_2-\mathrm{x}_2^2-\mathrm{x}_2^3=0$. For ease of calculation, the corrected off-the-manifold is given by
\begin{equation}
    \mathrm{x}_1+\mathrm{x}_2-\frac{1}{2}\mathrm{x}_2^2-\frac{1}{3}\mathrm{x}_2^3=0
\end{equation}
The virtual control law $\mathrm{x}_3=-\mathrm{x}_2-\mathrm{x}_2^2$  is responsible for stabilization of $\dt{\mathrm{x}}_2$.  In order to stabilize $\dt{\mathrm{x}}_1$ and $\dt{\mathrm{x}}_2$, the virtual law is modified as $\mathrm{x}_3=-\mathrm{x}_2-\mathrm{x}_2^2+v$. One step of forwarding to stabilize $\dt{\mathrm{x}}_2$ augmented by the top equation of  $\dt{\mathrm{x}}_1$ with new virtual law is employed to get
\begin{align}
    \begin{split}
        \dt{\mathrm{x}}_1&=\mathrm{x}_2-\mathrm{x}_2^2-\mathrm{x}_2^3+\mathrm{x}_2v\\
        \dt{\mathrm{x}}_2&=\mathrm{x}_2+v.
    \end{split}
\end{align}
The forwarding yields the Lyapunov function with the off-the-manifold
\begin{equation}\label{Lyapunov_Interlaced}
    \mathbb{S}_2=\mathbb{S}_1+\frac{1}{2}\underset{\mathrm{corrected \hspace{0.1cm}off-the-manifold}}{\underbrace{(\mathrm{x}_1+\mathrm{x}_2-\frac{1}{2}\mathrm{x}_2^2-\frac{1}{3}\mathrm{x}_2^3)^2}}
\end{equation}
The additional feedback $v$
\begin{equation}
    v=-(1-\mathrm{x}_2^2)(\mathrm{x}_1+\mathrm{x}_2-\frac{1}{2}\mathrm{x}_2^2-\frac{1}{3}\mathrm{x}_2^3)
\end{equation}
is obtained by solving 
\begin{equation*}
    \dt{\mathbb{S}}_2=\mathrm{x}_2\dt{\mathrm{x}}_2+(\mathrm{x}_1+\mathrm{x}_2-\frac{1}{2}\mathrm{x}_2^2-\frac{1}{3}\mathrm{x}_2^3)(\dt{\mathrm{x}}_1+\dt{\mathrm{x}}_2-\mathrm{x}_2\dt{\mathrm{x}}_2-\mathrm{x}_2^2\dt{\mathrm{x}}_2)
\end{equation*}
$\dt{\mathbb{S}}_2\leq0$ for (\ref{Lyapunov_Interlaced}).
The augmented virtual control law $\mathrm{x}_3$  
\begin{equation}
    \mathrm{x}_3=-\mathrm{x}_2-\mathrm{x}_2^2-(1-\mathrm{x}_2^2)(\mathrm{x}_1+\mathrm{x}_2-\frac{1}{2}\mathrm{x}_2^2-\frac{1}{3}\mathrm{x}_2^3)
\end{equation}
with additional feedback to stabilize $\dt{\mathrm{x}}_1$ and $\dt{\mathrm{x}}_2$ is used for the off-the-manifold as
\begin{equation}\label{offmanifoldnomultiplicative}
    \mathrm{x}_3+\mathrm{x}_2+\mathrm{x}_2^2+(1-\mathrm{x}_2^2)(\mathrm{x}_1+\mathrm{x}_2-\frac{1}{2}\mathrm{x}_2^2-\frac{1}{3}\mathrm{x}_2^3)=0
\end{equation}
to apply proposed P\&I approach (\ref{GammaPassive}-\ref{final control law}).
The control law 
\begin{small}
\begin{align}
    \begin{split}
        u&=-\mathrm{x}_1\mathrm{x}_2\mathrm{x}_3- [ \dt{\mathrm{x}}_2+2\mathrm{x}_2\dt{\mathrm{x}}_2+\left ( \dt{\mathrm{x}}_1+\dt{\mathrm{x}}_2-\mathrm{x}_2\dt{\mathrm{x}}_2-\mathrm{x}_2^2\dt{\mathrm{x}}_2 \right )\\&(1-\mathrm{x}_2^2) +(\mathrm{x}_1+\mathrm{x}_2-\frac{1}{2}\mathrm{x}_2^2-\frac{1}{3}\mathrm{x}_2^3)(-2\mathrm{x}_2\dt{\mathrm{x}}_2)+\\& \frac{\alpha}{2}(\mathrm{x}_3+\mathrm{x}_2+\mathrm{x}_2^2+(1-\mathrm{x}_2^2))(\mathrm{x}_1+\mathrm{x}_2-\frac{1}{2}\mathrm{x}_2^2-\frac{1}{3}\mathrm{x}_2^3)]
    \end{split}
\end{align}
\end{small}
applied to the system (\ref{interlaced system sel}) ensures GAS of the zero equilibrium.
\end{example}

In the above Example \ref{interexample}, the off-the-manifold (\ref{offmanifoldnomultiplicative}) consists of pure terms like $\mathrm{x_1}$, $\mathrm{x_2}$, $\mathrm{x_2}^2$, and $\mathrm{x_2}^3$. There are no nonlinear multiplicative terms involved in the storage function. Therefore, the explicit solution of $\dt{\mathbb{S}}\leq -\alpha\mathbb{S}$ is easily obtained.
%\section{The proposed P\&I approach and Parameter Independent Implicit Manifold (PIIM) for uncertain systems}
\section{Orbital Stabilization via the P\&I approach}\label{sec7}
In all the examples mentioned above, one deals with the problem of stabilization of equilibrium points in the desired
equilibrium for the system in the stabilization and adaptive control scenarios, or the zero equilibrium for the state
estimation error in observer design.  In some modern applications, for example,  DC-to-AC power, walking robots, electric motors
converters,  and oscillation mechanisms in biology the final objective is to induce a periodic orbit. The main objective of this section is to show that the proposed P\&I approach can also be applied to solve this new problem,
that is, the generation of attractive periodic solutions. The only modification required is in the definition of the target
dynamics that, instead of having an asymptotically stable equilibrium, should be chosen with attractive periodic
orbits.

After a change of coordinates and scaling of the input, the dynamic equations of the Inertia Wheel Pendulum (IWP) \cite{ortega2020orbital} are given by:
\begin{align}
    \begin{split}
         \mathrm{\dt{x}_{1}}&=\mathrm{x_3}\\
 \mathrm{\dt{x}_{2}}&=\mathrm{x_4}\\
 \mathrm{\dt{x}_{3}}&=\theta_{\mathrm{m}} \mathrm{sin}\left (\mathrm{x_{1}} \right )-\mathrm{bu}\\
 \mathrm{\dt{x}_{4}}&=\mathrm{u}.
    \end{split}
\end{align}
 
A simple undamped pendulum behavior for the target dynamics \cite{ortega2020orbital} with  a constant $a$ defined as:
\begin{align}
    \begin{split}
        &\mathrm{\dt{\xi}_{1}}=\mathrm{{\xi}_{2}}\\
        &\mathrm{\dt{\xi}_{2}}=\mathrm{-asin\left ( \xi_1 \right )}.
    \end{split}
\end{align}
Consider the choice $\mathrm{x_{1}}=\mathrm{\xi_1}$ and $\mathrm{x_{2}}= \mathrm{k\xi_1}$. This choice renders the   implicit manifold $\Psi_1=\mathrm{x_2}=\mathrm{kx_1}\Rightarrow \mathrm{x_2}-\mathrm{kx_1}=\mathrm{0}$. Differentiating the above equation results into the $\mathrm{\dt{x}_2}=\mathrm{k\dt{x}_1}$ $\Rightarrow$ $\mathrm{{x}_4}=\mathrm{k{x}_3}\Rightarrow$ implicit manifold $\Psi_2=\mathrm{x_4}-\mathrm{kx_3}=\mathrm{0}$    

Adding implicit manifold $\Psi_1$ and implicit manifold $\Psi_2$ to get the combined implicit manifold
\begin{align}
    \mathrm{\Psi\left ( x \right )}\Rightarrow \mathrm{x_2}-\mathrm{kx_1}+\mathrm{x_4}-\mathrm{kx_3}=0
\end{align}
Tangent space structure for control system with this combined off-the-manifold, the normal vector direction is given by $\mathrm{\nabla}\mathrm{\Psi} \left ( \mathrm{x} \right )$ and one can define a pseudo Riemannian metric R\\
\begin{align}
    \begin{split}
         \mathrm{R}={\triangledown {\Psi}\left ( \mathrm{x} \right )^{T}}{\triangledown \Psi\left ( \mathrm{x} \right )}=\begin{bmatrix}
\mathrm{-k}\\ 
\mathrm{1}\\ 
\mathrm{-k}\\
\mathrm{1} 
\end{bmatrix}
\begin{bmatrix}
\mathrm{-k} \mathrm{1}  &\mathrm{-k}  &\mathrm{1} 
\end{bmatrix}
=\begin{bmatrix}
\mathrm{k^{2}} & \mathrm{-k} & \mathrm{k^{2}} & \mathrm{-k}\\ 
\mathrm{-k} & \mathrm{1} & \mathrm{-k} & \mathrm{1}\\ 
\mathrm{k^{2}} & \mathrm{-k} & \mathrm{k^{2}} & \mathrm{-k}\\ 
 \mathrm{-k}& \mathrm{1} & \mathrm{-k} & \mathrm{1}
\end{bmatrix}
    \end{split}
\end{align}
The passive output and corresponding storage function are defined based on the combined implicit manifold and a pseudo-Riemannian metric.
\begin{figure}
    \centering
    \includegraphics[width=\linewidth]{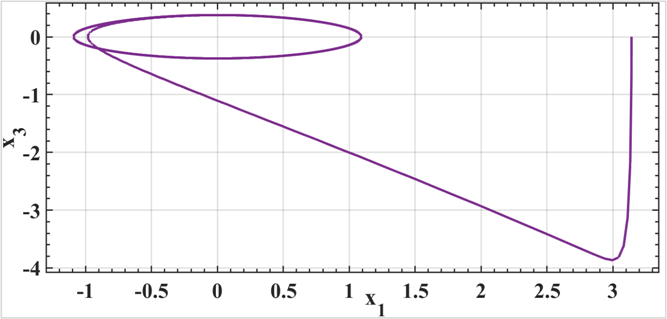}
        \caption{Plot of $\mathrm{x_1}$ vs $\mathrm{x_3}$ starting with the link hanging and lifting it to oscillate in the upper half-plane.}
    \label{orbital}
\end{figure}
\begin{figure}
    \centering
    \includegraphics[width=\linewidth]{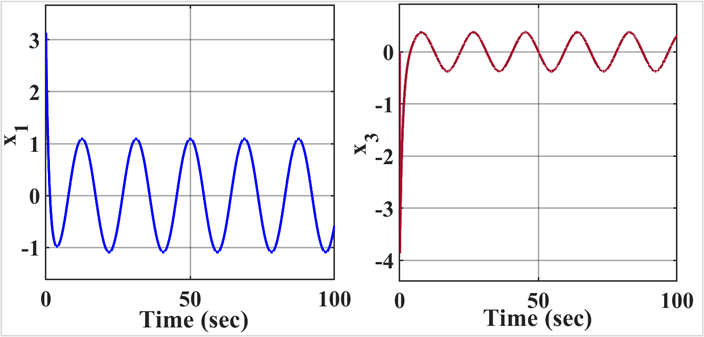}
        \caption{Plot of $\mathrm{x_1}$ and $\mathrm{x_3}$ vs time.}
    \label{orbital2}
\end{figure}
The convergence of the off-the-manifold dynamics to the implicit manifold at an exponential rate $\alpha$ is accompanied by selecting the condition (\ref{exponential laypunov}).
One can use the condition (\ref{exponential laypunov}) along with the storage function (\ref{Storage function}) and passive output (\ref{Gamma}) to get the final control law 
\begin{align}
    \begin{split}
        \mathrm{u}=\frac{-1}{\mathrm{\left ( 1+kb \right )}}\mathrm{\left ( x_{4}-kx_{3}-kmsinx_{1}+\frac{\alpha }{2} \left (x_{2}-kx_{1}+x_{4}-kx_{3}  \right )\right )}
    \end{split}
\end{align}
The plot of $\mathrm{x_1}$ vs $\mathrm{x_3}$ starting with the link hanging and lifting it to oscillate in the upper half plane is shown in Fig. \ref{orbital}. The time evolution of system trajectories is shown in Fig. \ref{orbital2}.
\section{Conclusions}\label{conclusion}
As an extended and enhanced version of the I\&I theory, the Passivity and Immersion based (P\&I) approach is proposed in this paper for the constructive stabilization and control of nonlinear systems based on the generation of a suitable passive output and corresponding storage function. The P\&I approach preserves many key properties of the existing control algorithms in the literature while being easier to interpret and implement. 

Target dynamics are set in a way that the parameter-independent combined implicit manifolds are obtained. It helps in getting the parameter-independent passive output and associated storage function. By adjusting the values of the flexible variables, the impact of the parametric changes in the system is mitigated. While keeping the system parameters to nominal values, the proposed P\&I controller exhibits system trajectories convergence with improved performance for both small and large parameter values in the controller without any parameter estimation. Being a constructive methodology, the various design paradigms such as Backstepping, Incremental Backstepping, Forwarding, I\&I, and the techniques based on the generation of CCM are unified in the P\&I methodology.

\textbf{The idea of the proposed P\&I approach is extended by adding a virtual control in the final equation of the  error dynamics of the classical gradient estimator in \cite{nayyer2022passivity}. The solution of the virtual control law is identified by the P\&I approach \cite{ShadabArxiv, nayyer2022passivitylcss}.  The P\&I approach \cite{ShadabArxiv} is based on the choice of an appropriate implicit manifold and the generation of a suitable passive output and a related storage function. This facilitates the virtual control law being obtained in a way that the parametric error converges asymptotically to zero.  Because the above ideas connect with the P\&I approach and GE, the developed methodology in \cite{nayyer2022passivity} is labeled the “passivity and immersion-based modified gradient estimator (MGE)”.}

The authors are also working on the development of the observer (state-estimator) via the proposed P\&I approach.

\section{Appendix 1: Theorem-Proof}\subsection{Splitting of Tangent Space}
Consider an n-dimensional manifold $\mathbb{M}$ with tangent bundle $\mathbb{T}_{\mathbb{M}}$, such that all $\mathrm{p}\in {\mathbb{M}}$, $\mathbb{T}_{\mathrm{p}}{\mathbb{M}}$ has the following structure
\begin{equation}
    \mathbb{T}_{\mathrm{p}}{\mathbb{M}} = \mathbb{H}_{\mathrm{p}} \oplus  \mathbb{V}_{\mathrm{p}}: \hspace{0.3cm}  \mathbb{H}_{\mathrm{p}} \cap   \mathbb{V}_{\mathrm{p}}=0.
\end{equation}
If $\mathbb{H}_{\mathrm{p}}$ horizontal space and $\mathbb{V}_{\mathrm{p}}$ the vertical space are considered then at all $\mathrm{p}\in\mathbb{M}$, $\mathbb{T}_{\mathrm{p}}{\mathbb{M}}$ is direct sum of $\mathbb{H}_{\mathrm{p}}$ and $\mathbb{V}_{\mathrm{p}}$. If $\mathbb{M}$ is coordinatized as $(\mathrm{x}, \lambda)$ with $\mathrm{x}\in \mathbb{R}^{\mathrm{k}}, \lambda \in \mathbb{R}^{\mathrm{n-k}}$, and $\mathrm{k}< \mathrm{n}$ then $\mathbb{T}_{\mathrm{p}}{\mathbb{M}}$ for any $\mathrm{p}\in\mathbb{M}$ is written as
\begin{equation}\label{TpM}
   \mathbb{T}_{\mathrm{p}}{\mathbb{M}} = \mathbb{H}_{\mathrm{p}} \oplus  \mathbb{V}_{\mathrm{p}}=(\dt{\mathrm{x}}, 0)\oplus (0, \dt{\lambda})=(\dt{\mathrm{x}},\dt{\lambda}).
\end{equation}
The interpretation of above decomposition  (\ref{TpM}) can be given in terms of $(\mathrm{M}, \chi)$, i.e., $\mathrm{M}$ is a Riemannian space with Identity $\chi$ as the Riemannian metric \cite{ShadabArxiv}. This gives
\begin{equation}
   {\left \langle \mathbb{H}_{\mathrm{p}},  \mathbb{V}_{\mathrm{p}} \right \rangle}_{\chi}= \begin{bmatrix}
\dt{\mathrm{x}} &0 
\end{bmatrix}\underset{\chi}{\underbrace{\begin{bmatrix}
\mathrm{I}^{\mathrm{k}\times \mathrm{k}} & 0\\ 
0 & \mathrm{I}^{\mathrm{n-k}\times \mathrm{n-k}}
\end{bmatrix}}}\begin{bmatrix}
0\\ 
\dt{\lambda}
\end{bmatrix}=0
\end{equation}

\begin{theorem}
For a given tangent vector $\mathrm{w}=(\dt{\mathrm{x}}, \dt{\lambda})\in \mathbb{T}_{\mathrm{p}}{\mathbb{M}}$, the orthogonality of  $\mathbb{H}_{\mathrm{p}}$ and $\mathbb{V}_{\mathrm{p}}$ is preserved under new pseudo-Riemannian metric $\mathrm{R}$.
\end{theorem}
\textbf{Proof:}   Suppose the metric $\chi$ is replaced with some other pseudo-Riemannian metric 
\begin{equation}
    \mathrm{R}=\begin{bmatrix}
\mathrm{\mathrm{m}}_{11} & \mathrm{\mathrm{m}}_{12}\\ 
 \mathrm{\mathrm{m}}_{21}&\mathrm{\mathrm{m}}_{22}
\end{bmatrix}
\end{equation}
(or (\ref{metric})) with $\mathrm{\mathrm{m}}_{12}=\mathrm{\mathrm{m}}_{21}$. Then to ensure that $\mathbb{T}_{\mathrm{p}}{\mathbb{M}} =\widetilde{\mathbb{H}}_{\mathrm{p}} \oplus \widetilde{\mathbb{V}}_{\mathrm{p}}$,  the following is the structure of $\widetilde{\mathbb{H}}_{\mathrm{p}}$ and $\widetilde{\mathbb{V}}_{\mathrm{p}}$ under $\mathrm{R}$
\begin{small}
\begin{equation}
    \mathbb{T}_{\mathrm{p}}{\mathbb{M}}\subset (\dt{\mathrm{x}},\dt{\lambda})=\left (\dt{\mathrm{x}}, -\mathrm{\mathrm{m}}_{22}^{-1}\mathrm{\mathrm{m}}_{21} \dt{\mathrm{x}} \right )\oplus \left (0, \dt{\lambda}+\mathrm{\mathrm{m}}_{22}^{-1}\mathrm{\mathrm{m}}_{21} \dt{\mathrm{x}} \right )
\end{equation}
\end{small}
becomes
\begin{small}
\begin{equation*}
    {\left \langle \mathbb{H}_{\mathrm{p}},  \mathbb{V}_{\mathrm{p}} \right \rangle}_\mathrm{R}= \begin{bmatrix}
\dt{\mathrm{x}} &-\mathrm{\mathrm{m}}_{22}^{-1}\mathrm{w}_{21}\dt{\mathrm{x}}
\end{bmatrix}\begin{bmatrix}
\mathrm{\mathrm{m}}_{11} & \mathrm{\mathrm{m}}_{12}\\ 
 \mathrm{\mathrm{m}}_{21}&\mathrm{\mathrm{m}}_{22}
\end{bmatrix}\begin{bmatrix}
0\\ 
\dt{\lambda}+\mathrm{\mathrm{m}}_{22}^{-1}\mathrm{\mathrm{m}}_{21}\dt{\mathrm{x}}
\end{bmatrix}
\end{equation*}
\end{small}
\begin{equation}
    = \mathrm{\mathrm{m}}_{21}\dt{\mathrm{x}}(\dt{\lambda}+\mathrm{\mathrm{m}}_{22}^{-1}\mathrm{\mathrm{m}}_{21}\dt{\mathrm{x}})-\mathrm{\mathrm{m}}_{21}\dt{\mathrm{x}}(\dt{\lambda}+\mathrm{\mathrm{m}}_{22}^{-1}\mathrm{\mathrm{m}}_{21}\dt{\mathrm{x}})=0
\end{equation}
\begin{remark}
The above splitting of the tangent space under a metric $\mathrm{R}$ is closely related to the theory of fiber bundles and Ehresmann connection.
\end{remark}
\begin{remark}
In differential geometry, a PR manifold is a differentiable manifold with a metric tensor that is everywhere nondegenerate. This is a generalization of a Riemannian manifold in which the requirement of positive-definiteness is relaxed.
\end{remark}
\subsection{Passive output and storage function}
\begin{lemma}
The obtained $\mathrm{y}=(\lambda+\mathrm{q(x)})$ in (\ref{Gamma}) is a passive output and associated function (\ref{Storage function}) is a storage function  with respect to  new input $\mathrm{v}$ and $\mathrm{y}$. 
\end{lemma}

\textbf{Proof:} The time-derivative of function (\ref{Storage function}) 
\begin{small}
\begin{equation}
   \dt{\mathbb{S}}= (\lambda+\mathrm{q(x)})(\dt{\lambda}+\frac{\partial \mathrm{q(x)}}{\partial \mathrm{x}}\dt{\mathrm{x}})=(\lambda+\mathrm{q(x)})(\mathrm{u}+\frac{\partial \mathrm{q(x)}}{\partial \mathrm{x}}\mathrm{f(x,\lambda)})
\end{equation}
\end{small}
is translated into
\begin{align}\label{khdsdfs}
    \begin{split}
         \dt{\mathbb{S}} %(\lambda+\mathrm{q(x)})\mathrm{u}+(\lambda+\mathrm{q(x)})\frac{\partial \mathrm{q(x)}}{\partial \mathrm{x}}\mathrm{f(x,\lambda)}\\
         =\mathrm{y^T}\mathrm{u}+\mathrm{y^T}\frac{\partial \mathrm{q(x)}}{\partial \mathrm{x}}\mathrm{f(x,\lambda)}
    \end{split}
\end{align}
with $\mathrm{y}=(\lambda+\mathrm{q(x)})$. Here, $\mathrm{y}\in \mathbb{R}^{\mathrm{n-k}}$ and the dimension of $\mathrm{y}, \mathrm{u}$, and $\mathrm{v}$ is same.   By substituting the control law 
\begin{equation}\label{final control law stable}
  \mathrm{u}=-\frac{\partial\mathrm{q(x)}}{\partial \mathrm{x}}\mathrm{f(x,\lambda)}+\mathrm{v}
\end{equation} 
in (\ref{khdsdfs}), it becomes passive with respect to  new input $\mathrm{v}$ and $\mathrm{y}$ due to 
\begin{equation}
    \dt{\mathbb{S}} \leq \mathrm{y^T v }. 
\end{equation}
Therefore, the $\mathbb{S}$ is called as the storage function with $\mathrm{y}=(\lambda+\mathrm{q(x)})$ as a passive output.

\begin{remark}
Since $\mathbb{S}$ is positive definite outside the manifold $\mathbb{M}$ and  $\mathbb{S}(0,0)=0$ on the manifold, the obtained storage function (\ref{Storage function}) can be used as the candidate Lyapunov function to prove the attractivity of the manifold.
\end{remark}

\section{Appendix 2: Preliminaries}\label{appen2}

\subsection{The I\&I based methodologies}
The I\&I exploits differential geometry theory to derive the (invariant) output zeroing manifold and then entails immersing the specified plant dynamics in a strictly lower-order target system. As the foundation of the I\&I approach relies on the concepts of system immersion and manifold invariance, therefore, does not necessitate the knowledge of a CLF. Two different approaches in the existing literature are reviewed briefly based on the I\&I notion.
\subsubsection{The classical I\&I \cite{Astolfi}}
Consider the system
\begin{equation}\label{eq1}
    \dt{\mathrm{x}}=\mathrm{f}(\mathrm{x})+\mathrm{g} (\mathrm{x})u,
\end{equation}
with the system states $\mathrm{x} \in \mathbb{R}^n$, inputs $u \in \mathbb{R}^m$, and $\mathrm{x}^*$  an equilibrium point to be stabilized. 
\begin{prop}\label{prop1}
Assume that there exist 
$\beta:\mathbb{R}^{\mathrm{h}}\rightarrow \mathbb{R}^{\mathrm{h}}$, $\nu: \mathbb{R}^{\mathrm{h}}\rightarrow \mathbb{R}^n$, $c: \mathbb{R}^{\mathrm{h}}\rightarrow \mathbb{R}^{\mathrm{m}}$,  $\Psi:\mathbb{R}^{\mathrm{n}}\rightarrow \mathbb{R}^{\mathrm{n-h}}$, $\pi:\mathbb{R}^{\mathrm{n}} \times \mathbb{R}^{\mathrm{n-h}}\rightarrow \mathbb{R}^{\mathrm{m}}$
with the $\mathrm{h < n}$ such that the following hold.
\\
$(S_1)$   The target dynamical system
\begin{equation}\label{target dynamics}
    \dt{\eta}=\beta(\eta)
\end{equation}
with some function $\beta$ and $\eta \in \mathbb{R}^{\mathrm{h}}$ has a Globally Exponentially Stable (GES) equilibrium at $\eta^* \in \mathbb{R}^{\mathrm{h}}$ and  $\mathrm{x}^* = \nu(\eta^*)$.\\
$(S_2)$  For all $\eta \in \mathbb{R}^{\mathrm{h}}$,  
\begin{equation}
   \mathrm{f}(\nu (\eta))+\mathrm{g}(\nu (\eta))c(\nu (\eta))=\frac{\partial \nu }{\partial \eta}\beta(\eta).
\end{equation}
\\
$(S_3)$ In the implicit manifold condition, the following set identity holds

\begin{align}
\begin{split}
 \mathrm{M}= & \{ \mathrm{x} \in \mathbb{R}^n  | \Psi (\mathrm{x})=0 \}=\{ \mathrm{x} \in \mathbb{R}^n | \mathrm{x} \\& = \nu (\eta) \mathrm{\hspace{0.09cm}for \hspace{0.09cm}some\hspace{0.09cm} \eta \in \mathbb{R}^h}.
 \end{split}
\end{align}
\\
$(S_4)$ For manifold attractivity and trajectory boundedness, $z=\Psi(\mathrm{x})$ is considered and all the solution trajectories of
\begin{equation}
     \dt{z}=\frac{\partial \Psi }{\partial \mathrm{x}}(\mathrm{f}(\mathrm{x})+\mathrm{g} (\mathrm{x}) \pi (\mathrm{x}, z))
\end{equation}
\begin{equation}
    \dt{\mathrm{x}}=\mathrm{f}(\mathrm{x})+\mathrm{g} (\mathrm{x}) \pi (\mathrm{x}, z)
\end{equation}
are bounded and satisfy $\underset{t\rightarrow \infty}{\mathrm{lim}}z(t)=0$. Then $\mathrm{x}^*$ is a GAS equilibrium of closed-loop system $ \dt{\mathrm{x}}=\mathrm{f}(\mathrm{x})+\mathrm{g} (\mathrm{x}) \pi (\mathrm{x}, z)$.
\end{prop}
\subsubsection{I\&I horizontal contraction procedure (I\&I HCP) \cite{WangTAC}}
With the three steps $(S_1)-(S_3)$ in proposition (\ref{prop1}),  the step $(S_4)$ is replaced by a horizontal contraction based design to ensure the attractivity of the  manifold $\mathrm{M}$.
\begin{prop}\label{prop2}
  An alternative procedure is adopted in \cite{WangTAC} to carry out the step $(S^{\dagger}_4)$ of the classical I\&I method mentioned below:
 It is assumed that the following mappings 
 
  $\Xi:\mathbb{R}^{\mathrm{n}}\rightarrow \mathbb{R}^{\mathrm{(n-h)\times(n-h)}}$ with $\Xi=\Xi^{\mathrm{T}}> 0$,
  
  $\Sigma :\mathbb{R}^n\rightarrow \mathbb{R}^{\mathrm{(n-h)\times n}}$, $\Upsilon:\mathbb{R}^n\rightarrow \mathbb{R}^m$, $\tau:\mathbb{R}^n\rightarrow \mathbb{R}$\\
 are exists such that the following holds. 
  \\
 $(S_4^{\dagger})$ For manifold attractivity via horizontal contraction, the following steps are mentioned
  \begin{itemize}
      \item $\forall \hspace{0.02cm}\mathrm{x} \in \mathbb{R}^n$, $\Sigma(\mathrm{x})$ is full rank and 
      \begin{equation}
          \Sigma(\nu(\eta))=\triangledown \Psi(\nu(\eta)), \hspace{0.5cm}\forall \eta \in \mathbb{R}^h
      \end{equation}
      \item For all $ \hspace{0.02cm}\eta \in \mathbb{R}^h$
      \begin{equation}
          \Upsilon(\nu(\eta))=c(\nu(\eta)).
      \end{equation}
      \item The candidate Finsler-Lyapunov Function (FLF) $\mathbb{V}:\mathbb{R}^n \times  \mathbb{R}^n \rightarrow \mathbb{R}_{\geq 0}$ is selected as
      \begin{equation}\label{flfwang}
          \mathbb{V}(\mathrm{x}, \delta \mathrm{x}):= \delta \mathrm{x}^{\mathrm{T}}\Sigma^{\mathrm{T}} (\mathrm{x})\Xi (\mathrm{x}) \Sigma(\mathrm{x})\delta \mathrm{x}
      \end{equation}
      satisfies 
      \begin{equation}\label{prol1}
          \dt{\mathbb{V}}(\mathrm{x}, \delta \mathrm{x})\leq -\tau (\mathrm{x}) {\mathbb{V}}(\mathrm{x}, \delta \mathrm{x})
      \end{equation}
      along the trajectories of the \textit{prolonged system}
      \begin{equation}\label{jgfsgf}
    \dt{\mathrm{x}}=\mathrm{f}(\mathrm{x})+\mathrm{g} (\mathrm{x}) \Upsilon  (\mathrm{x})
\end{equation}
\begin{equation}\label{prol22}
   \dt{\delta \mathrm{x}}= \underset{\sigma ( \mathrm{x})  }{\underbrace{\triangledown \left [\mathrm{f}(\mathrm{x})+\mathrm{g}(\mathrm{x}) \Upsilon  (\mathrm{x}) \right ]}}\delta \mathrm{x}\Rightarrow \frac{\mathrm{d} }{\mathrm{d} \mathrm{t}}\delta \mathrm{x}=\sigma ( \mathrm{x}) \delta \mathrm{x}
\end{equation}
\item The solution trajectories of the system (\ref{jgfsgf}) are bounded and $\mathrm{x}^*$ is a GAS equilibrium of (\ref{eq1}).
  \end{itemize}
\end{prop}
\subsection{Control Contraction Metrics (CCM)}
Constructing a distance-like function that must decrease over time in order to ensure stability is necessary for Lyapunov stability. Also, the design of a CLF entails addressing a non-convex optimization problem and is strongly dependent on the structural properties of the system in consideration \cite{slotinelopez2019contraction}. The contraction analysis for the stability of NLS is based on the notion of differential dynamics (such as a variational system), where stability is defined incrementally between two arbitrary trajectories \cite{IanManchester}. In \cite{IanManchester}, the concept of a CCM for NLS is proposed, and it is proved that the existence of a CCM is sufficient to guarantee universal exponential stabilizability. The derived criteria for the existence of CCM in \cite{IanManchester} can be framed as a convex feasibility problem, and the CCM stabilizability condition can be thought of as a differential version of the CLF condition \cite{IanManchester}.

Consider a system
\begin{equation}\label{CCMeq1}
    \dt{\mathrm{x}}=\mathrm{f}(\mathrm{x})+\mathrm{B} (\mathrm{x})u,
\end{equation}
with the system states $\mathrm{x} \in \mathbb{R}^n$, control input matrix $\mathrm{B} (\mathrm{x})$ with columns $\mathrm{b}_i$ for $i=1,...,m$,   inputs $u \in \mathbb{R}^m$, and $\mathrm{x}^*$  an equilibrium point to be stabilized. The associated differential dynamics
\begin{equation}\label{CCMeq2}
    \dt{\delta}_{\mathrm{x}}=\mathrm{A}(\mathrm{x},\mathrm{u})\delta_{\mathrm{x}}+\mathrm{B}(\mathrm{x})\delta_{\mathrm{u}},
\end{equation}
with  $\mathrm{A}(\mathrm{x},\mathrm{u})\delta_{\mathrm{x}}=\frac{\partial \mathrm{f}}{\partial x}+\sum_{i=1}^{m}\frac{\partial \mathrm{b}_i}{\partial x}\mathrm{u}_i$. It is considered that the function ${\delta}_v={\delta}_{\mathrm{x}}^T(\mathfrak{M}(\mathrm{x})){\delta}_{\mathrm{x}}$ is viewed as differential Riemannian energy at point $\mathrm{x}$ with $\mathfrak{M}(\mathrm{x})$ as a Riemannian metric. Differentiating and imposing that the differential Riemannian energy decreases with the rate $\alpha$, one gets
\begin{small}
\begin{equation}
    \dt{\delta}_v={\delta}_{\mathrm{x}}^T(\mathrm{A}^T\mathrm{R}+\mathrm{R}\mathrm{A}+\dt{\mathfrak{M}}){\delta}_{\mathrm{x}}+2{\delta}_{\mathrm{x}}^T {\mathrm{R}\mathrm{B}}{\delta}_{\mathrm{u}}\leq -2\alpha {\delta}_{\mathrm{x}}^T \mathfrak{M} {\delta}_{\mathrm{x}},
\end{equation}
\end{small}
where, $\dt{\mathfrak{M}}(\mathrm{x})=\frac{\partial \mathfrak{M}}{\partial t}+\sum_{i=1}^{n}\frac{\partial \mathfrak{M}}{\partial x_i}\mathrm{f}_i(\mathrm{x})$.
\begin{theorem} 
If there exist a uniformly bounded metric $\mathfrak{M}\mathrm{(x)}$ such that the following implication is true
\begin{equation}
    {\delta}_{\mathrm{x}}^T\mathfrak{M}\mathrm{B}=0\Rightarrow {\delta}_{\mathrm{x}}^T(\mathrm{A}^T\mathfrak{M}+\mathfrak{M}\mathrm{A}+\dt{\mathfrak{M}}+2 \alpha \mathfrak{M}){\delta}_{\mathrm{x}}\leq 0,
\end{equation}
for all ${\delta}_{\mathrm{x}}\neq 0$, $\mathrm{x}$, $\mathrm{u}$ then the system (\ref{CCMeq1}) is universally exponentially stabilizable via continuous feedback defined almost everywhere, and everywhere in neighborhood of the target trajectory \cite{IanManchester}.
\end{theorem}
Proof of Theorem 1:  Detailed proof can be found in \cite{IanManchester}.

\bibliography{references.bib}
\bibliographystyle{IEEEtran}

\end{document}